\documentclass[11pt,fleqn]{article}

\usepackage{a4}
\usepackage{amstext}
\usepackage{amsfonts}
\usepackage{amssymb}
\usepackage{booktabs}
\usepackage{color}
\usepackage{caption}         
\usepackage{cite}
\usepackage{dsfont}
\usepackage{epsfig}
\usepackage{mathtools}
\usepackage{multirow}
\usepackage{physics}
\usepackage{ulem}
\usepackage{subcaption}   
\usepackage{hyperref}
\usepackage{cases}
\usepackage{array}
\usepackage{comment}
\usepackage[
linecolor=orange!80!white, backgroundcolor=orange!10!white,
bordercolor=red, textsize=small,]{todonotes}
\usepackage{placeins}
\usepackage{colortbl}

\newcommand{\ltapprox}{\raisebox{-0.5ex}{$\,\stackrel{<}{\scriptstyle\sim}\,$}}
\newcommand{\RN}[1]{\uppercase\expandafter{\romannumeral#1}}
\newcolumntype{C}{>{\centering\arraybackslash}m{8.4em}}

\definecolor{col_tutorials}{rgb}{0.0,0.0,1.0}

\begin{document}

	% ********************
	% ********************
	% ********************
	% ********************
	% ********************

		\begin{center}
		
		{\huge \bf Gluelump masses and mass splittings from}
		
		{\huge \bf SU(3) lattice gauge theory}

		\vspace{0.5cm}
		
		\textbf{Jannis Herr$^{a}$, Carolin Schlosser$^{a,b}$, Marc Wagner$^{a,b}$}
		
		$^a$~Goethe-Universit\"at Frankfurt am Main, Institut f\"ur Theoretische Physik, Max-von-Laue-Stra{\ss}e 1, D-60438 Frankfurt am Main, Germany \\
		$^b$~Helmholtz Research Academy Hesse for FAIR, Campus Riedberg, Max-von-Laue-Stra{\ss}e 12, D-60438 Frankfurt am Main, Germany
		
		\vspace{0.5cm}
		
		June 16, 2023
		
	\end{center}
	
	\begin{tabular*}{16cm}{l@{\extracolsep{\fill}}r} \hline \end{tabular*}
	
	\vspace{-0.4cm}
	\begin{center} \textbf{Abstract} 
		
	\end{center}
	\vspace{-0.4cm}
	
	We compute gluelump masses and mass differences using SU(3) lattice gauge theory. We study states with total angular momentum up to $J = 3$, parity $P = +,-$ and charge conjugation $C = +,-$. Computations on four ensembles with rather fine lattice spacings in the range $0.040 \, \text{fm} \ldots 0.093 \, \text{fm}$ allow continuum extrapolations of gluelump mass differences. We complement existing results on hybrid static potentials with the obtained gluelump masses, which represent the limit of vanishing quark-antiquark separation. We also discuss the conversion of lattice gluelump masses to the Renormalon Subtracted scheme, which is e.g.\ important for studies of heavy hybrid mesons in the Born-Oppenheimer approximation.
	
	\begin{tabular*}{16cm}{l@{\extracolsep{\fill}}r} \hline \end{tabular*}
	
	\thispagestyle{empty}

	% ********************
	% ********************
	% ********************
	% ********************
	% ********************

	\newpage
	
	\setcounter{page}{1}
	
	\section{Introduction}
	
	Gluelumps are color-neutral states composed of a static adjoint color charge and gluons. Even though they do not seem to exist in nature, they are conceptually interesting and relevant in the context of certain QCD-calculations. An example of the latter is the Born-Oppenheimer effective field for heavy hybrid mesons, where gluelump masses are the non-perturbative matching coefficients for hybrid static potentials. Gluelump masses have to be provided as input, when computing the spectrum of heavy hybrid mesons (see e.g.\ Ref.~\cite{Berwein:2015vca}).
	Beyond the Standard Model, gluelumps are candidates for additional bound states, which contain gluinos, the counterparts of gluons in supersymmetric models.
	Because of this, gluelump masses help to investigate the spectrum of states in supersymmetric theories (see e.g.\ Ref.~\cite{ATLAS:2019duq}).
	
	Gluelumps were studied within models or using simplifying approximations of QCD, e.g.\ the bag model, potential models and Coulomb gauge QCD via the variational approach (see e.g.\ Refs.~\cite{Karl:1999wq,Guo:2007sm,Buisseret:2008pd,Simonov:2000ky,Mathieu:2005wc}). The resulting spectra, however, exhibit sizable discrepancies. Lattice gauge theory, on the other hand, is a first-principles approach, where the underlying quantum field theory, typically SU(3) gauge theory, is solved numerically on a hypercubic periodic spacetime lattice. The finite lattice spacing can be varied and the functional dependence of physical observables like gluelump masses or mass splittings is known, which allows trustworthy extrapolations to the continuum. Thus, lattice gauge theory is the ideal tool to compute the spectrum of gluelumps in a rigorous and reliable way.
	
	At the moment only few lattice computations of gluelump spectra exist in the literature.
	In Ref.~\cite{Foster:1998wu} masses of 10 gluelump states were computed in SU(3) gauge theory and 5 gluelump mass splittings were extrapolated to the continuum.
	While gluelump mass splittings are scheme independent, gluelump masses depend on the regularization scheme and the value of the regulator, e.g.\ in lattice gauge theory they diverge in the continuum limit.
	In Ref.~\cite{Bali:2003jq} the lightest gluelump mass, which is conventionally taken as reference mass, was, thus, converted to the Renormalon Subtracted (RS) scheme, which is common in perturbative calculations.
	The results of Refs.~\cite{Foster:1998wu,Bali:2003jq} are frequently used, for example to compare with model predictions or when computing heavy hybrid meson spectra in Born-Oppenheimer effective field theory.
	However, the precision of such applications is not only limited by perturbative systematics, but also by the somewhat outdated lattice data from Ref.~\cite{Foster:1998wu}, which was generated around 25 years ago.
	In a more recent lattice study \cite{Marsh:2013xsa} of the gluelump spectrum, masses of 20 gluelump states in the color octet representation were determined (and an even larger number in higher color representations).
	In that work, however, full QCD was used, not SU(3) gauge theory without dynamical quarks as in Refs.~\cite{Foster:1998wu,Bali:2003jq}. As a consequence, mixings with static adjoint mesons, which are states composed of a static adjoint color charge and a light quark-antiquark pair, are possible.
	Even though such mixings could be small, sea quarks are expected to have non-negligible effects on gluelump masses. Thus, even though technically more advanced, Ref.~\cite{Marsh:2013xsa} cannot be compared quantitatively to Ref.~\cite{Foster:1998wu}, nor can it replace Ref.~\cite{Foster:1998wu}.
	
	The aim of this work is to carry out an up-to-date precision computation of the gluelump spectrum in SU(3) lattice gauge theory, i.e.\ without dynamical quarks.
	We use four different small lattice spacings, compute for each of them 20 gluelump masses and are able to extrapolate 19 gluelump mass splittings to the continuum. 
	In addition to statistical errors, we also estimate systematic errors associated with extracting the asymptotic exponential behaviors of correlation functions as well as with the continuum extrapolations.
	We also repeat the conversion of the lowest lattice gluelump mass to the RS scheme following conceptually  Ref.~\cite{Bali:2003jq}, but using our new lattice data instead of the results from Ref.\cite{Foster:1998wu}.
	While this leads to higher precision, we show that the error of the RS gluelump mass is currently dominated by uncertainties on the perturbative side.
	
	This paper is structured in the following way. In Section~\ref{sec:quantumnumbers_and_correlationfcts} theoretical basics are discussed including gluelump quantum numbers, operators and correlation functions. Section~\ref{sec:computationaldetails} is devoted to the lattice setup and other computational details. In Section~\ref{sec:numericalresults}, which is the main section of this work, we present and discuss our results, in particular gluelump masses and gluelump mass splittings. For the latter we carry out continuum extrapolations. We also discuss in detail the assignment of continuum total angular momentum $J$, which is not obvious, because rotational symmetry is broken by the hypercubic lattice and there are cases of competing states with different $J$ in the same cubic representation. Moreover, we compare to existing lattice results \cite{Foster:1998wu,Marsh:2013xsa} and we complement our previous lattice results \cite{Capitani:2018rox,Schlosser:2021wnr} on hybrid static potentials, since gluelump masses can be interpreted as the limit of vanishing quark-antiquark separation of such potentials. Finally, we convert the lightest gluelump mass from the lattice to the RS scheme, as outlined in the previous paragraph. We conclude in Section~\ref{sec:conclusions}.

	% ********************
	% ********************
	% ********************
	% ********************
	% ********************

	\clearpage
	
	\section{Gluelump quantum numbers, operators, and correlation functions}\label{sec:quantumnumbers_and_correlationfcts}
	
	In the continuum, gluelumps are characterized by quantum numbers $J^{PC}$. $J$ is the total angular momentum of the gluons with respect to the position of the static adjoint quark and $P$ and $C$ denote parity and charge conjugation.

	A cubic lattice breaks rotational symmetry. The remaining symmetry group is the full cubic group $O_h$. The elements of this group are combinations of discrete $90^\circ$ rotations and spatial reflection.
	Lattice gluelumps can, thus, be classified according to the four 1-dimensional irreducible representations $\mathcal{R}^{PC} = A_1^{\pm}$, $A_2^{\pm}$ ,the two 2-dimensional irreducible representations, $\mathcal{R}^{PC} = E^{\pm}$, and the four 3-dimensional irreducible representations, $\mathcal{R}^{PC} = T_1^{\pm}$, $T_2^{\pm}$.
	Since each representation of the full cubic group corresponds to an infinite number of representations of the continuous rotation group, the identification and assignment of total angular momentum $J$ to gluelump states obtained by a lattice computation is a non-trivial task (see Section~\ref{sec:spinidentification} for a detailed discussion).
	
	We compute gluelump masses from temporal correlation functions
	\begin{equation}\label{eq:correlationfunction1}
		C_{\mathcal{R}^{PC}}(t_2-t_1) = H_{\mathcal{R}^{PC}}^{a}(\textbf{r}_Q;t_1) G^{a b} (\textbf{r}_Q;t_1,t_2) {H_{\mathcal{R}^{PC}}^{b \dagger}}(\textbf{r}_Q;t_2).
	\end{equation}
	$\textbf{r}_Q$ denotes the spatial position of the static quark.
	Due to translational invariance, the correlation function does not depend on $\textbf{r}_Q$. Numerically, we can, thus, average the right hand side of Eq.~\eqref{eq:correlationfunction1} over all possible quark positions to increase statistical precision. To keep the notation simple, we omit the spatial coordinate $\textbf{r}_Q$ from now on.
	
	$G$ denotes the static quark propagator in the adjoint representation. 
	It is given by a product of adjoint temporal gauge links (represented in SU(3) gauge theory by $8 \times 8$ matrices, where rows and columns are labeled by upper indices $a, b, c, \ldots = 1,\ldots, 8$) connecting time $t_1$ and time $t_2$,
	\begin{equation}
		G^{a b} (t_1,t_2) = U_{t}^{(8),a c}(t_1) 	U_{t}^{(8),c d}(t_1+a) U_{t}^{(8),d e}(t_1+2a)\dots U_{t}^{(8),f b}(t_2).
	\end{equation}
	Adjoint gauge links are related to ordinary gauge links in the fundamental representation via $U_{t}^{(8),a b}= \text{Tr}[T^a U_t T^b U_{t}^{\dagger}]$, where $T^a=\lambda^a/\sqrt{2}$ are the SU(3) generators with the Gell-Mann matrices $\lambda^a$.
	
	The operators  $H_{\mathcal{R}^{PC}}$ at time $t_1$ and time $t_2$ contain gauge links in the fundamental representation generating gluons with definite lattice quantum numbers, i.e.\ excite the gluon field according to one of the irreducible representations of $O_h$ (see above).
	We employ operators $H_{\mathcal{R}^{PC}}$ constructed and discussed in detail in Ref.~\cite{Marsh:2013xsa}.
	
	The operators are suitable linear combinations of closed gauge link paths.
	There are 24 basic building blocks $L_n$, $n = 1,\ldots,24$, which have a chair-like shape, i.e.\ are $1 \times 2$ rectangles bent by $\pi/2$:
	\begin{align}
		\nonumber
		L_{1\phantom{0}} 	&= U_{+x}^N U_{+y}^N U_{+z}^N U_{-x}^N U_{-z}^N U_{-y}^N \, , \, &L_{2\phantom{0}} 		&= U_{-y}^N U_{+x}^N U_{+z}^N U_{+y}^N U_{-z}^N U_{-x}^N \, , \\\nonumber
		L_{3\phantom{0}} 	&= U_{-x}^N U_{-y}^N U_{+z}^N U_{+x}^N U_{-z}^N U_{+y}^N \, , \, &L_{4\phantom{0}} 		&= U_{+y}^N U_{-x}^N U_{+z}^N U_{-y}^N U_{-z}^N U_{+x}^N \, , \\\nonumber
		L_{5\phantom{0}} 	&= U_{+y}^N U_{+z}^N U_{+x}^N U_{-y}^N U_{-x}^N U_{-z}^N \, , \, &L_{6\phantom{0}} 		&= U_{+x}^N U_{+z}^N U_{-y}^N U_{-x}^N U_{+y}^N U_{-z}^N \, , \\\nonumber
		L_{7\phantom{0}} 	&= U_{-y}^N U_{+z}^N U_{-x}^N U_{+y}^N U_{-x}^N U_{-z}^N \, , \, &L_{8\phantom{0}} 		&= U_{-x}^N U_{+z}^N U_{+y}^N U_{+x}^N U_{-y}^N U_{-z}^N \, , \\\nonumber
		L_{9\phantom{0}} 	&= U_{+z}^N U_{+x}^N U_{+y}^N U_{-z}^N U_{-y}^N U_{-x}^N \, , \, &L_{10} 	&= U_{+z}^N U_{-y}^N U_{+x}^N U_{-z}^N U_{-x}^N U_{+y}^N \, , \\\nonumber
		L_{11} 	&= U_{+z}^N U_{-x}^N U_{-y}^N U_{-z}^N U_{+y}^N U_{+x}^N \, , \, &L_{12} 	&= U_{+z}^N U_{+y}^N U_{-x}^N U_{-z}^N U_{+x}^N U_{-y}^N \, , \\\nonumber
		L_{13} 	&= U_{-y}^N U_{-x}^N U_{-z}^N U_{+y}^N U_{+z}^N U_{+x}^N \, , \, &L_{14} 	&= U_{-x}^N U_{+y}^N U_{-z}^N U_{+x}^N U_{+z}^N U_{-y}^N \, , \\\nonumber
		L_{15} 	&= U_{+y}^N U_{+x}^N U_{-z}^N U_{-y}^N U_{+z}^N U_{-x}^N \, , \, &L_{16} 	&= U_{+x}^N U_{-y}^N U_{-z}^N U_{-x}^N U_{+z}^N U_{+y}^N \, , \\\nonumber
		L_{17} 	&= U_{-z}^N U_{-y}^N U_{-x}^N U_{+z}^N U_{+x}^N U_{+y}^N \, , \, &L_{18} 	&= U_{-z}^N U_{-x}^N U_{+y}^N U_{+z}^N U_{-y}^N U_{+x}^N \, , \\\nonumber
		L_{19} 	&= U_{-z}^N U_{+y}^N U_{+x}^N U_{+z}^N U_{-x}^N U_{-y}^N \, , \, &L_{20} 	&= U_{-z}^N U_{+x}^N U_{-y}^N U_{+z}^N U_{+y}^N U_{-x}^N \, , \\\nonumber
		L_{21} 	&= U_{-x}^N U_{-z}^N U_{-y}^N U_{+x}^N U_{+y}^N U_{+z}^N \, , \, &L_{22} 	&= U_{+y}^N U_{-z}^N U_{-x}^N U_{-y}^N U_{+x}^N U_{+z}^N \, , \\
		L_{23} 	&= U_{+x}^N U_{-z}^N U_{+y}^N U_{-x}^N U_{-y}^N U_{+z}^N \, , \, &L_{24} 	&= U_{-y}^N U_{-z}^N U_{+x}^N U_{+y}^N U_{-x}^N U_{+z}^N \, , \label{eq:L_ns}
	\end{align}
	with $U_{\pm j}^{N}$ denoting a product of $N$ gauge links in the fundamental representation in $\pm j$-direction. All $24$ chair-like building blocks are also defined in a graphical way in Figure~1 of Ref.~\cite{Marsh:2013xsa} (the red chair-shaped paths).
	Applying parity $\mathcal{P}$ (i.e.\ spatial reflections) and/or charge conjugation $\mathcal{C}$ (i.e.\ Hermitian conjugation) to these 24 building blocks leads to a total of 96 building blocks.
	
	Linear combinations of $L_n$ that correspond to the five representations, $A_1$, $A_2$, $T_1$, $T_2$ and $E$, have been worked out in Ref.~\cite{Marsh:2013xsa} and are given by
	\begin{align}
		H^a_{A_1} &= \left(\tilde{H}_{A_1}\right)_{\alpha \beta} T_{\alpha \beta}^a = \bigg(\sum_{n=1}^{24} L_n\bigg)_{\alpha \beta} T_{\alpha \beta}^a,\\
		H^a_{A_2} &= \left(\tilde{H}_{A_2}\right)_{\alpha \beta} T_{\alpha \beta}^a = \bigg(\sum_{n=1}^{12} (-1)^a L_n - \sum_{n=13}^{24} (-1)^a L_n\bigg)_{\alpha \beta} T_{\alpha \beta}^a\\
		H^a_{T_1^x} &= \left(\tilde{H}_{T_1^x}\right)_{\alpha \beta} T_{\alpha \beta}^a = \left( L_{6}+ L_{20}+ L_{21}+ L_{11}- L_{18}- L_{8}- L_{9}- L_{23} \right)_{\alpha \beta} T_{\alpha \beta}^a\\
		H^a_{T_1^y} &= \left(\tilde{H}_{T_1^y}\right)_{\alpha \beta} T_{\alpha \beta}^a = \left(L_{5}+ L_{19}+ L_{24}+ L_{10}- L_{17}- L_{7}- L_{12}- L_{22} \right)_{\alpha \beta} T_{\alpha \beta}^a\\
		H^a_{T_1^z} &= \left(\tilde{H}_{T_1^z}\right)_{\alpha \beta} T_{\alpha \beta}^a = \left( L_{1}+ L_{2}+ L_{3}+ L_{4}- L_{13}- L_{14}- L_{15}- L_{16} \right)_{\alpha \beta} T_{\alpha \beta}^a\\
		H^a_{T_2^x} &= \left(\tilde{H}_{T_2^x}\right)_{\alpha \beta} T_{\alpha \beta}^a = \left(L_{6}- L_{20}+ L_{21}- L_{11}+ L_{18}- L_{8}+ L_{9}- L_{23} \right)_{\alpha \beta} T_{\alpha \beta}^a\\
		H^a_{T_2^y} &= \left(\tilde{H}_{T_2^y}\right)_{\alpha \beta} T_{\alpha \beta}^a = \left(L_{5}- L_{19}+ L_{24}- L_{10}+ L_{17}- L_{7}+ L_{12}- L_{22} \right)_{\alpha \beta} T_{\alpha \beta}^a\\
		H^a_{T_2^z} &= \left(\tilde{H}_{T_2^z}\right)_{\alpha \beta} T_{\alpha \beta}^a = \left(L_{1}- L_{2}+ L_{3}- L_{4}+ L_{13}- L_{14}+ L_{15}- L_{16} \right)_{\alpha \beta} T_{\alpha \beta}^a\\
		H^a_{E_1} &= \left(\tilde{H}_{E^1}\right)_{\alpha \beta} T_{\alpha \beta}^a = \left(v^x-v^y \right)_{\alpha \beta} T_{\alpha \beta}^a\\
		H^a_{E_2} &= \left(\tilde{H}_{E^2}\right)_{\alpha \beta} T_{\alpha \beta}^a = \left(v^x+v^y-2v^z \right)_{\alpha \beta} T_{\alpha \beta}^a
	\end{align}
	with
	\begin{align}
		& v^x= L_{6}+ L_{20}+ L_{21}+ L_{11}+ L_{18}+ L_{8}+ L_{9}+ L_{23} \\
		& v^y=L_{5}+ L_{19}+ L_{24}+ L_{10}+ L_{17}+ L_{7}+ L_{12}+ L_{22} \\
		& v^z=L_{1}+ L_{2}+ L_{3}+ L_{4}+ L_{13}+ L_{14}+ L_{15}+ L_{16} ,
	\end{align}
	where lower indices $\alpha,\beta = 1,\ldots,3$ refer to the rows and columns of the $3 \times 3$ matrices $L_n$, which are defined in Eq.~(\ref{eq:L_ns}).
	An operator generating a state, which has also definite parity and charge conjugation, is given by
	\begin{eqnarray}
		H^a_{\mathcal{R}^{PC}} = H^a_{\mathcal{R}^{{\pm}{\pm}}} = \frac{1}{4} \Big(
		\Big(H^a_{\mathcal{R}} {\pm} ({\mathcal{P}} H^a_{\mathcal{R}}) \Big)
		\pm \mathcal{C} \Big(H^a_{\mathcal{R}} {\pm} ({\mathcal{P}} H^a_{\mathcal{R}}) \Big)
		\Big) .
	\end{eqnarray}
	
	The correlation function (\ref{eq:correlationfunction1}) can be simplified analytically,
	\begin{eqnarray}\label{eq:latticecorrelationfct}
		\nonumber & & \hspace{-0.7cm} C_{\mathcal{R}^{PC}}(t_2-t_1) = \\
		& & = \Tr \left[ \tilde{H}_{\mathcal{R}^{PC}}(t_1)Q(t_1,t_2)  \tilde{H}_{\mathcal{R}^{PC}}^{\dagger}(t_2) (Q(t_1,t_2))^{\dagger} \right]
		- \frac{1}{3}\Tr [ \tilde{H}_{\mathcal{R}^{PC}}(t_1) ]\Tr [\tilde{H}_{\mathcal{R}^{PC}}^{\dagger}(t_2) ] ,
	\end{eqnarray}
	by exploiting $T^a_{\alpha \beta}T^a_{\gamma \delta} = \delta_{\alpha \delta} \delta_{\beta \gamma} - \delta_{\alpha \beta} \delta_{\gamma \delta} / 3$.
	$Q(t_1,t_2) $ denotes a product of temporal gauge links in the fundamental representation connecting time $t_1$ and time $t_2$.
	
	To optimize the groundstate overlaps, we chose $N = 2$ (see Eq.~\eqref{eq:L_ns}) and apply APE smearing to the spatial gauge links appearing in the operators $\tilde{H}_{\mathcal{R}^{PC}}$ (see e.g.\ Ref.~\cite{Jansen:2008si} for detailed equations).
	The number of smearing steps was optimized on ensemble $B$ (see Table~\ref{tab:latticesetups4}) in Ref.~\cite{JH2022}.
	The APE step numbers $N_\text{APE}$ applied for computations on the other three ensembles were chosen according to a similar optimization carried out in Ref.~\cite{Schlosser:2021wnr} (see Table~$6$ in Appendix~A in that reference).
	In summary, we use $N_\text{APE}=33,\,82,\,115$ and $164$ for ensembles $A,\,B,\,C$ and $D$, respectively.

	% ********************
	% ********************
	% ********************
	% ********************
	% ********************

%	\newpage
\clearpage

	\section{Computational setup and details}\label{sec:computationaldetails}
	The gluelump correlation functions (\ref{eq:latticecorrelationfct}) were computed on four ensembles of SU(3) gauge link configurations with gauge couplings $\beta=6.594,\,6.451,\,6.284,\,6.000$.
	The configurations were generated with the CL2QCD software package~\cite{Philipsen:2014mra} in the context of a previous project~\cite{Schlosser:2021wnr}.
	Physical units are introduced by setting $r_0 = 0.5 \, \text{fm}$, which is a simple and common choice in pure gauge theory.
	Details concerning these gauge link ensembles, which we label by $A$, $B$, $C$ and $D$, can be found in Table~\ref{tab:latticesetups4} and in Section~3 of Ref.~\cite{Schlosser:2021wnr}.

	\begin{table}[htb]
		\begin{center}
			\def\arraystretch{1.2}
			\begin{tabular}{cccccccccc}
				\hline
				ensemble & $\beta$ & $a$ in $\text{fm}$ \cite{Necco:2001xg} & $(L/a)^3 \times T/a$ & $N_{\text{sim}}$ & $N_{\text{total}}$ & $N_{\text{or}}$ & $N_{\text{therm}}$ & $N_{\text{sep}}$ & $N_{\text{meas}}$ \\
				\hline
				$A$ & $6.000$ & $0.093$ & $12^3\times 26$ & $4$ & 
				$60 000$ & $\phantom{0}4$ & $20 000$ & $\phantom{0}50$ & $3200$ \\
				$B$ & $6.284$ & $0.060$ & $20^3\times 40$ & $4$ & 
				$60 000$ & $12$ & $20 000$ & $100$ & $\phantom{0}1600$ \\
				$C$ & $6.451$ & $0.048$ & $26^3\times 50$ & $4$ & 
				$80 000$ & $15$ & $40 000$ & $200$ & $\phantom{0}800$ \\ 
				$D$ & $6.594$ & $0.040$ & $30^3\times 60$ & $4$ & 
				$80 000$ & $15$ & $40 000$ & $200$ & $\phantom{0}800$ \\
				\hline
			\end{tabular}
		\end{center}
		\caption{Gauge link ensembles.}
		\label{tab:latticesetups4}
	\end{table}
	
	We used the multilevel algorithm \cite{Luscher:2001up} to reduce statistical errors in the gluelump correlation functions.
	Since we applied the multilevel algorithm already in previous projects, we refer for technical details to the corresponding references \cite{Brambilla:2021wqs,Schlosser:2021wnr}.
	We employed a single level of time-slice partitioning, a regular pattern with time-slice thickness $p_1=p_2=\dots = p_{n_{ts}}=a$ and $n_m=10$ sublattice configurations, which are separated by $n_u=30$ standard heatbath sweeps.
	These parameters were optimized specifically for gluelump computations in Ref.~\cite{JH2022}.
	
	We carried out two computations of gluelump correlation functions, one with unsmeared temporal links and the other with HYP2 smeared temporal links \cite{Hasenfratz:2001tw,DellaMorte:2003mn,DellaMorte:2005nwx}.
	HYP2 smearing leads to a reduced self-energy of the static adjoint quark and, consequently, to smaller statistical errors.
	However, the computation without HYP2 smearing is equally important, because it allows to complement our previous results for hybrid static potentials from Ref.~\cite{Schlosser:2021wnr}, which were computed with unsmeared temporal links (see Sec~\ref{sec:hybridpotentials_and_gluelumps}).
	Moreover, the conversion of gluelump masses from the lattice scheme to the RS scheme as in Ref.~\cite{Bali:2003jq} requires results with unsmeared temporal links (see Sec~\ref{sec:RSschemegluelump}).
	
	Statistical errors of results corresponding to individual ensembles were determined using the jackknife method. 
	For continuum extrapolations (see Section~\ref{sec:continuumextrapolated_masssplittings}), where we had to combine data from several ensembles, the bootstrap method was applied.
	This procedure is equivalent to the one used and explained in Ref.~\cite{Schlosser:2021wnr}.
	Here we use $N^A = 640$, $N^B=320$ and $N^C = N^D = 160$ reduced jackknife bins and $K = 10000$ bootstrap samples.

	% ********************
	% ********************
	% ********************
	% ********************
	% ********************

%\newpage
\clearpage

	\section{\label{sec:numericalresults}Numerical results}
	In the following we determine lattice gluelump masses and gluelump mass splittings. The methods we use are based on the asymptotic exponential falloff in $t$ of correlation functions $C_{\mathcal{R}^{PC}}(t)$ defined in Eq.~\eqref{eq:latticecorrelationfct}.
	
	Numerically, the asymptotic $t$ region is approximated by large values of $t$. We assume in the following that the numerically extracted asymptotic exponential falloff corresponds to a single state with mass $m_{\mathcal{R}^{PC}}$, the ground state in the $\mathcal{R}^{PC}$ representation. In other words, we assume that the data points of each correlation function $C_{\mathcal{R}^{PC}}(t)$ in the numerically accessible large-$t$ region are proportional to $e^{-m_{\mathcal{R}^{PC}} t}$. For certain $\mathcal{R}^{PC}$ representations one can expect that this assumption is fulfilled, but for other $\mathcal{R}^{PC}$ representations this is questionable. In the latter situation there might be two gluelump states with similar masses, which have the same lattice quantum numbers $\mathcal{R}^{PC}$, but different continuum total angular momenta $J$. 
	Then one might extract a mass somewhere between the masses of the two states.
	In Section~\ref{sec:spinidentification}, where we try to assign continuum total angular momenta $J$ to the extracted lattice gluelump mass splittings, we discuss in detail, which of our results are solid and trustworthy and which of them should be treated with caution.
	
		% ********************
		
		\subsection{\label{sec:masses_at_finite_a}Gluelump masses at finite values of the lattice spacing}
		
		A straightforward approach to determine gluelump masses is to compute effective masses
		\begin{equation}\label{eq:effpotential}
			m^{e,s}_{\text{eff};\mathcal{R}^{PC}}(t) = \frac{1}{a} \ln(\frac{C^{e,s}_{\mathcal{R}^{PC}}(t)}{C^{e,s}_{\mathcal{R}^{PC}}(t+a)}) .
		\end{equation}
		The large-$t$ limit
		\begin{equation}
			\label{eq:effectivemass_largetimelimit} m^{e,s}_{\mathcal{R}^{PC}} = \lim_{t \rightarrow \infty} m^{e,s}_{\text{eff};\mathcal{R}^{PC}}(t) 
		\end{equation}
		is obtained numerically from a fit of a constant to $m^{e,s}_{\text{eff};\mathcal{R}^{PC}}(t)$ in the range $t^\prime_\text{min} \le t \le t^\prime_\text{max}$, where $m^{e,s}_{\text{eff};\mathcal{R}^{PC}}(t)$ exhibits a plateau within statistical errors. This provides a gluelump mass $m^{e,s}_{\mathcal{R}^{PC}}$ for each representation $\mathcal{R}^{PC} \in \{A_1^{\pm\pm} , A_2^{\pm\pm} , E^{\pm\pm} , T_1^{\pm\pm} , T_2^{\pm\pm} \}$, each ensemble $e \in \{ A , B , C , D  \}$ and both unsmeared and HYP2 smeared temporal links indicated by labels $s \in \{ \text{none} , \text{HYP2} \}$.
		
		The fitting range is chosen individually for each gluelump mass $m^{e,s}_{\mathcal{R}^{PC}}$ by an algorithm already used in previous related work \cite{Capitani:2018rox,Schlosser:2021wnr}:
		\begin{itemize}
			\item $t_{\text{min}}$ is defined as the minimal $t$, where the values of $m^{e,s}_{\text{eff};\mathcal{R}^{PC}}(t) a$ and $m^{e,s}_{\text{eff};\mathcal{R}^{PC}}(t+a) a$ differ by less than $2 \sigma$.
			
			\item $t_{\text{max}}$ is the maximal $t$, where $C^{e,s}_{\mathcal{R}^{PC}}(t+a)$ has been computed, i.e.\ $t_{\text{max}} = 11 a , \, 19 a , \, 19 a , \, 19 a$ for ensembles $A , \, B , \, C , \, D$, respectively.
			
			\item Fits to $m^{e,s}_{\text{eff};\mathcal{R}^{PC}}(t) a$ are performed for all ranges $t^\prime_\text{min} \le t \le t^\prime_\text{max}$ with $t_\text{min} \le t^\prime_\text{min}$, \\ $t^\prime_\text{max} \le t_\text{max}$ and $t^\prime_\text{max} - t^\prime_\text{min} \ge 3 a$.
			
			\item The result of the fit with the longest plateau and $\chi_\text{red}^2 \le 1$ is taken as result for $m^{e,s}_{\mathcal{R}^{PC}} a$, where
		\begin{equation}
			\chi_\text{red}^2 = \frac{a}{t^\prime_\text{max} - t^\prime_\text{min}} \sum_{t=t^\prime_\text{min}, t^\prime_\text{min}+a, \ldots,t^\prime_\text{max}} \frac{\Big(m^{e,s}_{\text{eff};\mathcal{R}^{PC}}(t) a- m^{e,s}_{\mathcal{R}^{PC}} a\Big)^2}{\Big(\sigma[m^{e,s}_{\text{eff};\mathcal{R}^{PC}}](t) a\Big)^2}
		\end{equation}
		with $\sigma[m^{e,s}_{\text{eff};\mathcal{R}^{PC}}](t) a$ denoting the statistical error of $m^{e,s}_{\text{eff};\mathcal{R}^{PC}}(t) a$.
		\end{itemize}
		For around ten percent of the fits the fitting range was adjusted manually to correct for non-ideal or unreasonable fitting ranges. As a cross-check, we compared the resulting masses to masses obtained by analogous fits in the range $t^\prime_\text{min} + a \le t \le t^\prime_\text{max}$ and found agreement within statistical errors. For $\mathcal{R}^{PC} = A_2^{--}$ and unsmeared temporal links ($s=\text{none}$) statistical errors are rather large and the identification of effective mass plateaus is not possible. Therefore, we do not quote gluelump masses for that particular case. Moreover, we do not use the corresponding correlator data for the remainder of this work.
		
\begin{figure}
			\centering
			\includegraphics[width=0.49\linewidth,page=1]{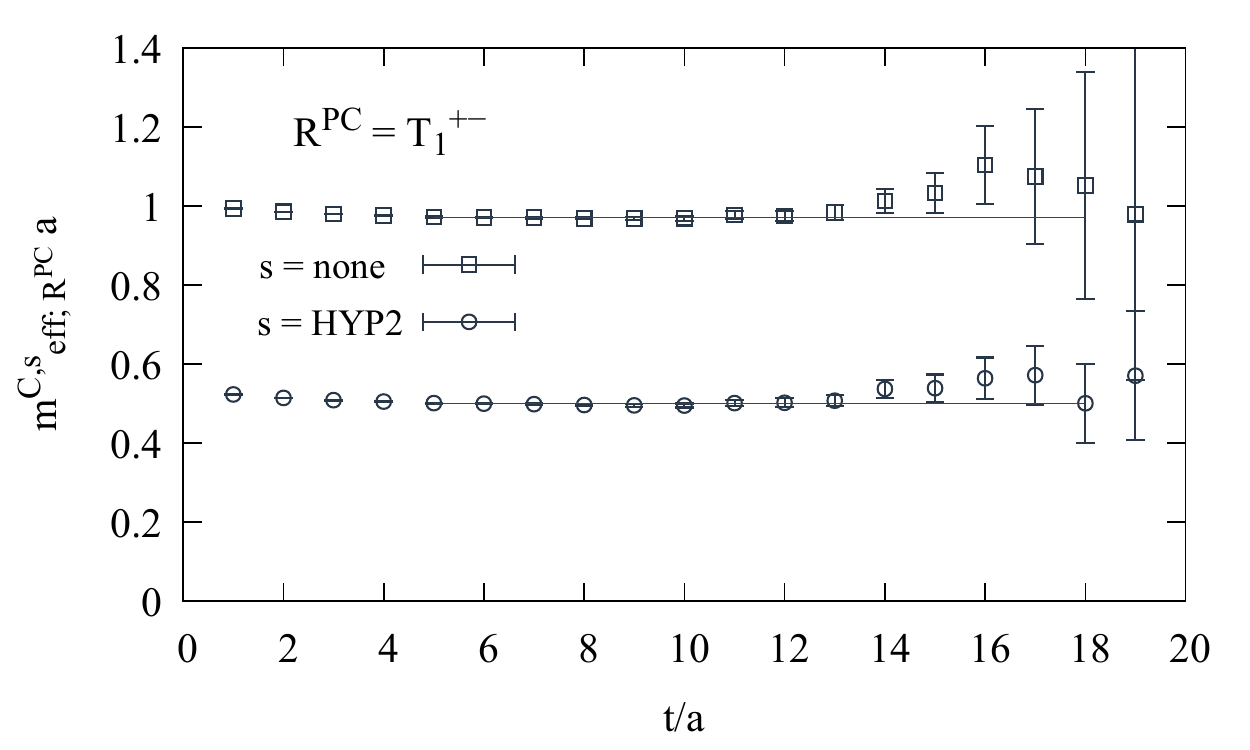}
			\includegraphics[width=0.49\linewidth,page=1]{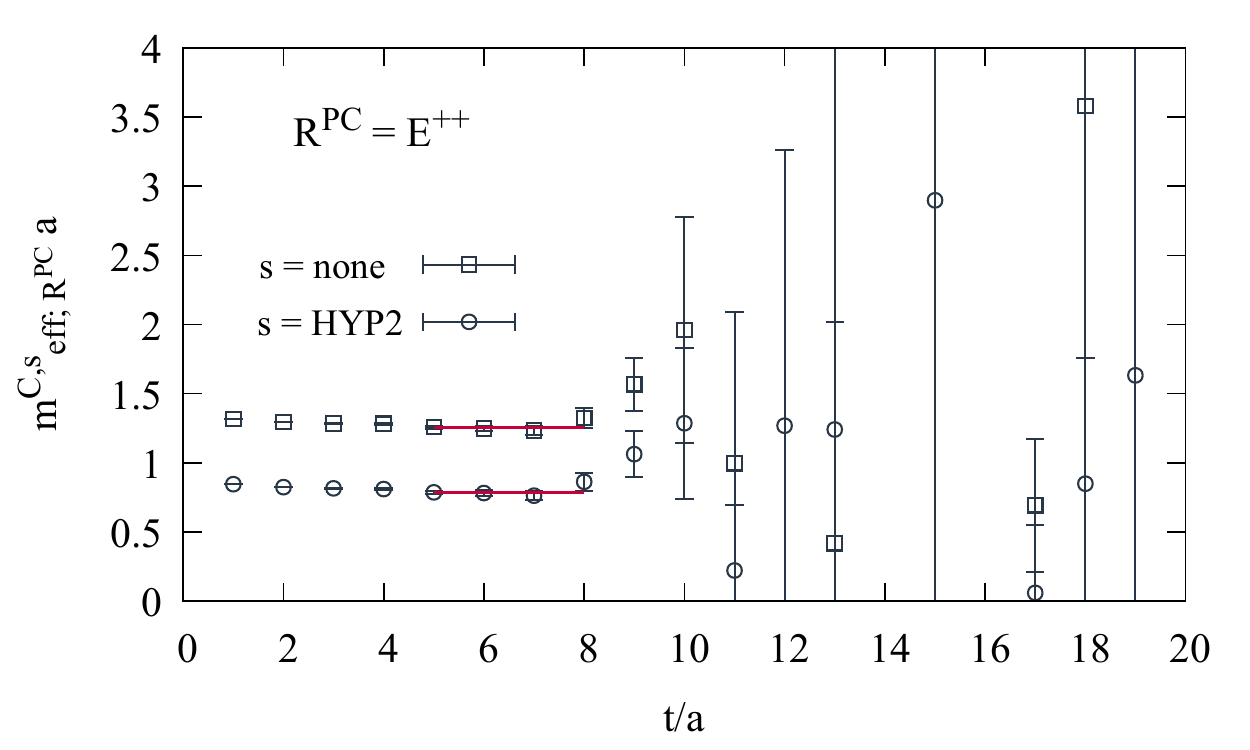}
			\caption{\label{fig:effpotentials0}Exemplary plots of effective masses $m^{C,s}_{\text{eff};\mathcal{R}^{PC}}(t) a$ and corresponding plateau fits $m^{C,s}_{\mathcal{R}^{PC}} a$. {(Left)}~$\mathcal{R}^{PC} = T_1^{+-}$, {(right)}~$\mathcal{R}^{PC} = E^{++}$.}
		\end{figure}

		The quality of our lattice data is exemplified by Figure~\ref{fig:effpotentials0}, where we present two typical effective mass plots and the corresponding plateau fits for representations $\mathcal{R}^{PC} = T_1^{+-}$ (left plot) and $\mathcal{R}^{PC} = E^{++}$ (right plot), $e = C$ and both $s = \text{none}$ and $s =\text{HYP2}$.
		
		The complete set of resulting gluelump masses $m^{e,s}_{\mathcal{R}^{PC}} a$ (i.e.\ for all 20 $\mathcal{R}^{PC}$ representations, the four ensembles from Table~\ref{tab:latticesetups4} and computations with unsmeared und with smeared temporal links) are collected in Appendix \ref{app:lattice_gluelump_masses}, Table~\ref{tab:amRPC}.
		
		Lattice gluelump masses contain an $a$-dependent self-energy, which originates from the static adjoint quark and diverges in the continuum limit. This self-energy is reduced, when using HYP2 smeared temporal links, which correspond to a less localized static charge. Because of the divergent self-energy, one cannot carry out meaningful continuum extrapolations. Nevertheless, such lattice gluelump masses at several finite values of the lattice spacing are important. They can, for example, be converted into the renormalon subtracted (RS) scheme (see Section~\ref{sec:RSschemegluelump} and Ref.~\cite{Bali:2003jq}) and then be used to fix the energy scale in Born-Oppenheimer effective field theory determinations of the spectra of heavy hybrid mesons (see e.g.\ Ref.~\cite{Berwein:2015vca}). Moreover, they complement lattice results on hybrid static potentials computed within the same lattice setup (see Section~\ref{sec:hybridpotentials_and_gluelumps} and Ref.~\cite{Capitani:2018rox,Schlosser:2021wnr}).
		
		% ********************
		
		\subsection{Continuum extrapolated gluelump mass splittings}\label{sec:continuumextrapolated_masssplittings}
		
		As discussed in the previous subsection, the $a$-dependent divergent self-energy is a consequence of the static adjoint quark. It is, thus, independent of $\mathcal{R}^{PC}$ and the same for all gluelumps. Consequently, for gluelump mass splittings
		\begin{equation}
			\label{EQN102} \Delta m_{\mathcal{R}^{PC}} = m_{\mathcal{R}^{PC}} - m_{T_1^{+-}}
		\end{equation}
		the self-energy cancels and a continuum extrapolation is possible and should lead to a finite mass difference. As previously done by other authors (see e.g.\ Refs.~\cite{Foster:1998wu,Marsh:2013xsa}), we use the mass of the lightest gluelump, which has $J^{PC} = 1^{+-}$ corresponding to $\mathcal{R}^{PC} = T_1^{+-}$, as reference mass.
		
		% **********
		
		\subsubsection{\label{sec:continuumextrapolation_effmassfit}Method~1: continuum extrapolated gluelump mass splittings from gluelump masses}
		
		A straightforward approach to compute gluelump mass splittings $\Delta m^{e,s}_{\mathcal{R}^{PC}}$ for each $e,s$ is to use the gluelump masses from Section~\ref{sec:masses_at_finite_a} extracted from effective mass plateaus. There are correlations between $m^{e,s}_{\mathcal{R}^{PC}}$ and $m^{e,s}_{T_1^{+-}}$, because both quantities are evaluated on the same set of gauge link configurations, but these correlations are taken into account by a proper error analysis as stated in Section~\ref{sec:computationaldetails}.
		
		The complete set of resulting gluelump mass splittings $\Delta m^{e,s}_{\mathcal{R}^{PC}} a$ (i.e.\ for all 19 $\mathcal{R}^{PC}$ representations, the four ensembles from Table~\ref{tab:latticesetups4} and computations with unsmeared und with smeared temporal links) are collected in Appendix~\ref{app:gluelump_mass_splittings}, Table~\ref{tab:Delta amRPC}. Results from the computations with and without HYP2 smeared temporal links are mostly consistent within statistical errors. Moreover, the gluelump mass splittings obtained for different lattice spacings, i.e.\ ensembles $A$, $B$, $C$ and $D$ are very similar, when not expressed in units of $a$, but in physical units, e.g.\ in $\text{GeV}$. This supports the above statement that there is no divergent self-energy in gluelump mass splittings and that continuum extrapolations are possible and should lead to finite mass differences.
		
\begin{figure}[htb]\centering
			\includegraphics[width=0.49\linewidth,page=1]{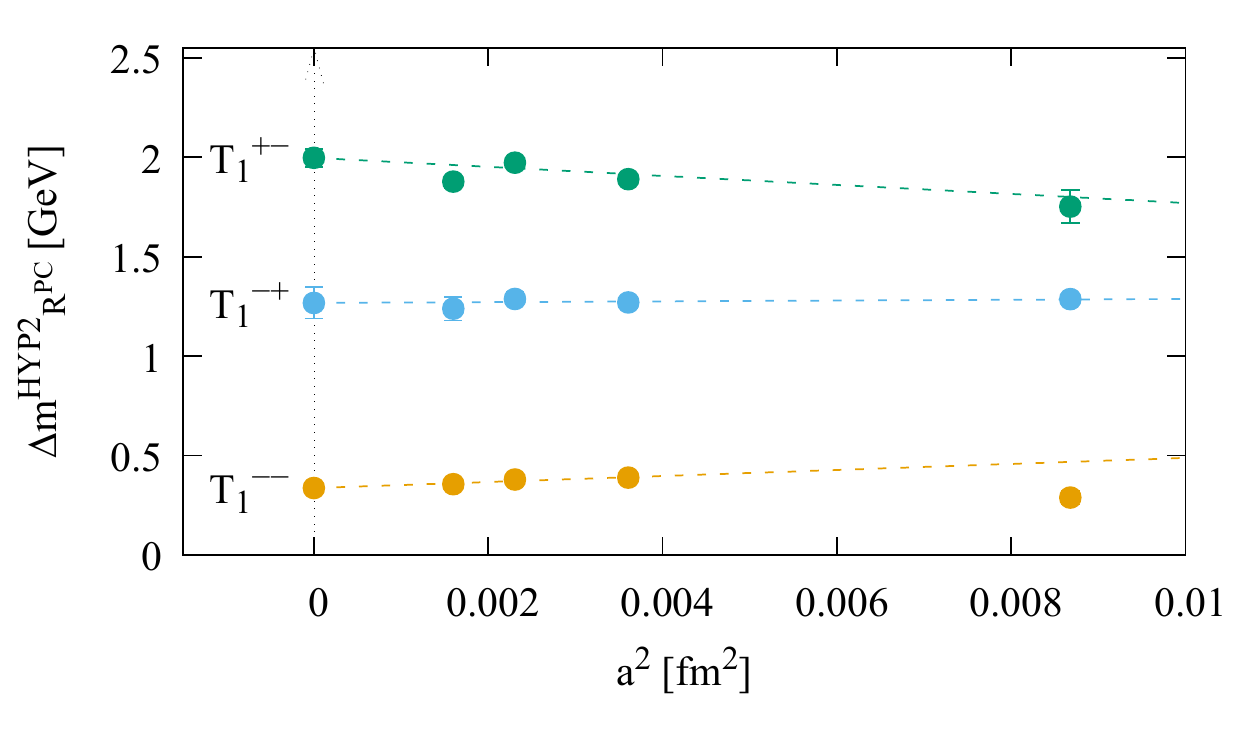}
			\includegraphics[width=0.49\linewidth,page=2]{Figure2}
			\includegraphics[width=0.49\linewidth,page=3]{Figure2}
			\includegraphics[width=0.49\linewidth,page=4]{Figure2}
			\includegraphics[width=0.49\linewidth,page=5]{Figure2}
		\caption{\label{fig:continuumextrapolation}Continuum extrapolations of gluelump mass splittings $\Delta m_{\mathcal{R}^{PC}}$ for HYP2 smeared temporal links.}
		\end{figure}

		In Figure~\ref{fig:continuumextrapolation} we show the gluelump mass splittings $\Delta m^{e,s}_{\mathcal{R}^{PC}}$ as functions of $a^2$ obtained from computations with HYP2 smeared temporal links (plots for computations with unsmeared temporal links are quite similar, but exhibit somewhat larger statistical errors). The observed $a$-dependencies for the three smaller lattice spacings ($e \in \{ B , C , D \}$) are consistent with linear behaviors in $a^2$. This is expected for the Wilson plaquette action. Thus, to extrapolate to $a = 0$, we use the function
		\begin{equation}
			\label{eq:continuumextrapolation_mass_finite_a} \Delta m^{\text{fit},s}_{\mathcal{R}^{PC}}(a) = \Delta m^s_{\mathcal{R}^{PC},\text{cont}} + c^s_{\mathcal{R}^{PC}} a^2
		\end{equation}
		and carry out a $\chi^2$-minimizing fit for each representation $\mathcal{R}^{PC}$ to the corresponding three data points (see the dashed lines in Figure~\ref{fig:continuumextrapolation}). The fit parameters are $c^s_{\mathcal{R}^{PC}}$ and $\Delta m^s_{\mathcal{R}^{PC},\text{cont}}$, where the latter represents the continuum limit of the gluelump mass splitting. Most of the $\chi_\text{red}^2$ values are of $\order{1}$, which indicate reasonable fits. We do not use data points from ensemble $A$ for the fits, because this ensemble has the largest lattice spacing and Figure~\ref{fig:continuumextrapolation} indicates that non-linear contributions in $a^2$ are already sizable.
		
		The complete set of continuum extrapolated gluelump mass splittings $\Delta m^s_{\mathcal{R}^{PC},\text{cont}}$ (i.e.\ for all 19 $\mathcal{R}^{PC}$ representations and computations with unsmeared und with smeared temporal links) are collected in Table~\ref{tab:continuumsmRPC}.
		For the majority of $\mathcal{R}^{PC}$ representations the resulting continuum extrapolations $\Delta m^{\text{none}}_{\mathcal{R}^{PC},\text{cont}}$ and $\Delta m^{\text{HYP2}}_{\mathcal{R}^{PC},\text{cont}}$ obtained with unsmeared and with HYP2 smeared temporal links are consistent within statistical errors (the statistical error is the first of the two errors provided in Table~\ref{tab:continuumsmRPC}).
		
		\begin{table}[htb]
		\centering
		\begin {tabular}{cll}%
\toprule $\mathcal {R}^{PC}$&$\Delta m_{\mathcal {R}^{PC}, \text {cont}}^\text {none}\,$ [GeV]&$\Delta m_{\mathcal {R}^{PC}, \text {cont}}^\text {HYP2}\,$ [GeV]\\\midrule %
$T_1^{++}$&\pgfmathprintnumber [fixed,fixed zerofill,precision=3]{3.8589848e-1}(\pgfmathprintnumber [fixed,fixed zerofill,precision=0]{1.39339722e1})(\pgfmathprintnumber [fixed,fixed zerofill,precision=0]{1.6922638e1})&\pgfmathprintnumber [fixed,fixed zerofill,precision=3]{3.9408269e-1}(\pgfmathprintnumber [fixed,fixed zerofill,precision=0]{8.8176938e0})(\pgfmathprintnumber [fixed,fixed zerofill,precision=0]{7.6332703e0})\\%
\rowcolor [gray]{1.0}$T_1^{+-}$&0&0\\%
$T_1^{-+}$&\pgfmathprintnumber [fixed,fixed zerofill,precision=3]{2.8224014e-1}(\pgfmathprintnumber [fixed,fixed zerofill,precision=0]{7.2029146e0})(\pgfmathprintnumber [fixed,fixed zerofill,precision=0]{7.7889297e0})&\pgfmathprintnumber [fixed,fixed zerofill,precision=3]{2.5008888e-1}(\pgfmathprintnumber [fixed,fixed zerofill,precision=0]{1.54994054e1})(\pgfmathprintnumber [fixed,fixed zerofill,precision=0]{1.1644699e0})\\%
$T_1^{--}$&\pgfmathprintnumber [fixed,fixed zerofill,precision=3]{6.6239058e-2}(\pgfmathprintnumber [fixed,fixed zerofill,precision=0]{4.0846263e0})(\pgfmathprintnumber [fixed,fixed zerofill,precision=0]{5.8910599e0})&\pgfmathprintnumber [fixed,fixed zerofill,precision=3]{6.6442844e-2}(\pgfmathprintnumber [fixed,fixed zerofill,precision=0]{4.0418913e0})(\pgfmathprintnumber [fixed,fixed zerofill,precision=0]{6.5478745e0})\\%
$T_2^{++}$&\pgfmathprintnumber [fixed,fixed zerofill,precision=3]{3.6268378e-1}(\pgfmathprintnumber [fixed,fixed zerofill,precision=0]{9.7853896e0})(\pgfmathprintnumber [fixed,fixed zerofill,precision=0]{6.8818405e0})&\pgfmathprintnumber [fixed,fixed zerofill,precision=3]{3.616638e-1}(\pgfmathprintnumber [fixed,fixed zerofill,precision=0]{9.5987772e0})(\pgfmathprintnumber [fixed,fixed zerofill,precision=0]{6.5148499e0})\\%
$T_2^{+-}$&\pgfmathprintnumber [fixed,fixed zerofill,precision=3]{1.85400243e-1}(\pgfmathprintnumber [fixed,fixed zerofill,precision=0]{6.686414e0})(\pgfmathprintnumber [fixed,fixed zerofill,precision=0]{1.95074e0})&\pgfmathprintnumber [fixed,fixed zerofill,precision=3]{1.8637145e-1}(\pgfmathprintnumber [fixed,fixed zerofill,precision=0]{6.3197643e0})(\pgfmathprintnumber [fixed,fixed zerofill,precision=0]{8.4339996e-1})\\%
$T_2^{-+}$&\pgfmathprintnumber [fixed,fixed zerofill,precision=3]{3.4524722e-1}(\pgfmathprintnumber [fixed,fixed zerofill,precision=0]{1.24063457e1})(\pgfmathprintnumber [fixed,fixed zerofill,precision=0]{8.61772e0})&\pgfmathprintnumber [fixed,fixed zerofill,precision=3]{3.3066016e-1}(\pgfmathprintnumber [fixed,fixed zerofill,precision=0]{1.12142126e1})(\pgfmathprintnumber [fixed,fixed zerofill,precision=0]{7.8163101e0})\\%
$T_2^{--}$&\pgfmathprintnumber [fixed,fixed zerofill,precision=3]{1.04837187e-1}(\pgfmathprintnumber [fixed,fixed zerofill,precision=0]{6.8609422e0})(\pgfmathprintnumber [fixed,fixed zerofill,precision=0]{4.2399994e-3})&\pgfmathprintnumber [fixed,fixed zerofill,precision=3]{1.10917673e-1}(\pgfmathprintnumber [fixed,fixed zerofill,precision=0]{6.9477757e0})(\pgfmathprintnumber [fixed,fixed zerofill,precision=0]{7.0319397e0})\\%
$A_1^{++}$&\pgfmathprintnumber [fixed,fixed zerofill,precision=3]{1.96342148e-1}(\pgfmathprintnumber [fixed,fixed zerofill,precision=0]{6.7142949e0})(\pgfmathprintnumber [fixed,fixed zerofill,precision=0]{6.6106003e-1})&\pgfmathprintnumber [fixed,fixed zerofill,precision=3]{1.9658817e-1}(\pgfmathprintnumber [fixed,fixed zerofill,precision=0]{6.5365846e0})(\pgfmathprintnumber [fixed,fixed zerofill,precision=0]{6.3279999e-1})\\%
$A_1^{+-}$&\pgfmathprintnumber [fixed,fixed zerofill,precision=3]{4.3863023e-1}(\pgfmathprintnumber [fixed,fixed zerofill,precision=0]{7.8939702e0})(\pgfmathprintnumber [fixed,fixed zerofill,precision=0]{2.6523102e0})&\pgfmathprintnumber [fixed,fixed zerofill,precision=3]{3.9409685e-1}(\pgfmathprintnumber [fixed,fixed zerofill,precision=0]{1.5531528e1})(\pgfmathprintnumber [fixed,fixed zerofill,precision=0]{2.1443909e1})\\%
$A_1^{-+}$&\pgfmathprintnumber [fixed,fixed zerofill,precision=3]{4.6680423e-1}(\pgfmathprintnumber [fixed,fixed zerofill,precision=0]{1.21901695e1})(\pgfmathprintnumber [fixed,fixed zerofill,precision=0]{2.4763336e1})&\pgfmathprintnumber [fixed,fixed zerofill,precision=3]{4.774267e-1}(\pgfmathprintnumber [fixed,fixed zerofill,precision=0]{1.51752345e1})(\pgfmathprintnumber [fixed,fixed zerofill,precision=0]{6.0882599e0})\\%
$A_1^{--}$&\pgfmathprintnumber [fixed,fixed zerofill,precision=3]{3.806291e-1}(\pgfmathprintnumber [fixed,fixed zerofill,precision=0]{3.1266034e1})(\pgfmathprintnumber [fixed,fixed zerofill,precision=0]{6.3047424e1})&\pgfmathprintnumber [fixed,fixed zerofill,precision=3]{2.9190825e-1}(\pgfmathprintnumber [fixed,fixed zerofill,precision=0]{9.1107867e0})(\pgfmathprintnumber [fixed,fixed zerofill,precision=0]{1.6464798e0})\\%
$A_2^{++}$&\pgfmathprintnumber [fixed,fixed zerofill,precision=3]{4.3724538e-1}(\pgfmathprintnumber [fixed,fixed zerofill,precision=0]{1.48689583e1})(\pgfmathprintnumber [fixed,fixed zerofill,precision=0]{4.7593796e0})&\pgfmathprintnumber [fixed,fixed zerofill,precision=3]{4.3823163e-1}(\pgfmathprintnumber [fixed,fixed zerofill,precision=0]{1.42837901e1})(\pgfmathprintnumber [fixed,fixed zerofill,precision=0]{5.74655e0})\\%
$A_2^{+-}$&\pgfmathprintnumber [fixed,fixed zerofill,precision=3]{3.2724588e-1}(\pgfmathprintnumber [fixed,fixed zerofill,precision=0]{2.4959697e1})(\pgfmathprintnumber [fixed,fixed zerofill,precision=0]{2.2224243e1})&\pgfmathprintnumber [fixed,fixed zerofill,precision=3]{2.726497e-1}(\pgfmathprintnumber [fixed,fixed zerofill,precision=0]{2.4184567e1})(\pgfmathprintnumber [fixed,fixed zerofill,precision=0]{1.5201828e1})\\%
$A_2^{-+}$&\pgfmathprintnumber [fixed,fixed zerofill,precision=3]{2.7752832e-1}(\pgfmathprintnumber [fixed,fixed zerofill,precision=0]{1.2143837e1})(\pgfmathprintnumber [fixed,fixed zerofill,precision=0]{1.6411758e1})&\pgfmathprintnumber [fixed,fixed zerofill,precision=3]{2.8899704e-1}(\pgfmathprintnumber [fixed,fixed zerofill,precision=0]{1.29445973e1})(\pgfmathprintnumber [fixed,fixed zerofill,precision=0]{8.9389496e0})\\%
$A_2^{--}$&-&\pgfmathprintnumber [fixed,fixed zerofill,precision=3]{4.5698862e-1}(\pgfmathprintnumber [fixed,fixed zerofill,precision=0]{1.56302142e1})(\pgfmathprintnumber [fixed,fixed zerofill,precision=0]{2.0376663e1})\\%
$E^{++}$&\pgfmathprintnumber [fixed,fixed zerofill,precision=3]{2.5296307e-1}(\pgfmathprintnumber [fixed,fixed zerofill,precision=0]{5.4800864e0})(\pgfmathprintnumber [fixed,fixed zerofill,precision=0]{2.8248093e0})&\pgfmathprintnumber [fixed,fixed zerofill,precision=3]{2.5278966e-1}(\pgfmathprintnumber [fixed,fixed zerofill,precision=0]{5.3505872e0})(\pgfmathprintnumber [fixed,fixed zerofill,precision=0]{2.3753403e0})\\%
$E^{+-}$&\pgfmathprintnumber [fixed,fixed zerofill,precision=3]{1.73806252e-1}(\pgfmathprintnumber [fixed,fixed zerofill,precision=0]{4.8226164e0})(\pgfmathprintnumber [fixed,fixed zerofill,precision=0]{3.3641495e0})&\pgfmathprintnumber [fixed,fixed zerofill,precision=3]{1.74114894e-1}(\pgfmathprintnumber [fixed,fixed zerofill,precision=0]{4.686765e0})(\pgfmathprintnumber [fixed,fixed zerofill,precision=0]{3.3041199e0})\\%
$E^{-+}$&\pgfmathprintnumber [fixed,fixed zerofill,precision=3]{3.1745224e-1}(\pgfmathprintnumber [fixed,fixed zerofill,precision=0]{1.01033195e1})(\pgfmathprintnumber [fixed,fixed zerofill,precision=0]{1.0709152e1})&\pgfmathprintnumber [fixed,fixed zerofill,precision=3]{3.1433932e-1}(\pgfmathprintnumber [fixed,fixed zerofill,precision=0]{9.9401443e0})(\pgfmathprintnumber [fixed,fixed zerofill,precision=0]{9.2782394e0})\\%
$E^{--}$&\pgfmathprintnumber [fixed,fixed zerofill,precision=3]{1.18363326e-1}(\pgfmathprintnumber [fixed,fixed zerofill,precision=0]{5.3575266e0})(\pgfmathprintnumber [fixed,fixed zerofill,precision=0]{5.8168503e0})&\pgfmathprintnumber [fixed,fixed zerofill,precision=3]{1.22096538e-1}(\pgfmathprintnumber [fixed,fixed zerofill,precision=0]{8.0036267e0})(\pgfmathprintnumber [fixed,fixed zerofill,precision=0]{7.9111206e0})\\\bottomrule %
\end {tabular}%

		\caption{\label{tab:continuumsmRPC}Continuum extrapolated gluelump mass splittings $\Delta m^s_{\mathcal{R}^{PC},\text{cont}}$ obtained by using the gluelump masses from Table~\ref{tab:amRPC} and a fit function linear in $a^2$ (see Eq.~\eqref{eq:continuumextrapolation_mass_finite_a}). The first error is the statistical error, while the second error is a systematic error representing the difference between an $a^2$ and an $a^4$ ansatz for the continuum extrapolation (see text for details).}
		\end{table}
		
		We also checked the validity and stability of our continuum extrapolations by extending the fit function (\ref{eq:continuumextrapolation_mass_finite_a}) by a term proportional to $a^4$ and at the same time including the data points from ensemble $A$ with the coarsest lattice spacing in the fits. Again the resulting continuum extrapolated gluelump mass splittings are mostly consistent with those listed in Table~\ref{tab:continuumsmRPC} within statistical errors. We use the differences as an estimate of the systematic error (the second of the two errors provided in Table~\ref{tab:continuumsmRPC}).
		
		% **********
		
		\subsubsection{\label{sec:continuumextrapolation_method2}Method~2: continuum extrapolated gluelump mass splittings from simultaneous fits to correlator data from several ensembles}
		
		Each of the continuum extrapolations carried out in the previous subsection is based on just three data points, where some have rather large statistical errors. Moreover, the data points are differences of gluelump masses, where each gluelump mass is the result of a fit to a few effective mass values consistent with a plateau. Some of these effective mass values also exhibit large statistical errors and there are cases, where clear plateau identifications are difficult. Because of these problems, we present and employ another method in the following. The method is based on simultaneous fits to several correlation functions (\ref{eq:latticecorrelationfct}) computed on different ensembles both with unsmeared and with HYP2 smeared temporal links. As one might expect, we find that this method is more stable than that of the previous Section~\ref{sec:continuumextrapolation_effmassfit} and we consider our results for continuum extrapolated gluelump mass splittings presented in this section (Table~\ref{tab:jointfits_continuummass}) to be superior to those presented in the previous section (Table~\ref{tab:continuumsmRPC}). Within statistical errors they are, however, identical.
		
		The basic idea is that the gluelump mass splitting $\Delta m^{e,s}_{\mathcal{R}^{PC}}$ can be extracted from the asymptotic behavior of
		\begin{equation}
			\label{eq:def_correlatorratios} \tilde{C}^{e,s}_{\mathcal{R}^{PC}}(t) = \frac{{C}^{e,s}_{\mathcal{R}^{PC}}(t)}{{C}^{e,s}_{{T}_1^{+-}}(t)} \,\, \sim_{t \to \infty} \,\, A^{e,s}_{\mathcal{R}^{PC}}\exp(-\Delta m^{e,s}_{\mathcal{R}^{PC}} t) ,
		\end{equation}
		which is a ratio of temporal correlation functions $C^{e,s}_{\mathcal{R}^{PC}}(t)$ defined in Eq.~(\ref{eq:latticecorrelationfct}). As discussed in Section~\ref{sec:continuumextrapolation_effmassfit} the dependence of $\Delta m^e_{\mathcal{R}^{PC}}$ on the lattice spacing is expected to be linear in $a^2$ at leading order (see Eq.~\eqref{eq:continuumextrapolation_mass_finite_a}). For each representation $\mathcal{R}^{PC}$ we, thus, carry out a simultaneous $9$-parameter fit of
		\begin{equation}
			\label{eq:joint_9_parameters_fit}
			\tilde{C}^{\text{fit},e,s}_{\mathcal{R}^{PC}}(t) = A^{e,s}_{\mathcal{R}^{PC}} \exp\Big(-(\Delta m_{\mathcal{R}^{PC},\text{cont}} + c^s_{\mathcal{R}^{PC}} a^2) t\Big)
		\end{equation}
		to the correlator data from ensembles $B$, $C$ and $D$ for unsmeared and HYP2 smeared temporal links. Numerically, we find $A^{e,\text{none}}_{\mathcal{R}^{PC}} = A^{e,\text{HYP2}}_{\mathcal{R}^{PC}}$ and $c^\text{none}_{\mathcal{R}^{PC}} = c^\text{HYP2}_{\mathcal{R}^{PC}}$ within statistical errors. Thus we reduce the number of fit parameters from 9 to 5 and repeat all fits using the fit function
		\begin{equation}
			\label{eq:joint_5_parameters_fit}
			\tilde{C}^{\text{fit},e,s}_{\mathcal{R}^{PC}}(t) = A^e_{\mathcal{R}^{PC}} \exp\Big(-(\Delta m_{\mathcal{R}^{PC},\text{cont}} + c_{\mathcal{R}^{PC}} a^2) t\Big) .
		\end{equation}
		
		The fitting range $t_\text{min} \leq t \le t_\text{max}$ is chosen individually for each representation $\mathcal{R}^{PC}$ in the following way:
		\begin{itemize}
		\item For each $e,s$ we define $t^{e,s}_\text{min} = \tilde{t} - a/2$ with $\tilde{t}$ denoting the smallest value of $t$, where the effective mass
		\begin{equation}
		\label{eq:def_effmass} \tilde{m}^{e,s}_{\text{eff},\mathcal{R}^{PC}}(t) = \frac{1}{a} \ln(\frac{\tilde{C}^{e,s}_{\mathcal{R}^{PC}}(t)}{\tilde{C}^{e,s}_{\mathcal{R}^{PC}}(t+a)})
		\end{equation}
		satisfies $|\tilde{m}^{e,s}_{\text{eff},\mathcal{R}^{PC}}(t) a - \tilde{m}^{e,s}_{\text{eff},\mathcal{R}^{PC}}(t+a) a| < 2 \sigma$.
		
		\item $t_\text{min}$ is the largest $t^{e,s}_\text{min}$ from the previous item, i.e.\ we start the fit for all ensembles at the same temporal separation in physical units.
		
		\item $t_\text{max}$ is the largest $t$, where the correlation functions $C^{e,s}_{\mathcal{R}^{PC}}(t)$ have been computed, i.e.\ $t_{\max} = 20 a(\beta = 6.284) = 1.20 \, \text{fm}$.
		\end{itemize}
		Note that this procedure to select the fit range resembles that used previously in Section~\ref{sec:masses_at_finite_a}.
		
		For $\mathcal{R}^{PC} = A_2^{--}$ we only include correlator data obtained with smeared temporal links \\ ($s=\text{HYP2}$) in the fit since we already saw in Section~\ref{sec:masses_at_finite_a} that a clear plateau identification in the corresponding effective masses obtained with unsmeared temporal links ($s=\text{none}$) is not possible and statistical errors are large.
		
		The complete set of continuum extrapolated gluelump mass splittings $\Delta m_{\mathcal{R}^{PC},\text{cont}}$ (i.e.\ for all 19 $\mathcal{R}^{PC}$ representations), which are the main results of this subsection, are collected in Table~\ref{tab:jointfits_continuummass}.
		The corresponding $\chi^2_\text{red}$ values are $\mathcal{O}(1)$ indicating reasonable fits.
			
		\begin{table}[htb]\centering
			\begin {tabular}{llc}%
\toprule $\mathcal {R}^{PC}$&$\Delta m_{\mathcal {R}^{PC}, \text {cont}}\,$ [GeV]&$\chi ^2_{\text {red}}$\\\midrule %
$T_1^{++}$&\pgfmathprintnumber [fixed,fixed zerofill,precision=3]{1.79312321e0}(\pgfmathprintnumber [fixed,fixed zerofill,precision=0]{9.4154762e1})(\pgfmathprintnumber [fixed,fixed zerofill,precision=0]{3.4779007e1})(\pgfmathprintnumber [fixed,fixed zerofill,precision=0]{4.2371652e1})&\pgfutilensuremath {0.88}\\%
$T_1^{-+}$&\pgfmathprintnumber [fixed,fixed zerofill,precision=3]{1.21343687e0}(\pgfmathprintnumber [fixed,fixed zerofill,precision=0]{5.8552019e1})(\pgfmathprintnumber [fixed,fixed zerofill,precision=0]{3.2815994e0})(\pgfmathprintnumber [fixed,fixed zerofill,precision=0]{2.4026875e1})&\pgfutilensuremath {1.05}\\%
$T_1^{--}$&\pgfmathprintnumber [fixed,fixed zerofill,precision=3]{3.4241158e-1}(\pgfmathprintnumber [fixed,fixed zerofill,precision=0]{1.93710408e1})(\pgfmathprintnumber [fixed,fixed zerofill,precision=0]{2.161731e1})(\pgfmathprintnumber [fixed,fixed zerofill,precision=0]{2.1012492e1})&\pgfutilensuremath {0.43}\\%
$T_2^{++}$&\pgfmathprintnumber [fixed,fixed zerofill,precision=3]{1.77109978e0}(\pgfmathprintnumber [fixed,fixed zerofill,precision=0]{8.4780003e1})(\pgfmathprintnumber [fixed,fixed zerofill,precision=0]{6.0441696e1})(\pgfmathprintnumber [fixed,fixed zerofill,precision=0]{3.5340222e1})&\pgfutilensuremath {0.39}\\%
$T_2^{+-}$&\pgfmathprintnumber [fixed,fixed zerofill,precision=3]{9.6605688e-1}(\pgfmathprintnumber [fixed,fixed zerofill,precision=0]{2.9100056e1})(\pgfmathprintnumber [fixed,fixed zerofill,precision=0]{2.1697098e0})(\pgfmathprintnumber [fixed,fixed zerofill,precision=0]{1.28568095e1})&\pgfutilensuremath {0.64}\\%
$T_2^{-+}$&\pgfmathprintnumber [fixed,fixed zerofill,precision=3]{1.63831519e0}(\pgfmathprintnumber [fixed,fixed zerofill,precision=0]{7.2801955e1})(\pgfmathprintnumber [fixed,fixed zerofill,precision=0]{7.8239304e1})(\pgfmathprintnumber [fixed,fixed zerofill,precision=0]{3.0488617e1})&\pgfutilensuremath {1.38}\\%
$T_2^{--}$&\pgfmathprintnumber [fixed,fixed zerofill,precision=3]{5.034903e-1}(\pgfmathprintnumber [fixed,fixed zerofill,precision=0]{1.16366452e1})(\pgfmathprintnumber [fixed,fixed zerofill,precision=0]{5.4138794e0})(\pgfmathprintnumber [fixed,fixed zerofill,precision=0]{4.5993835e0})&\pgfutilensuremath {1.34}\\%
$A_1^{++}$&\pgfmathprintnumber [fixed,fixed zerofill,precision=3]{9.7855312e-1}(\pgfmathprintnumber [fixed,fixed zerofill,precision=0]{2.5754454e1})(\pgfmathprintnumber [fixed,fixed zerofill,precision=0]{2.1057953e1})(\pgfmathprintnumber [fixed,fixed zerofill,precision=0]{1.40144023e1})&\pgfutilensuremath {1.00}\\%
$A_1^{+-}$&\pgfmathprintnumber [fixed,fixed zerofill,precision=3]{2.0879065e0}(\pgfmathprintnumber [fixed,fixed zerofill,precision=0]{5.1217485e1})(\pgfmathprintnumber [fixed,fixed zerofill,precision=0]{1.2290604e2})(\pgfmathprintnumber [fixed,fixed zerofill,precision=0]{3.6184111e1})&\pgfutilensuremath {0.82}\\%
$A_1^{-+}$&\pgfmathprintnumber [fixed,fixed zerofill,precision=3]{2.3542535e0}(\pgfmathprintnumber [fixed,fixed zerofill,precision=0]{5.2517656e1})(\pgfmathprintnumber [fixed,fixed zerofill,precision=0]{2.7432297e1})(\pgfmathprintnumber [fixed,fixed zerofill,precision=0]{1.06216837e2})&\pgfutilensuremath {0.92}\\%
$A_1^{--}$&\pgfmathprintnumber [fixed,fixed zerofill,precision=3]{1.43317242e0}(\pgfmathprintnumber [fixed,fixed zerofill,precision=0]{3.0578662e1})(\pgfmathprintnumber [fixed,fixed zerofill,precision=0]{3.1376297e1})(\pgfmathprintnumber [fixed,fixed zerofill,precision=0]{1.62902911e1})&\pgfutilensuremath {0.91}\\%
$A_2^{++}$&\pgfmathprintnumber [fixed,fixed zerofill,precision=3]{2.2100388e0}(\pgfmathprintnumber [fixed,fixed zerofill,precision=0]{6.621955e1})(\pgfmathprintnumber [fixed,fixed zerofill,precision=0]{5.7414398e1})(\pgfmathprintnumber [fixed,fixed zerofill,precision=0]{3.7892001e1})&\pgfutilensuremath {0.75}\\%
$A_2^{+-}$&\pgfmathprintnumber [fixed,fixed zerofill,precision=3]{1.37643182e0}(\pgfmathprintnumber [fixed,fixed zerofill,precision=0]{1.28170702e2})(\pgfmathprintnumber [fixed,fixed zerofill,precision=0]{1.5539322e2})(\pgfmathprintnumber [fixed,fixed zerofill,precision=0]{6.0236938e1})&\pgfutilensuremath {0.92}\\%
$A_2^{-+}$&\pgfmathprintnumber [fixed,fixed zerofill,precision=3]{1.495724e0}(\pgfmathprintnumber [fixed,fixed zerofill,precision=0]{3.245391e1})(\pgfmathprintnumber [fixed,fixed zerofill,precision=0]{1.0943954e2})(\pgfmathprintnumber [fixed,fixed zerofill,precision=0]{2.1105128e1})&\pgfutilensuremath {0.48}\\%
${A_2^{--}}^{\star }$&\pgfmathprintnumber [fixed,fixed zerofill,precision=3]{2.1494815e0}(\pgfmathprintnumber [fixed,fixed zerofill,precision=0]{3.4040413e2})(\pgfmathprintnumber [fixed,fixed zerofill,precision=0]{6.7046005e0})(\pgfmathprintnumber [fixed,fixed zerofill,precision=0]{1.33302289e2})&\pgfutilensuremath {1.34}\\%
$E^{++}$&\pgfmathprintnumber [fixed,fixed zerofill,precision=3]{1.25829149e0}(\pgfmathprintnumber [fixed,fixed zerofill,precision=0]{1.86427686e1})(\pgfmathprintnumber [fixed,fixed zerofill,precision=0]{1.8087997e0})(\pgfmathprintnumber [fixed,fixed zerofill,precision=0]{1.5055306e1})&\pgfutilensuremath {0.83}\\%
$E^{+-}$&\pgfmathprintnumber [fixed,fixed zerofill,precision=3]{8.576539e-1}(\pgfmathprintnumber [fixed,fixed zerofill,precision=0]{2.0514667e1})(\pgfmathprintnumber [fixed,fixed zerofill,precision=0]{2.271437e1})(\pgfmathprintnumber [fixed,fixed zerofill,precision=0]{1.8456981e1})&\pgfutilensuremath {0.50}\\%
$E^{-+}$&\pgfmathprintnumber [fixed,fixed zerofill,precision=3]{1.51110373e0}(\pgfmathprintnumber [fixed,fixed zerofill,precision=0]{1.61582993e2})(\pgfmathprintnumber [fixed,fixed zerofill,precision=0]{4.4355194e1})(\pgfmathprintnumber [fixed,fixed zerofill,precision=0]{8.0901381e1})&\pgfutilensuremath {1.18}\\%
$E^{--}$&\pgfmathprintnumber [fixed,fixed zerofill,precision=3]{5.5853746e-1}(\pgfmathprintnumber [fixed,fixed zerofill,precision=0]{1.24835634e1})(\pgfmathprintnumber [fixed,fixed zerofill,precision=0]{4.3640152e1})(\pgfmathprintnumber [fixed,fixed zerofill,precision=0]{1.05293437e1})&\pgfutilensuremath {0.87}\\\bottomrule %
\end {tabular}%

			\\ \vspace{0.1cm}\scriptsize{${}^\star$~For $\mathcal{R}^{PC} = A_2^{--}$ we exclude correlator data obtained with unsmeared temporal links ($s=\text{none}$) from the fit (see the discussion in Section~\ref{sec:masses_at_finite_a}).}
			\caption{\label{tab:jointfits_continuummass}Continuum extrapolated gluelump mass splittings $\Delta m_{\mathcal{R}^{PC},\text{cont}}$ obtained from 5-param\-eter fits of the fit function (\ref{eq:joint_5_parameters_fit}) to correlator data from ensembles $B$, $C$ and $D$. The first error is the statistical error, while the second error is a systematic error representing the difference between an $a^2$ and an $a^4$ ansatz for the continuum extrapolation, respectively,
			and the third error represents the systematic error
			coming from the choice of fitting range (see text for details).}
		\end{table}
		
		To check the stability of the resulting continuum extrapolated gluelump mass splittings with respect to a variation of the fitting range, we repeat all fits using the range $t'_\text{min} \leq t \le t_\text{max}$. $t'_\text{min}$ is defined in a similar way as $t_\text{min}$ with the difference that the condition below Eq.~\eqref{eq:def_effmass} is replaced by $|\tilde{m}^{e,s}_{\text{eff},\mathcal{R}^{PC}}(t-a) a - \tilde{m}^{e,s}_{\text{eff},\mathcal{R}^{PC}}(t) a| < 2 \sigma$, i.e.\ a more restrictive condition shifted by $a$. This leads to a more conservative fitting range with $t'_\text{min} > t_\text{min}$. When using $t'_\text{min}$ instead of $t_\text{min}$, statistical errors are increased by around $50 \%$. Such an increase, is expected, because the signal-to-noise ratio of correlation functions becomes worse with increasing $t$. Most importantly, mass splittings obtained with $t_\text{min}$ (i.e.\ those collected in Table~\ref{tab:jointfits_continuummass}) and with $t'_\text{min}$ are consistent within statistical errors. There is also no clear systematic trend, i.e.\ $t'_\text{min}$ mass splittings are not generally smaller, but in several cases also larger than $t_\text{min}$ mass splittings. We interpret this as indication that excited states are strongly suppressed and that their effect is small compared to statistical errors.
		We quote the differences between those two sets of results as systematic errors (the third of the three errors provided in Table~\ref{tab:jointfits_continuummass}).
		
		As in Section~\ref{sec:continuumextrapolation_effmassfit} we also checked the validity and stability of our continuum extrapolation by extending the fit function~\eqref{eq:joint_5_parameters_fit} by a term proportional to $a^4$,
		\begin{equation}
			\label{eq:joint_7_parameters_fit}
			\tilde{C}^{\text{fit},e,s}_{\mathcal{R}^{PC}}(t) = A^e_{\mathcal{R}^{PC}} \exp\Big(-(\Delta m_{\mathcal{R}^{PC},\text{cont}} + c_{\mathcal{R}^{PC}} a^2 + d_{\mathcal{R}^{PC}} a^4) t\Big) ,
		\end{equation}
		and at the same time including correlator data from ensemble $A$ with the coarsest lattice spacing in the fits. All 19 resulting continuum extrapolated gluelump mass splittings obtained from such 7-parameter fits are consistent with those listed in Table~\ref{tab:jointfits_continuummass} (obtained from 5-parameter fits) within statistical errors. We quote the differences between those two sets of results as systematic errors (the second of the three errors provided in Table~\ref{tab:jointfits_continuummass}).
		
\begin{figure}[p]
			\includegraphics[width=0.5\linewidth,page=1]{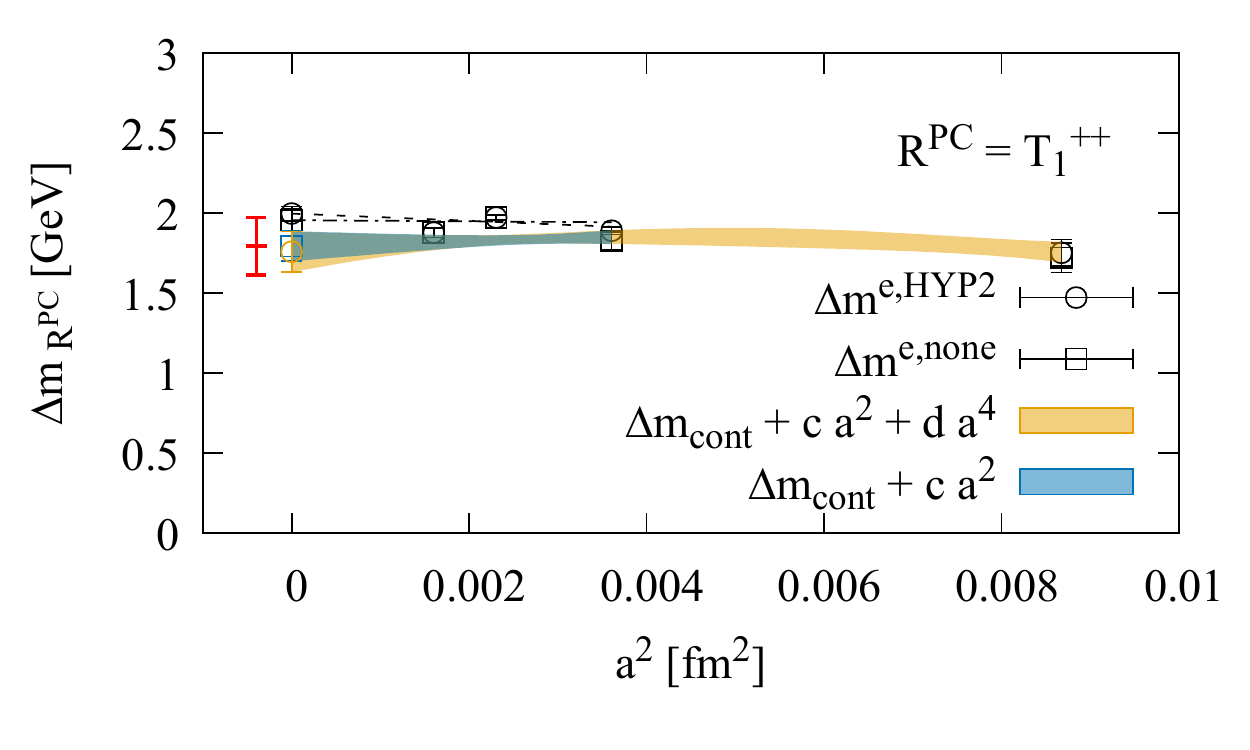}\\
			\includegraphics[width=0.5\linewidth,page=2]{Figure3_4_5}
			\includegraphics[width=0.5\linewidth,page=3]{Figure3_4_5}
			\vspace{-1.0cm}
			
			\rule{\textwidth}{0.4pt}
			\includegraphics[width=0.5\linewidth,page=4]{Figure3_4_5}
			\includegraphics[width=0.5\linewidth,page=5]{Figure3_4_5}
			\includegraphics[width=0.5\linewidth,page=6]{Figure3_4_5}
			\includegraphics[width=0.5\linewidth,page=7]{Figure3_4_5}
			\caption{\label{fig:m_a_comparison_fits1}Summary of the results on gluelump mass splittings from Section~\ref{sec:continuumextrapolation_effmassfit} and Section~\ref{sec:continuumextrapolation_method2} for representations $T_1^{PC}$ and $T_2^{PC}$ (see text for details). The final result for each representation is represented by the red data point at $a^2 = 0$ (also provided in Table~\ref{tab:jointfits_continuummass}).}
		\end{figure}

\begin{figure}[p]
			\includegraphics[width=0.5\linewidth,page=8]{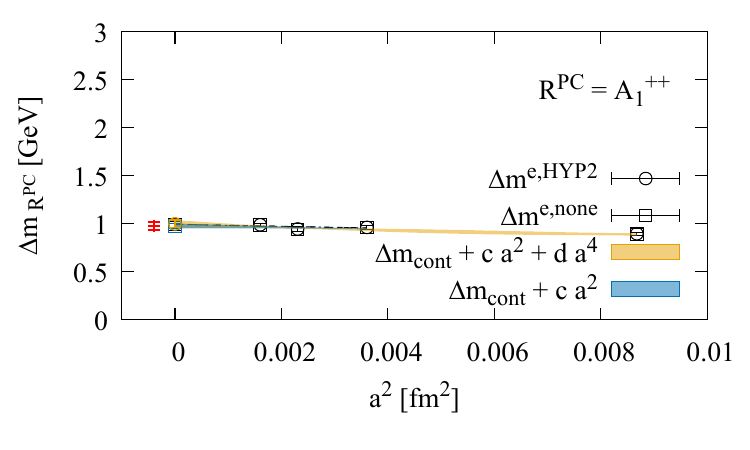}
			\includegraphics[width=0.5\linewidth,page=9]{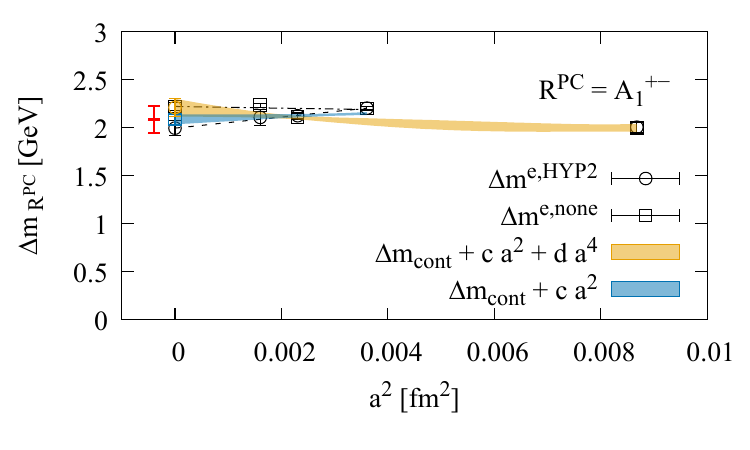}
			\includegraphics[width=0.5\linewidth,page=10]{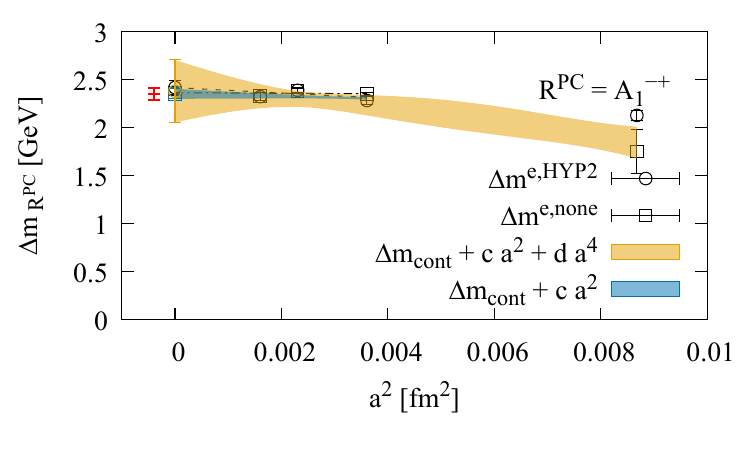}
			\includegraphics[width=0.5\linewidth,page=11]{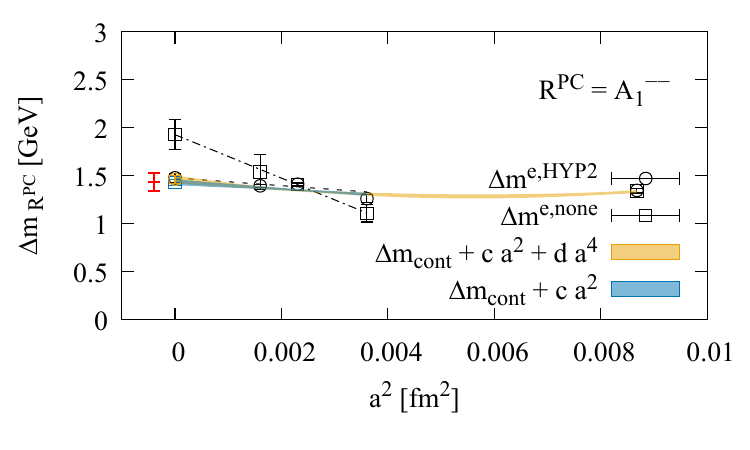}
			\vspace{-1.0cm}

			\rule{\textwidth}{0.4pt}
			\includegraphics[width=0.5\linewidth,page=12]{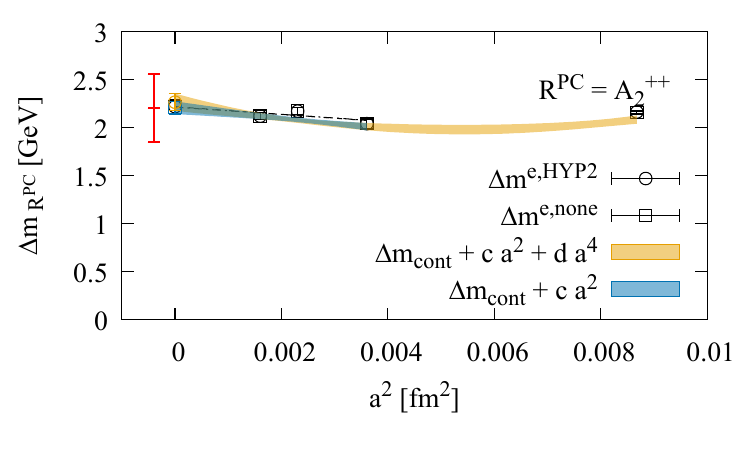}
			\includegraphics[width=0.5\linewidth,page=13]{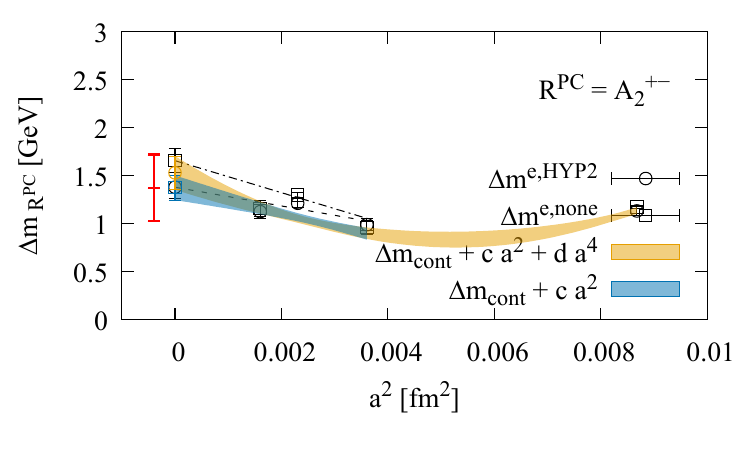}
			\includegraphics[width=0.5\linewidth,page=14]{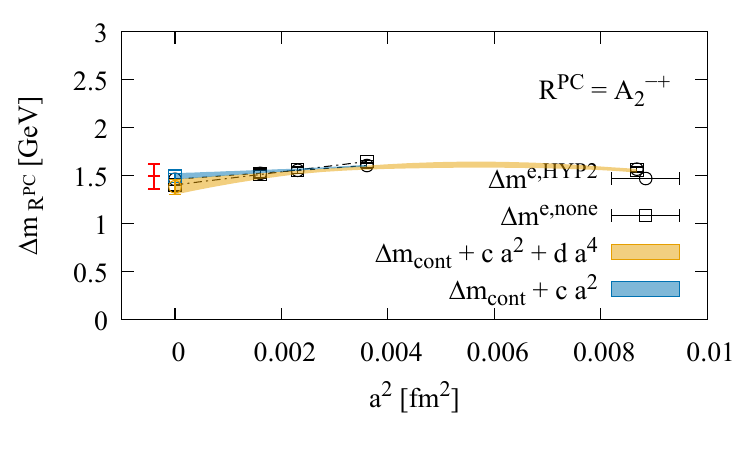}
			\includegraphics[width=0.5\linewidth,page=15]{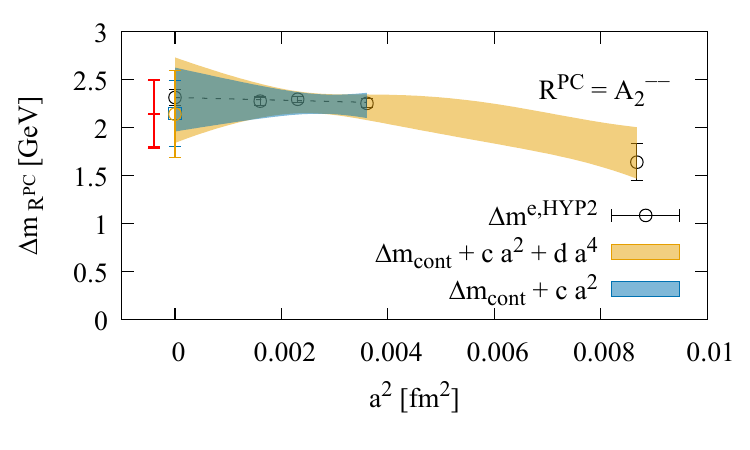}
			\caption{\label{fig:m_a_comparison_fits2}Summary of the results on gluelump mass splittings from Section~\ref{sec:continuumextrapolation_effmassfit} and Section~\ref{sec:continuumextrapolation_method2} for representations $A_1^{PC}$ and $A_2^{PC}$ (see text for details). The final result for each representation is represented by the red data point at $a^2 = 0$ (also provided in Table~\ref{tab:jointfits_continuummass}).}
		\end{figure}

\begin{figure}[p]
			\includegraphics[width=0.5\linewidth,page=16]{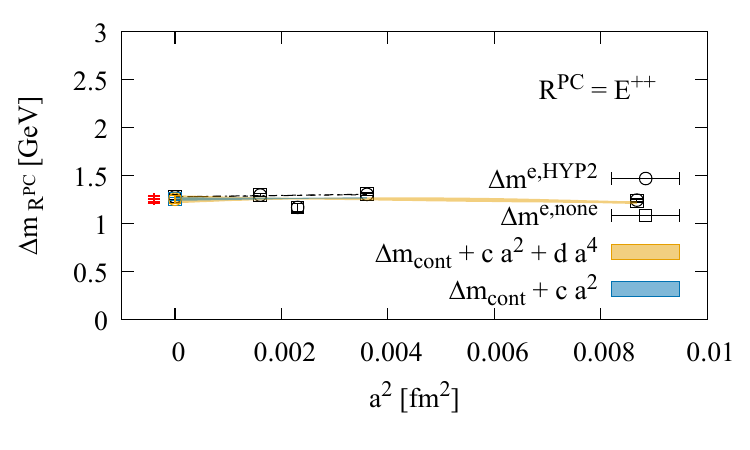}
			\includegraphics[width=0.5\linewidth,page=17]{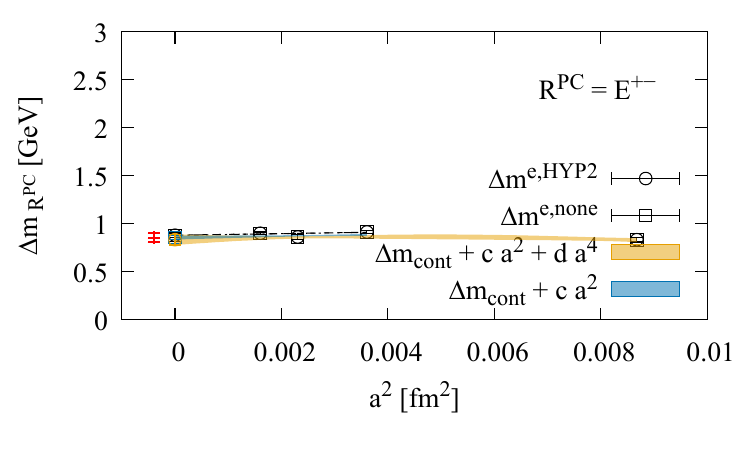}
			\includegraphics[width=0.5\linewidth,page=18]{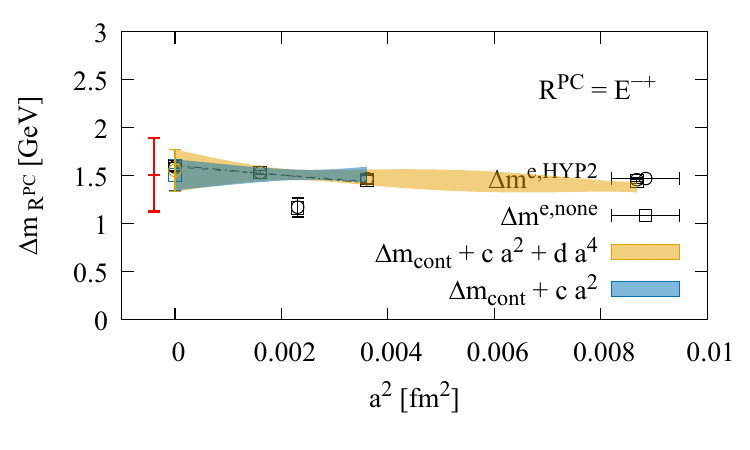}
			\includegraphics[width=0.5\linewidth,page=19]{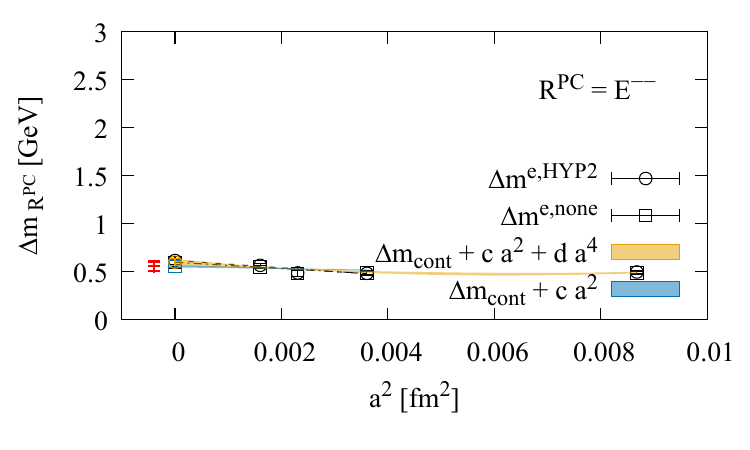}
		\caption{\label{fig:m_a_comparison_fits3}Summary of the results on gluelump mass splittings from Section~\ref{sec:continuumextrapolation_effmassfit} and Section~\ref{sec:continuumextrapolation_method2} for representations $E^{PC}$ (see text for details). The final result for each representation is represented by the red data point at $a^2 = 0$ (also provided in Table~\ref{tab:jointfits_continuummass}).}
		\end{figure}

		In Figure~\ref{fig:m_a_comparison_fits1} to Figure~\ref{fig:m_a_comparison_fits3} we summarize our results on gluelump mass splittings from Section~\ref{sec:continuumextrapolation_effmassfit} and from this subsection.
		\begin{itemize}
		\item The black curves and data points represent results from Section~\ref{sec:continuumextrapolation_effmassfit}.
			\begin{itemize}
				\item The black data points show the gluelump mass splittings, $\Delta m^{e,s}_{\mathcal{R}^{PC}}$ corresponding to finite lattice spacing (see Table~\ref{tab:Delta amRPC} in App.~\ref{app:gluelump_mass_splittings}), as well as the continuum extrapolations $\Delta m^{s}_{\mathcal{R}^{PC},\text{cont}}$ (see Table~\ref{tab:continuumsmRPC}).
				\item The black dashed curves correspond to the fit function~\eqref{eq:continuumextrapolation_mass_finite_a},
					$\Delta m^{\text{fit},s}_{\mathcal{R}^{PC}}(a)$ (see Section~\ref{sec:continuumextrapolation_effmassfit}).
			\end{itemize}
		
		% ---
		
		\item The blue curves and data points represent fit results obtained with the 5-parameter fit function~\eqref{eq:joint_5_parameters_fit}:
		\begin{itemize}
		\item The error bands show
		\begin{equation}
			\label{EQN962} \Delta m_{\mathcal{R}^{PC},\text{cont}} + c_{\mathcal{R}^{PC}} a^2
		\end{equation}
		appearing in the exponent of the fit function~\eqref{eq:joint_5_parameters_fit} and its statistical uncertainty.
		
		\item The data points at $a^2 = 0$ correspond to $\Delta m_{\mathcal{R}^{PC},\text{cont}}$ and their statistical error (see Table~\ref{tab:jointfits_continuummass}). 
		\end{itemize}
		
		\item The orange curves and data points represent fit results obtained with the 7-parameter fit function~\eqref{eq:joint_7_parameters_fit}:
		\begin{itemize}
		\item The error bands show
		\begin{equation}
			\label{EQN963} \Delta m_{\mathcal{R}^{PC},\text{cont}} + c_{\mathcal{R}^{PC}} a^2 + d_{\mathcal{R}^{PC}} a^4
		\end{equation}
		appearing in the exponent of the fit function~\eqref{eq:joint_7_parameters_fit} with its statistical uncertainty.
		
		\item The data points at $a^2 = 0$ correspond to $\Delta m_{\mathcal{R}^{PC},\text{cont}}$ and its statistical error.
		\end{itemize}
		
		\item The red data points represent the main results of this work, continuum extrapolated gluelump mass splittings $\Delta m_{\mathcal{R}^{PC},\text{cont}}$ obtained with the 5-parameter fit function~\eqref{eq:joint_5_parameters_fit}. Uncertainties include both the statistical and the two systematic errors, as quoted in Table~\ref{tab:jointfits_continuummass}, added quadratically.
		
		% ---
		
		\end{itemize}
		
		% **********
		
%		\newpage
		
		\subsubsection{\label{sec:spinidentification}Assigning continuum total angular momentum}
		
		In the following we try to assign the correct continuum total angular momenta $J$ to the lattice gluelump masses computed in the previous sections. We start by repeating our cautionary remarks made at the beginning of Section~\ref{sec:numericalresults}. On a cubic spatial lattice each of the five irreducible representations of the cubic group, denoted by $\mathcal{R}$, contains an infinite number of continuum total angular momenta $J$,
		\begin{eqnarray}
			\begin{array}{lcl}
				A_1 & \leftrightarrow & 0, 4, 6, 8, \ldots \\
				T_1 & \leftrightarrow & 1, 3, 4, 5, \ldots \\
				T_2 & \leftrightarrow & 2, 3, 4, 5, \ldots \\
				E   & \leftrightarrow & 2, 4, 5, 6, \ldots \\
				A_2 & \leftrightarrow & 3, 6, 7, 9, \ldots
			\end{array}
		\end{eqnarray}
		(see e.g.\ Ref.~\cite{Johnson:1982yq}). Moreover, there are cases in the gluelump spectra, where states with the same lattice quantum numbers $\mathcal{R}^{P C}$, but different continuum $J$ could have similar masses. In such cases it is not obvious, which is the correct $J$ for such a state. Two competing states with similar masses may also generate a fake effective mass plateau within statistical errors and one might extract an energy somewhere between the masses of the two states.
		
\begin{figure}[htb]\centering
			\includegraphics[width=0.7\linewidth,page=1]{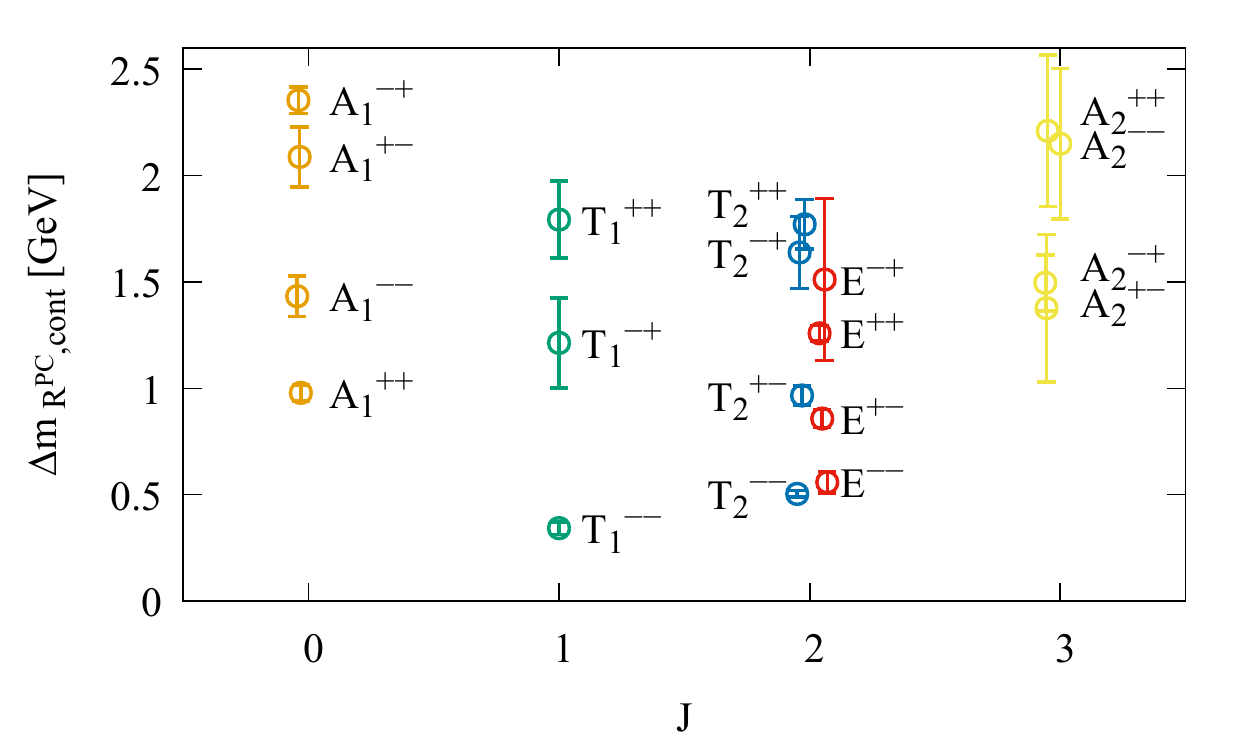}
			\caption{\label{fig:lowestspincontent}Summary of continuum extrapolated gluelump mass splittings. The horizontal axis indicates the lowest continuum total angular momentum $J$ appearing in the corresponding representation of the cubic group.
				The error bars denote statistical and  systematic errors.}
		\end{figure}

		A simple strategy to assign continuum total angular momenta $J$, which was used e.g.\ in Ref.~\cite{Marsh:2013xsa}, is to assume that the lowest state in a cubic representation $\mathcal{R}$ has the smallest allowed $J$ value, i.e.\ $J = 0$ for $A_1$, $J = 1$ for $T_1$, $J = 2$ for $T_2$ and $E$ and $J = 3$ for $A_2$. It is possible to check this assumption to some extent, because the majority of $J$ values appear in more than one cubic representation and one should observe a corresponding pattern in the extracted lattice spectra. In particular, $J = 2$ is the lowest continuum total angular momentum in both $T_2$ and $E$ and, thus, one expects a degeneracy of the lowest energy levels in these cubic representations within uncertainties. There is no obvious contradiction to this assumption in our numerical results, which are collected in Table~\ref{tab:jointfits_continuummass} and summarized graphically in Figure~\ref{fig:lowestspincontent}. 
		On the other hand, there are other possible $J$ assignments, which are also plausible or might in some cases even be more likely. 
		In the following we discuss this individually for all $\mathcal{R}^{PC}$ representations.
		\begin{itemize}
		% *****
		\item $A_1$ states:
		\begin{itemize}
		\item $A_1^{PC}$ $\ \rightarrow \ $ probably $J = 0$: \\ The two lowest $J$ values contained in $A_1$ are $J = 0$ and $J = 4$. Since significantly larger angular momenta are typically associated with larger energies, it seems natural to assign $J = 0$ to the $A_1$ energy levels. In particular, for the lighter $A_1$ states, $A_1^{++}$ and $A_1^{--}$, $J = 0$ seems to be the only plausible assignment.
		\end{itemize}
		
		% *****
		
		\item $T_1$ states:
		\begin{itemize}
		\item $T_1^{+-}$ $\ \rightarrow \ $ $J = 1$: \\ Besides $J = 1$ the next-lowest $J$ value contained in $T_1$ is $J = 3$, which is also part of $A_2$. The corresponding energy level for $A_2^{+-}$ is, however, significantly larger than its $T_1$ counterpart, which is a clear sign that $T_1^{+-}$ has $J = 1$.
		
		\item $T_1^{--}$ $\ \rightarrow \ $ $J = 1$: \\ Explanation as for $T_1^{+-}$ (see previous item).
		
		\item $T_1^{-+}$ $\ \rightarrow \ $ could be $J = 1$, but also $J = 3$: \\ The energy level for $T_1^{-+}$ is consistent with the energy level for $A_2^{-+}$, which might be a sign that they correspond to the same $J = 3$ state. On the other hand, the $T_1^{-+}$ energy level has a rather large error and it could also be that $T_1^{-+}$ has $J = 1$ and and $A_2^{-+}$ has $J = 3$, where both states are in the same energy region, $1.0 \, \text{GeV} \ldots 1.5 \, \text{GeV}$ above the lightest $1^{-+}$ gluelump.
		
		\item $T_1^{++}$ $\ \rightarrow \ $ could be $J = 1$, but also $J = 3$: \\ Explanation as for $T_1^{-+}$ (see previous item).
		\end{itemize}
		
		% *****
		
		\item $T_2$ and $E$ states:
		\begin{itemize}
		\item $T_2^{--}$, $E^{--}$ $\ \rightarrow \ $ $J = 2$: \\ Both $T_2$ and $E$ contain $J = 2$. The energy levels for $T_2^{--}$ and $E^{--}$ are degenerate within errors, which provides some indication that they correspond to the same state, which has $J = 2$. In principle, a state appearing in both $T_2$ and $E$ could also have $J = 4$, but this seems unlikely, because the $J = 3$ state is already quite heavy, as indicated by the energy level for $A_2^{--}$, and the energy level for $A_1^{--}$, which can be considered as a lower bound for the $J = 4$ energy, is significantly above the $T_2^{--}$ and $E^{--}$ energies. Thus, $J = 2$ seems to be the only plausible assignment.
		
		\item $T_2^{+-}$, $E^{+-}$ $\ \rightarrow \ $ $J = 2$: \\ Explanation as for $T_2^{--}$, $E^{--}$ (see previous item).
		
		\item $E^{++}$ $\ \rightarrow \ $ $J = 2$ (discard $T_2^{++}$): \\ The $T_2^{++}$ energy level is around $3 \sigma$ above the $E^{++}$ energy level. This is surprising, because there is no small $J$ value in $E^{++}$, which is not as well part of $T_2^{++}$. It could be that this discrepancy is just a statistical fluctuation. Another possible explanation is that the overlaps generated by the $T_2^{++}$ operator are not favorable for an extraction of the lowest state in this sector. For example, there could be a large overlap to a rather heavy $J = 3$ state (the $A_2^{++}$ energy level indicates that $3^{++}$ is quite heavy), which generates a fake effective mass plateau within errors. In any case, the assignment of $J = 2$ to the $E^{++}$ energy level seems to be plausible, while the interpretation of the $T_2^{++}$ energy level is less clear and, thus, should be discarded.
		
		\item $T_2^{-+}$, $E^{-+}$ $\ \rightarrow \ $ not inconsistent with $J = 2$: \\ Both $T_2$ and $E$ contain $J = 2$. The energy levels for $T_2^{-+}$ and $E^{-+}$ are degenerate within errors, indicating that they could correspond to the same state, probably with $J = 2$. However, their errors, in particular that of $E^{-+}$ is very large, such that other $J$ assignments cannot be ruled out.
		\end{itemize}
		
		% *****
		
		\item $A_2$ states:
		\begin{itemize}
		\item $A_2^{PC}$ $\ \rightarrow \ $ probably $J = 3$: \\ The two lowest $J$ values contained in $A_2$ are $J = 3$ and $J = 6$. Following the same argument as previously for the $A_1$ states, $J = 3$ seems rather plausible and $J = 6$ rather unlikely. For $A_2^{-+}$ and $A_2^{++}$ this is supported by our discussion of the $T_1^{-+}$, and $T_1^{++}$ states above, where we have presented scenarios, which imply $J = 3$ for $A_2^{-+}$ and $A_2^{++}$.
		\end{itemize}
		
		% *****
		
		\end{itemize}
		
		We generate final results for $J^{PC} = 2^{+-}$, $J^{PC} = 2^{-+}$ and $J^{PC} = 2^{--}$ by carrying out additional combined fits for the representations $T_2^{+-}$ and $E^{+-}$, $T_2^{-+}$ and $E^{-+}$ and $T_2^{--}$ and $E^{--}$, respectively.
		In detail, these are $\chi^2$-minimizing 9-parameter fits of Eq.~\eqref{eq:joint_5_parameters_fit} with a single fit parameter $\Delta m_{\mathcal{R}^{PC},\text{cont}}$ linking the two representations $E^{PC}$ and $T_2^{PC}$, i.e.\ we set \\ $\Delta m_{E^{PC},\text{cont}} = \Delta m_{T_2^{PC},\text{cont}} = \Delta m_{J^{PC}=2^{PC}, \text{cont}}$. 
		The resulting $\chi_\text{red}^2$ values are of $\order{1}$ indicating consistency for these combined fits.
		
		We summarize our final results for gluelump mass splittings with quantum numbers $J^{PC}$ in Table~\ref{tab:final_results}. Energy levels, where the assignment of continuum total angular momentum $J$ is a plausible scenario, but not fully established, are shaded in gray.
		
		\begin{table}
			 \def\arraystretch{1.2}
			\begin{center}
				\begin{tabular}{c|c|c||c|c|c}
					\hline
					$J^{PC}$ & $\Delta m_{{J^{PC}}}$ in GeV & $\mathcal{R}^{PC}$   &   $J^{PC}$ & $\Delta m_{{J^{PC}}}$ in GeV & $\mathcal{R}^{PC}$ \\
					\hline
					$0^{++}$ & $0.979(36)$ & $A_1^{++}$   &   $2^{++}$ & $1.258(24)$ & $E^{++}$ \\
					\cellcolor{lightgray} $0^{+-}$ & \cellcolor{lightgray} $2.088(138)$ & \cellcolor{lightgray} $A_1^{+-}$   &   $2^{+-}$ & $0.925(24)(2)(34) $ & $T_2^{+-}\, \&\, E^{+-} $ combined fit \\
					\cellcolor{lightgray} $0^{-+}$ & \cellcolor{lightgray} $2.354(122)$ & \cellcolor{lightgray} $A_1^{-+}$   &   \cellcolor{lightgray} $2^{-+}$ & \cellcolor{lightgray} $1.664(107)(126)(195)$ & \cellcolor{lightgray}$T_2^{-+}\, \&\, E^{-+} $ combined fit \\
					$0^{--}$ & $1.433(47)$ & $A_1^{--}$   &   $2^{--}$ & $0.523(9)(7)(1)$ & $T_2^{--}\, \&\, E^{--} $ combined fit \\
					\hline
					\cellcolor{lightgray} $1^{++}$ & \cellcolor{lightgray} $1.793(108)$ & \cellcolor{lightgray} $T_1^{++}$   &   \cellcolor{lightgray} $3^{++}$ & \cellcolor{lightgray} $2.210(95)$ & \cellcolor{lightgray} $A_2^{++}$ \\
					$1^{+-}$ & $0$         & $-$          &   $3^{+-}$ & $1.376(210)$ & $A_2^{+-}$ \\
					\cellcolor{lightgray} $1^{-+}$ & \cellcolor{lightgray} $1.213(64)$ & \cellcolor{lightgray} $T_1^{-+}$   &   $3^{-+}$ & $1.496(116)$ & $A_2^{-+}$ \\
					$1^{--}$ & $0.342(36)$ & $T_1^{--}$   &   \cellcolor{lightgray} $3^{--}$ & \cellcolor{lightgray} $2.149(340)$ & \cellcolor{lightgray} ${A_2^{--}}^{\star}$ \\
					\hline
				\end{tabular}
			\end{center}
			\caption{\label{tab:final_results}Final results for gluelump mass splittings with quantum numbers $J^{PC}$. The errors include statistical as well as systematic errors (added in quadrature). The column $\mathcal{R}^{PC}$ indicates, from which cubic representation the result was taken.
			For $J=2$ and $PC = +-, -+, --$ we generate the final results by carrying out additional combined fits (see text for details).
			Energy levels, where the assignment of continuum total angular momentum $J$ is a plausible scenario, but not fully established, are shaded in gray.}
		\end{table}
		
		There are possibilities to check, whether there are close-by competing states in certain $\mathcal{R}^{PC}$ representations with continuum quantum numbers $J_1 \neq J_2$, and to resolve and determine the masses of both states reliably. For example one could design not just one but several operators generating $\mathcal{R}^{PC}$ trial states. If some of these trial states are similar to continuum states with $J_1$ and others to continuum states with $J_2$, the corresponding correlation matrix should allow to determine both energy levels. If this is done e.g.\ by solving a generalized eigenvalue problem, the eigenvector components should provide information concerning the continuum total angular momenta of the extracted states. While this seems to be non-trivial for gluelumps and has not been attempted previously in the literature, it might be an interesting direction for future work.

		% **********
		
		\subsubsection{Comparison of gluelump mass splittings to existing lattice results}
				
\begin{figure}
			\includegraphics[width=0.49\linewidth,page=1]{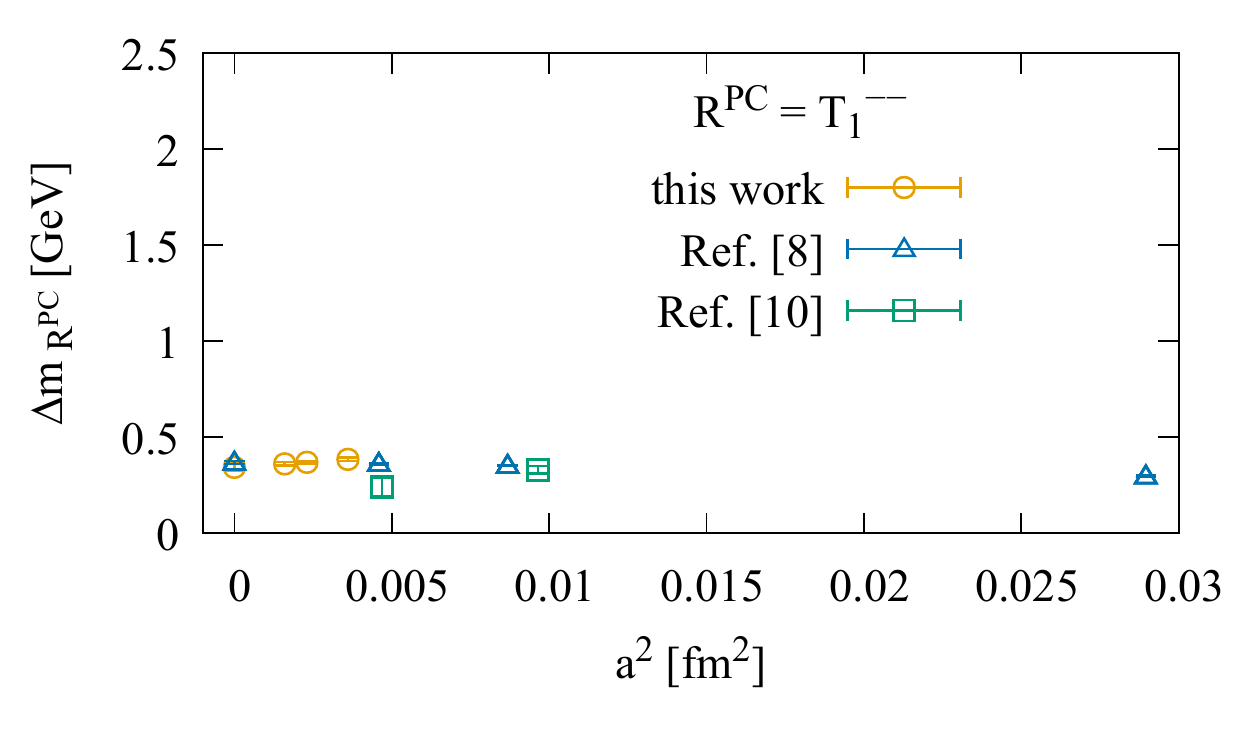}
			\includegraphics[width=0.49\linewidth,page=2]{Figure7}
			\includegraphics[width=0.49\linewidth,page=3]{Figure7}
			\includegraphics[width=0.49\linewidth,page=4]{Figure7}
			\includegraphics[width=0.49\linewidth,page=5]{Figure7}
		\caption{\label{fig:comparison}Comparison of our results for gluelump mass splittings to results from Refs.~\cite{Foster:1998wu,Marsh:2013xsa}. Error bars represent exclusively statistical errors.}
		\end{figure}

		In Figure~\ref{fig:comparison} we compare our results for gluelump mass splittings $\Delta m_{\mathcal{R}^{PC}}$ to results from similar computations from Refs.~\cite{Foster:1998wu,Marsh:2013xsa}. We show plots for the cubic representations \\ $\mathcal{R}^{PC} = T_1^{--} , T_2^{--} , A_1^{++} , A_2^{+-} , E^{+-}$, for which continuum extrapolations were carried out in Ref.~\cite{Foster:1998wu}.
		\begin{itemize}		
		\item The orange data points represent our results, generated by evaluating $\Delta m_{\mathcal{R}^{PC}, \text{cont}} + c_{\mathcal{R}^{PC}} a^2$ at $a = 0$ and at our three smallest lattice spacings $a = 0.040 \,\text{fm} , \, 0.048 \, \text{fm} , \, 0.060 \, \text{fm}$ (see Section~\ref{sec:continuumextrapolation_method2}, in particular Table~\ref{tab:jointfits_continuummass}). 
		
		\item The blue data points are the results from Ref.~\cite{Foster:1998wu} for three lattice spacings \\ $a = 0.068\,\text{fm} ,\, 0.095\,\text{fm} ,\, 0.170 \,\text{fm} $ and a corresponding continuum extrapolation linear in $a^2$, which is similar to the method we used in Section~\ref{sec:continuumextrapolation_effmassfit}. Ref.~\cite{Foster:1998wu} is as well a computation in pure SU(3) gauge theory, i.e.\ without dynamical quarks. Thus the continuum extrapolated gluelump mass splittings should be directly comparable to our work. While there is qualitative agreement for the five shown representations, one can observe quantitative discrepancies of up to $\approx 30 \%$. Since we extract the continuum mass splittings from a combined fit to a large number of correlator data points (see Section~\ref{sec:continuumextrapolation_method2}) instead of extrapolating to $a=0$ with just a three data points as done in Ref.~\cite{Foster:1998wu}, and since our lattice spacings are significantly smaller than those from Ref.~\cite{Foster:1998wu}, we consider our continuum extrapolations superior and more trustworthy than those from Ref.~\cite{Foster:1998wu}.
		
		\item The green data points are the results from Ref.~\cite{Marsh:2013xsa} for two lattice spacings \\ $a = 0.0685 \,\text{fm} , \, 0.0982 \, \text{fm}$. A continuum extrapolation was not carried out in Ref.~\cite{Marsh:2013xsa}. Since the computations in Ref.~\cite{Marsh:2013xsa} were done in full QCD, i.e.\ with dynamical quarks (the corresponding pion mass is around $3.5$ times heavier than its physical value), a quantitative comparison might exhibit certain discrepancies. Still, one can expect qualitative agreement, because gluelumps are extracted from purely gluonic correlation functions. This expectation is reflected by the plots in Figure~\ref{fig:comparison}.
		\end{itemize}

		% ********************
		% ********************
		% ********************
		
%		\newpage

		\subsection{\label{sec:hybridpotentials_and_gluelumps}Gluelumps as the $r \rightarrow 0$ limit of hybrid static potentials}
		
		Hybrid static potentials in the continuum are typically characterized by quantum numbers $\Lambda_\eta^\epsilon$, where $\Lambda = 0, 1, 2, \ldots \equiv \Sigma , \Pi , \Delta, \ldots$ is the absolute value of total angular momentum with respect to the axis of separation of the static charges, $\eta$ corresponds to $\mathcal{P} \circ \mathcal{C}$ and $\epsilon$ denotes the behavior under reflection with respect to an axis perpendicular to the separation axis (for details see e.g.\ Ref.~\cite{Capitani:2018rox}). Because of the separation of the static charges, a particular axis is singled out and, consequently, the symmetry group is different from that of gluelumps. In the limit of vanishing charge separation $r$, rotational symmetry is, however, restored. Since the static quark-antiquark pair in the fundamental representation (for hybrid static potentials) is then equivalent to a static quark in the adjoint representation (needed for gluelumps), gluelumps can be interpreted as the $r \rightarrow 0$ limit of hybrid static potentials. The correspondence between gluelump quantum numbers $J^{PC}$ and hybrid static potential quantum numbers $\Lambda_\eta^\epsilon$ in the limit $r \rightarrow 0$ is discussed e.g.\ in Ref.\cite{Berwein:2015vca} and summarized in Table~\ref{tab:correspondence_gluelumps_hybridpotentials}.
		
		\begin{table}[h]\centering
			\def\arraystretch{1.2}
			\begin{tabular}{l|l}
				$J^{PC}$ & $\Lambda_{\eta}^{\epsilon}$ \\ \hline
				$1^{+-}$ & $\Pi_u,\, \Sigma_u^-$       \\
				$1^{--}$ & $\Pi_g,\, \Sigma_g^{+ \prime}$       \\
				$2^{--}$ & $\Pi_g^{\prime},\, \Sigma_g^{-},\, \Delta_g$       \\
				$2^{+-}$ & $\Pi_u^{\prime},\, \Sigma_u^{+},\, \Delta_u$       \\
			\end{tabular}
		\caption{Correspondence between gluelump quantum numbers $J^{PC}$ and hybrid static potential quantum numbers $\Lambda_\eta^\epsilon$ in the limit $r \rightarrow 0$.}
		\label{tab:correspondence_gluelumps_hybridpotentials}
		\end{table}
		
\begin{figure}[h]
		\centering\includegraphics[width=0.66\linewidth,page=1]{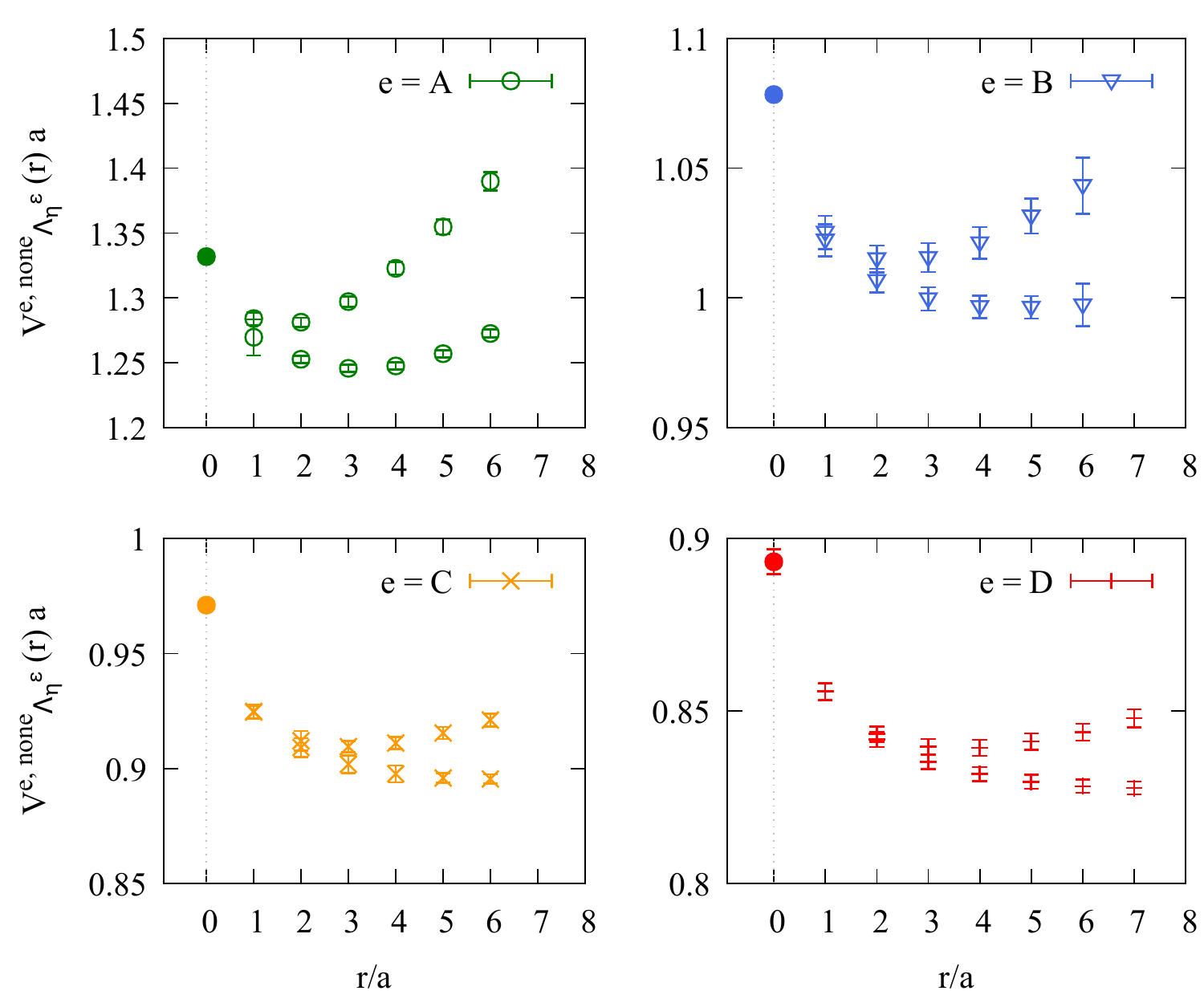}
		\caption{\label{fig:lowesthybridpotentials_and_gluelump_ABCD}
			Hybrid static potentials $V^{e,\text{none}}_{\Lambda_\eta^\epsilon}(r) a$ with $\Lambda_{\eta}^{\epsilon} = \Pi_u$ (lower curves) and $\Lambda_{\eta}^{\epsilon} = \Sigma_u^-$ (upper curves) for $r/a \geq 1$ and gluelump masses $m^{e,\text{none}}_{T_1^{+-}} a$ at $r/a = 0$.
			The hybrid static potential data points at $r/a=1$ were computed in the context of this work, while those for $r/a \geq 2$ were taken from Ref.~\cite{Schlosser:2021wnr}.}
		\end{figure}

		In Refs.~\cite{Capitani:2018rox,Schlosser:2021wnr} we have computed hybrid static potentials using the same lattice setup as for this work (see Section~\ref{sec:computationaldetails}). We complement our lattice results from Refs.~\cite{Schlosser:2021wnr} in Table~\ref{tab:aV R1}, where we quote $T_1^{+-}$ gluelump masses from Table~\ref{tab:amRPC} with $s = \text{none}$ as $r \rightarrow 0$ limits of $\Pi_u$ and $\Sigma_u^-$ hybrid static potentials. We also provide previously unpublished results for $r = a$ computed as discussed in Ref.~\cite{Schlosser:2021wnr}. In Figure~\ref{fig:lowesthybridpotentials_and_gluelump_ABCD} we plot these $\Pi_u$ and $\Sigma_u^-$ hybrid static potentials for each of our four lattice spacings, i.e.\ $e \in \{ A , B , C , D \}$, together with the $T_1^{+-}$ gluelump masses at $r = 0$. 
		
		\begin{table}[htb]\centering
			\begin {tabular}{cllll}%
\toprule ensemble&$r/a$&$V^{e,\text {none}}_{\Sigma _g^+ }\, a$&$V^{e,\text {none}}_{\Pi _u^{\phantom {+}} }\, a$&$V^{e,\text {none}}_{\Sigma _u^- }\, a$\\\midrule %
\multirow {2}{*}{$A$}&\pgfutilensuremath {0}&-&\pgfmathprintnumber [fixed,fixed zerofill,precision=4]{1.33190446543e+00}(\pgfmathprintnumber [fixed,fixed zerofill,precision=0]{2.2199646e1})&\pgfmathprintnumber [fixed,fixed zerofill,precision=4]{1.33190446543e+00}(\pgfmathprintnumber [fixed,fixed zerofill,precision=0]{2.2199646e1})\\%
&\pgfutilensuremath {1}&\pgfmathprintnumber [fixed,fixed zerofill,precision=6]{4.11037565602e-01}(\pgfmathprintnumber [fixed,fixed zerofill, set thousands separator={},precision=0]{2.6894943e1})&\pgfmathprintnumber [fixed,fixed zerofill,precision=4]{1.26966225673e+00}(\pgfmathprintnumber [fixed,fixed zerofill,precision=0]{1.3907059e2})&\pgfmathprintnumber [fixed,fixed zerofill,precision=4]{1.28406796119e+00}(\pgfmathprintnumber [fixed,fixed zerofill,precision=0]{4.7954681e1})\\%
\midrule \multirow {2}{*}{$B$}&\pgfutilensuremath {0}&-&\pgfmathprintnumber [fixed,fixed zerofill,precision=4]{1.07774148529e+00}(\pgfmathprintnumber [fixed,fixed zerofill,precision=0]{2.0461624e1})&\pgfmathprintnumber [fixed,fixed zerofill,precision=4]{1.07774148529e+00}(\pgfmathprintnumber [fixed,fixed zerofill,precision=0]{2.0461624e1})\\%
&\pgfutilensuremath {1}&\pgfmathprintnumber [fixed,fixed zerofill,precision=6]{3.65472291336e-01}(\pgfmathprintnumber [fixed,fixed zerofill, set thousands separator={},precision=0]{8.6986908e0})&\pgfmathprintnumber [fixed,fixed zerofill,precision=4]{1.02218677970e+00}(\pgfmathprintnumber [fixed,fixed zerofill,precision=0]{6.2658585e1})&\pgfmathprintnumber [fixed,fixed zerofill,precision=4]{1.02527902455e+00}(\pgfmathprintnumber [fixed,fixed zerofill,precision=0]{6.3938538e1})\\%
\midrule \multirow {2}{*}{$C$}&\pgfutilensuremath {0}&-&\pgfmathprintnumber [fixed,fixed zerofill,precision=4]{9.71009515125e-01}(\pgfmathprintnumber [fixed,fixed zerofill,precision=0]{1.5022705e1})&\pgfmathprintnumber [fixed,fixed zerofill,precision=4]{9.71009515125e-01}(\pgfmathprintnumber [fixed,fixed zerofill,precision=0]{1.5022705e1})\\%
&\pgfutilensuremath {1}&\pgfmathprintnumber [fixed,fixed zerofill,precision=6]{3.45080991088e-01}(\pgfmathprintnumber [fixed,fixed zerofill, set thousands separator={},precision=0]{3.6797363e0})&\pgfmathprintnumber [fixed,fixed zerofill,precision=4]{9.24504083915e-01}(\pgfmathprintnumber [fixed,fixed zerofill,precision=0]{2.8032364e1})&\pgfmathprintnumber [fixed,fixed zerofill,precision=4]{9.24759791082e-01}(\pgfmathprintnumber [fixed,fixed zerofill,precision=0]{2.9150635e1})\\%
\midrule \multirow {2}{*}{$D$}&\pgfutilensuremath {0}&-&\pgfmathprintnumber [fixed,fixed zerofill,precision=4]{8.97840115443e-01}(\pgfmathprintnumber [fixed,fixed zerofill,precision=0]{1.6568146e1})&\pgfmathprintnumber [fixed,fixed zerofill,precision=4]{8.97840115443e-01}(\pgfmathprintnumber [fixed,fixed zerofill,precision=0]{1.6568146e1})\\%
&\pgfutilensuremath {1}&\pgfmathprintnumber [fixed,fixed zerofill,precision=6]{3.29925020271e-01}(\pgfmathprintnumber [fixed,fixed zerofill, set thousands separator={},precision=0]{2.132338e0})&\pgfmathprintnumber [fixed,fixed zerofill,precision=4]{8.55574550272e-01}(\pgfmathprintnumber [fixed,fixed zerofill,precision=0]{2.4225159e1})&\pgfmathprintnumber [fixed,fixed zerofill,precision=4]{8.55680920987e-01}(\pgfmathprintnumber [fixed,fixed zerofill,precision=0]{2.3741959e1})\\\bottomrule %
\end {tabular}%

			\caption{Addendum to Table~6 from Ref~\cite{Schlosser:2021wnr}. $\Pi_u$ and $\Sigma_u^-$ hybrid static potentials for $r = 0$ (equivalent to $T_1^{+-}$ gluelump masses) and $\Sigma_g^+$, $\Pi_u$ and $\Sigma_u^-$ (hybrid) static potentials for $r = a$.}
			\label{tab:aV R1}
		\end{table}

\begin{figure}[h]
			\centering\includegraphics[width=0.66\linewidth,page=1]{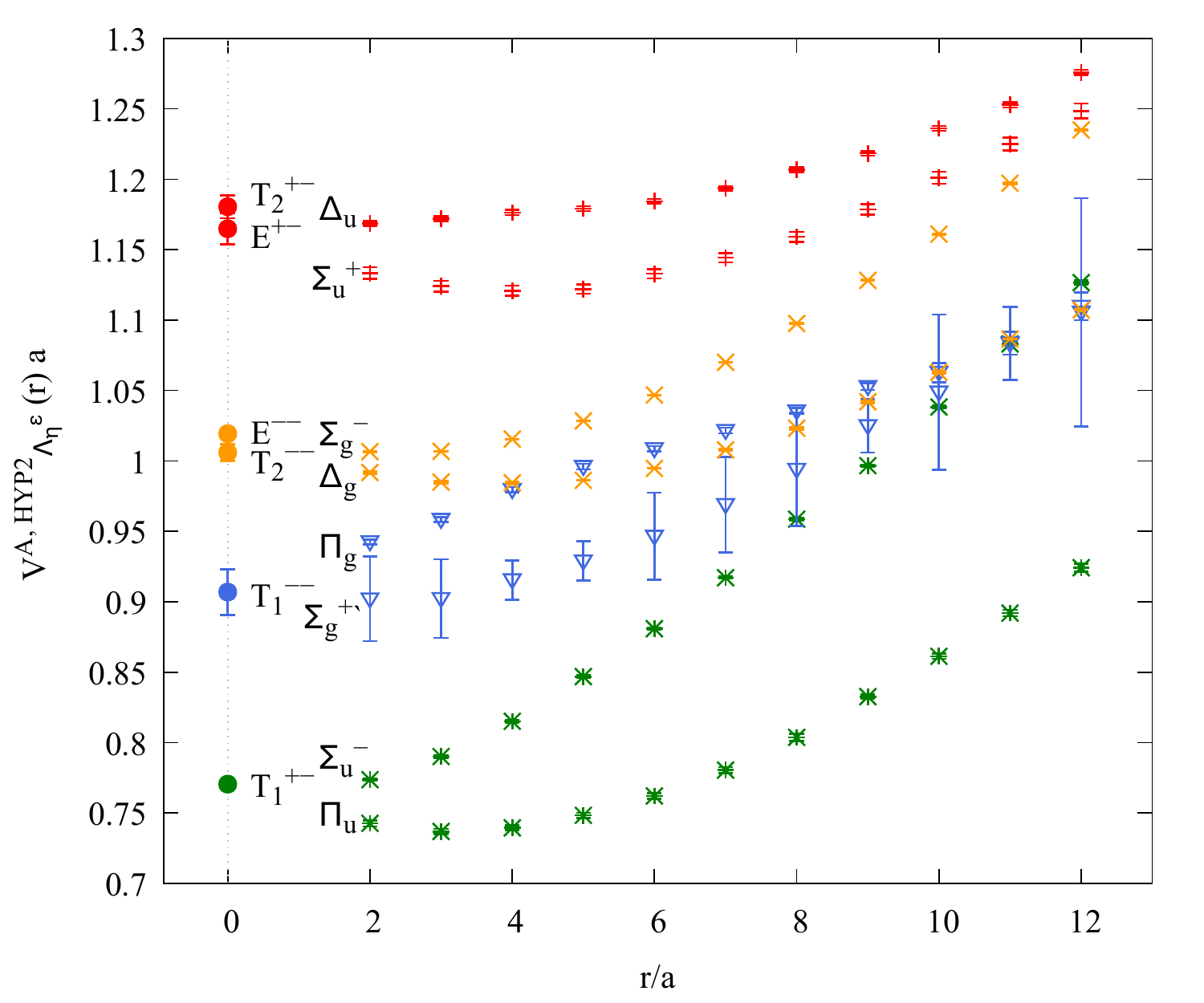}
				\caption{Hybrid static potentials $V^{A,\text{HYP2}}_{\Lambda_\eta^\epsilon}(r) a$ with $\Lambda_{\eta}^{\epsilon} = \Pi_u, \Sigma_u^-, \Sigma_g^{+\prime},\Pi_g,\Delta_g,\Sigma_g^-,\Sigma_u^+,\Delta_u$ for $r/a \geq 2$ and gluelump masses $m^{A,\text{HYP2}}_{\mathcal{R}^{PC}} a$ with $\mathcal{R}^{PC}= T_1^{+-},T_1^{--},T_2^{--},E^{--},T_2^{+-},E^{+-}$ at \\
					 $r/a = 0$.
					The hybrid static potential data points were taken from Ref.~\cite{Capitani:2018rox}.}
				\label{fig:hybridpotentials_and_gluelumps_AHYP2}
		\end{figure}

		In Figure~\ref{fig:hybridpotentials_and_gluelumps_AHYP2} we show even higher hybrid static potentials from our previous work\cite{Capitani:2018rox} computed with a lattice spacing equal to the one of ensemble $A$ with HYP2-smeared temporal links together with the corresponding gluelump masses obtained from ensemble $A$ with $s=\text{HYP2}$. 
		In both Figure~\ref{fig:lowesthybridpotentials_and_gluelump_ABCD} and Figure~\ref{fig:hybridpotentials_and_gluelumps_AHYP2} hybrid static potential and gluelump data points are consistent with smooth curves, which is a valuable cross-check of this work as well as of our previous work \cite{Capitani:2018rox,Schlosser:2021wnr} on hybrid static potentials.

		% ********************
		% ********************
		% ********************
		% ********************
		% ********************

\clearpage

		\subsection{Conversion of $1^{+-}$ gluelump masses from the lattice to the RS scheme}\label{sec:RSschemegluelump}
		
		In this subsection we convert our lattice results for the $1^{+-}$ gluelump mass obtained at several values of $a$ into the renormalon subtracted (RS) scheme at a specific scale $2.5/r_0 \approx 1\,\text{GeV}$. 
		The result is an essential input for Born-Oppenheimer effective field theory predictions of heavy hybrid meson masses~\cite{Berwein:2015vca,Brambilla:2018pyn} (the scale $\nu_f = 1 \, \text{GeV}$ was chosen, because it can be interpreted as a cut-off scale fitting in the hierachy of scales of this effective field theory \cite{Pineda:2002se}).
		The accuracy of such predictions is currently limited by the precision of this $1^{+-}$ gluelump mass in the RS scheme.
		We follow the same procedure discussed and employed in Ref.~\cite{Bali:2003jq} using our up-to-date precision lattice data on gluelump masses as input.
		Our aim is to clarify the impact of this more accurate lattice data on the current uncertainty of the $1^{+-}$ gluelump mass in the RS scheme.
		
		% **********
		
			\subsubsection{Method}
			
			We start by summarizing the method of conversion of gluelump masses from the lattice to the RS scheme proposed and used in Ref.~\cite{Bali:2003jq}.
			The key equation is
			\begin{eqnarray}
				\label{EQN004} \Lambda_B^\text{RS}(\nu_f) = \Lambda_B^L(a) - \Big(\delta \Lambda_B^L(a) + \delta \Lambda_B^\text{RS}(\nu_f)\Big) .
			\end{eqnarray}
			$\Lambda_B^L(a) \equiv m_{T_1^{+-}}^{e,\text{none}}$ is the lattice result for the $T_1^{+-}$ gluelump mass obtained with unsmeared temporal links at one of our four lattice spacings, i.e.\ $e = A, B, C, D$ (see Table~\ref{tab:latticesetups4}).
			Numerical values are listed in Table~\ref{tab:amRPC}.	
			$\Lambda_B^\text{RS}(\nu_f)$ is the corresponding scale dependent gluelump mass in the RS scheme.
			The remaining two terms are perturbative expressions, which are discussed below.
			$a$ and $\nu_f$ are independent, but in practice it is advantageous to choose $\nu_f \approx 1 / a$, to avoid large logarithms.
			
			The lattice self-energy $\delta \Lambda^L(a)$ is  given by
			\begin{eqnarray}
				\label{EQN001} \delta \Lambda_B^L(a) = \frac{1}{a} \sum_{n = 0}^{\infty} c_n^{(8,0)} (\alpha_L(a))^{n+1} ,
			\end{eqnarray}
			where $\alpha_L(a)$ is the lattice coupling (see below).
			The coefficients $c_n^{(8,0)}$ were computed in Refs.~\cite{Bali:2013pla,Bali:2013qla} up to $n = 19$ (the label $(8,0)$ indicates a static charge in the adjoint representation and refers to the standard Wilson plaquette action and a static propagator with unsmeared temporal links;
			we use the improved determinations of $c_n^{(8,0)}$ from Ref.~\cite{Bali:2013qla}).
			
			$\delta \Lambda_B^\text{RS}(\nu_f)$ is given by
			\begin{eqnarray}
				\label{EQN002} \delta \Lambda_B^\text{RS}(\nu_f) = \sum_{n = 1}^{\infty} \nu_f\left(\tilde{V}_{s,n}^\text{RS}-\tilde{V}_{o,n}^\text{RS}\right) (\alpha_{\overline{\text{MS}}}(\nu_f))^{n+1}
			\end{eqnarray}
			(see Ref.~\cite{Bali:2003jq}), where $\alpha_{\overline{\text{MS}}}(\nu_f)$ is the $\overline{\text{MS}}$ coupling.
			The coefficients $\tilde{V}_{s,n}^\text{RS}$ and $\tilde{V}_{o,n}^\text{RS}$ are known exactly for $n = 0, 1, 2$ and were estimated for $n = 3, 4$ (see Table~2 in Ref.~\cite{Bali:2003jq} and references therein).
			
\begin{figure}\centering
				\includegraphics[width=0.75\linewidth,page=1]{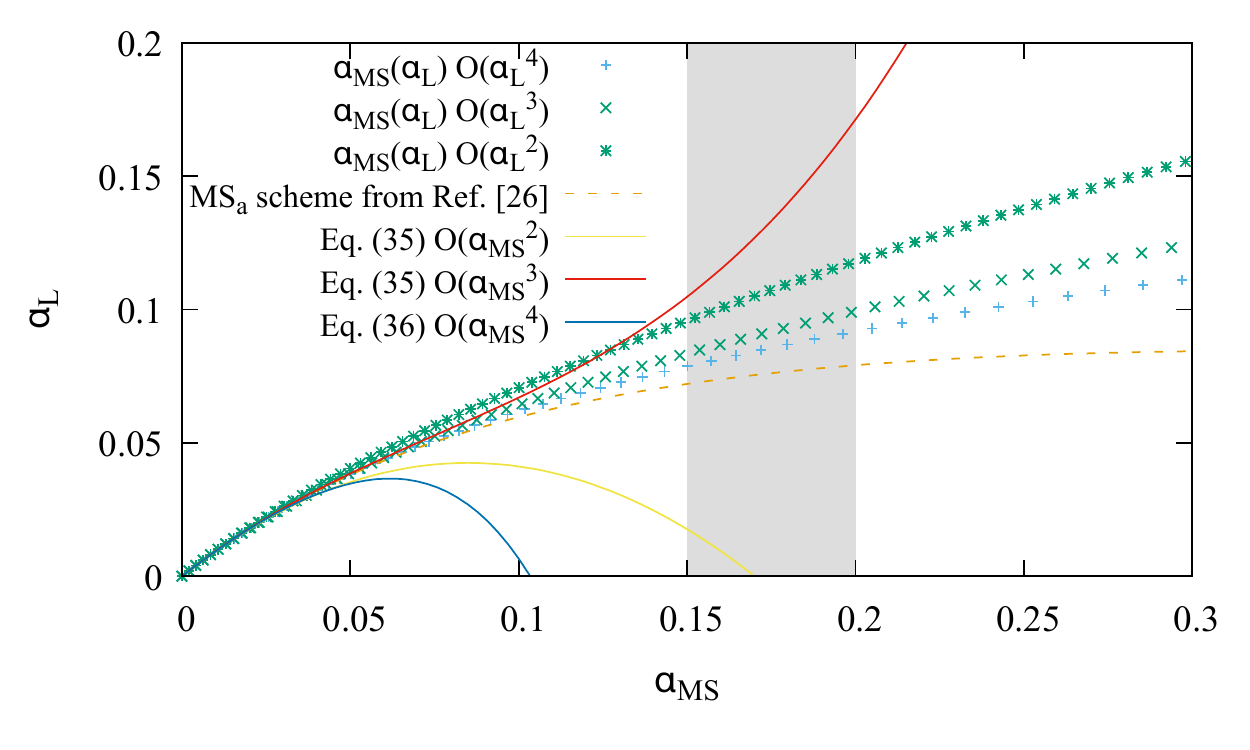}
				\caption{$\alpha_L(a)$ as a function of $\alpha_{\overline{\text{MS}}}(1/a)$.
					The three solid lines represent the polynomial expression \eqref{EQN003} (or equivalently \eqref{EQN003_}) up to $\alpha_{\overline{\text{MS}}}^n$, $n=2,3,4$, with higher orders discarded.
					Similarly, the data points represent a polynomial expansion of $\alpha_{\overline{\text{MS}}}$ in terms of $\alpha_L$, i.e.\ \\ $ \alpha_{\overline{\text{MS}}}=\alpha_L(1+\sum_{j = 0}^{n}d_j\alpha_L^j)$ up to $n=1,2,3$, respectively (see Ref.~\cite{Bali:2013qla} and references therein). 
					The dashed line represents the $\overline{\text{MS}}_a$ conversion scheme (see Eq.~(99) of Ref.~\cite{Bali:2013qla}).
					The shaded region shows the range of $\alpha_{\overline{\text{MS}}}(1/a)$ corresponding to the lattice spacings $0.040 \, \text{fm} \ldots 0.093 \, \text{fm}$ used in this work.
				}
				\label{fig:alpha_L_plot}
			\end{figure}

			The lattice coupling $\alpha_L$ and the $\overline{\text{MS}}$ coupling $\alpha_{\overline{\text{MS}}}$ can be related perturbatively.
			For that we use
			\begin{eqnarray}
				\label{EQN003} \alpha_L(a) = \alpha_{\overline{\text{MS}}}(1/a) \Big(1 - d_1 \alpha_{\overline{\text{MS}}}(1/a) + (2 d_1^2 - d_2) (\alpha_{\overline{\text{MS}}}(1/a))^2\Big)
			\end{eqnarray}
			with $d_1=5.883\ldots$ and $d_2=43.407\ldots$
			% $d_1=5.88359144663707$ and $d_2=43.4073028$
			(see Refs.~\cite{Bali:2003jq,Bali:2013pla} and references therein), which was also used in Ref.~\cite{Bali:2003jq} for the conversion of the gluelump mass.
			We note that in Ref.~\cite{Bali:2013pla} an alternative relation between $\alpha_L$ and $\alpha_{\overline{\text{MS}}}$ is discussed (identical in the leading orders in $\alpha_{\overline{\text{MS}}}$ but different in higher orders), denoted as $\overline{\text{MS}}_a$, which turned out to be superior in the context of that reference. 
			Moreover, in Ref.\ \cite{Bali:2013pla} the estimate $d_3= 352$ is provided such that Eq.~\eqref{EQN003} can be extended by another order in $\alpha_{\overline{\text{MS}}}$,
			\begin{eqnarray}
				\nonumber & & \hspace{-0.7cm} \alpha_L(a) = \alpha_{\overline{\text{MS}}}(1/a) \Big(1 - d_1 \alpha_{\overline{\text{MS}}}(1/a) + (2 d_1^2 - d_2) (\alpha_{\overline{\text{MS}}}(1/a))^2 \\
				\label{EQN003_} & & + (-5d_1^3+3d_1d_2-d_3) (\alpha_{\overline{\text{MS}}}(1/a))^3\Big) .
			\end{eqnarray}
			In Figure~\ref{fig:alpha_L_plot} we compare several truncations of the perturbative expansion of $\alpha_L$ in terms of $\alpha_{\overline{\text{MS}}}$. As expected, there is almost perfect agreement for small $\alpha_{\overline{\text{MS}}} \ltapprox 0.05$.
			For larger values of $\alpha_{\overline{\text{MS}}}$, however, there are sizable discrepancies. This concerns in particular the region \\ $0.15 \ltapprox \alpha_{\overline{\text{MS}}}(1/a) \ltapprox 0.20$, which corresponds to typical lattice spacings $0.040 \, \text{fm} \ldots 0.093 \, \text{fm}$ as used in this work (the region shaded in gray in Figure~\ref{fig:alpha_L_plot}).
			This paragraph and Figure~\ref{fig:alpha_L_plot} is intended as a cautionary remark that systematic errors in the conversion of a gluelump mass due to the perturbative relation between $\alpha_L$ and $\alpha_{\overline{\text{MS}}}$ might be large.
			We leave a future more detailed investigation and discussion of these systematics to experts in the field of perturbation theory. Our aim in the following is to use exactly the same method as in Ref.~\cite{Bali:2003jq}, i.e.\ Eq.\ (\ref{EQN003}), but with our updated and more accurate lattice gluelump masses, to clarify how this improved data affects the final uncertainty of $\Lambda_B^\text{RS}(\nu_f \approx 1 \, \text{GeV})$ quoted in Ref.~\cite{Bali:2003jq}.
			
			Numerical values for $\alpha_{\overline{\text{MS}}}$ are generated via the five-loop running coupling from Ref.~\cite{Baikov:2016tgj} using $r_0 \Lambda_{\overline{\text{MS}}}^{(0)} = 0.624(36)$ \cite{FlavourLatticeAveragingGroupFLAG:2021npn} and $r_0 = 0.5 \, \text{fm}$. In Table~\ref{TAB001} we list both $\alpha_{\overline{\text{MS}}}(1/a)$ and $\alpha_L(a)$ for $a = 0.040 \, \text{fm} \, , \, 0.048 \, \text{fm} \, , \, 0.060 \, \text{fm} \, , \, 0.093 \, \text{fm}$, i.e.\ the four lattice spacings used in our simulations.
			
			\begin{table}[htb]
				\begin{center}
					\begin{tabular}{c|c|c|c|c}
						$\beta$ & $a$ in $\text{fm}$ & $1/a$ in $\text{GeV}$ & $\alpha_{\overline{\text{MS}}}(1/a)$ & $\alpha_L(a)$ (Eq.~\eqref{EQN003}) \\%&$\alpha_L(a)$ (Eq.~\eqref{eq:Msa}) \\
						\hline
						6.000 & 0.093 & 2.118 & 0.200 & 0.172 \\%& 0.0792 \\
						6.284 & 0.060 & 3.285 & 0.170 & 0.127 \\%& 0.0754 \\
						6.451 & 0.048 & 4.108 & 0.158 & 0.113 \\%& 0.0735 \\
						6.594 & 0.040 & 4.932 & 0.150 & 0.104 %& 0.0720 
					\end{tabular}
				\end{center}
				\caption{\label{TAB001}$\alpha_{\overline{\text{MS}}}(1/a)$ from the five-loop running coupling from Ref.~\cite{Baikov:2016tgj} and $\alpha_L(a)$ according to Eq.~\eqref{EQN003} for the four lattice spacings used in our simulations.}
			\end{table}
			
			% **********
			
			\subsubsection{Numerical results}
			
			To convert the $T_1^{+-} \equiv 1^{+-}$ lattice gluelump mass obtained with unsmeared temporal links at lattice spacing $a$ into the RS scheme at scale $\nu_f = 1/a$ we use Eq.~\eqref{EQN004}, where we insert Eq.~\eqref{EQN001} and Eq.~\eqref{EQN002} and eliminate $\alpha_L(a)$ in favor of $\alpha_{\overline{\text{MS}}}(1/a)$ via Eq.~\eqref{EQN003}.
			The expression $\delta \Lambda_B^L(a) + \delta \Lambda_B^\text{RS}(1/a)$ on the right hand side of Eq.~\eqref{EQN001} is then a power series in $\alpha_{\overline{\text{MS}}}(1/a)$.
			We truncate this power series, i.e.\ keep all terms proportional to $\alpha_{\overline{\text{MS}}}(1/a)^n$ with $n \leq n_{\text{max}}$ and discard all remaining terms corresponding to $n > n_{\text{max}}$. 
			For $n_{\text{max}} = 3$, for example, Eq.~\eqref{EQN004} becomes
			\begin{eqnarray}
				\nonumber & & \hspace{-0.7cm} \Lambda_B^\text{RS}(\nu_f) = \Lambda_B^L(a) - \frac{1}{a}c_0^{(8,0)} \alpha_{\overline{\text{MS}}}(\nu_f) \\
				\nonumber & & + \bigg(  \frac{1}{a} c_1^{(8,0)} + \frac{1}{a} c_0^{(8,0)} \left[-d_1 + \frac{2\beta_0}{4\pi}\ln(\nu_f a)\right] + \nu_f\left(\tilde{V}_{s,1}^\text{RS}-\tilde{V}_{o,1}^\text{RS}\right) \bigg)(\alpha_{\overline{\text{MS}}}(\nu_f))^2 \\
				\nonumber & &
				+ \bigg(  \frac{1}{a} c_2^{(8,0)} ~
				+ \frac{2}{a} c_1^{(8,0)} \left[-d_1 + \frac{2\beta_0}{4\pi}\ln(\nu_f a)\right]
				+\frac{1}{a} c_0^{(8,0)} \left[-d_2 + \frac{2\beta_1}{(4\pi)^2}\ln(\nu_f a) + d_1^2\right] \\
				\label{eq:expanded_conversion_formula} & & \phantom{+\bigg(}
				+ \frac{2}{a} c_0^{(8,0)} \left[-d_1 + \frac{2\beta_0}{4\pi}\ln(\nu_f a)\right]^2
				+ \nu_f\left(\tilde{V}_{s,2}^\text{RS}-\tilde{V}_{o,2}^\text{RS}\right) \bigg)(\alpha_{\overline{\text{MS}}}(\nu_f))^3 .
			\end{eqnarray}
			For $n_\text{max}=2$ this equation is identical to Eq.~(70) in Ref.~\cite{Bali:2003jq}.
			
			As in Ref.~\cite{Bali:2003jq} we use $n_{\text{max}} = 0, 1, 2, 3$ denoted as LO, NLO, NNLO and NNNLO. 
			In Ref.~\cite{Bali:2003jq} the coefficient $c_2^{(8,0)}$ appearing in the $n_{\text{max}} = 3$ expression was estimated,
			$c_2^{(8,0)} = 193.8(2.8)$.
			% $c_2^{(8,0)} = (C_A/32 \pi^2) v_3 = (C_A/32 \pi^2) \times 20.4(3) \times 10^3 = 193.8(2.8)$.
			% > 3 / (32 * pi^2) * 1000 * 20.4
			% [1] 193.7768
			% > 3 / (32 * pi^2) * 1000 * 0.3
			% [1] 2.849658			
			Meanwhile, it is now known quite accurately,
			$c_2^{(8,0)} = 193.2(3)$
			% $c_2^{(8,0)} = 85.87(14) \times C_A/C_F = 193.2(3)$
			\cite{Bali:2013qla}.
			% C_A = N_c = 3 , C_A/C_F = 9/4
			% > 10 * 8.587 * 9/4
			% [1] 193.2075
			% > 10 * 0.014 * 9/4
			% [1] 0.315
			We use this more accurate value, but since the difference between the two values is almost negligible, we do not expect a significant impact on the final result for the gluelump mass in the RS scheme.
			Moreover, as noted above, we use the five-loop running coupling to generate numerical values for $\alpha_{\overline{\text{MS}}}(1/a)$, which is an improvement compared to Ref.~\cite{Bali:2003jq}, where the four-loop running coupling was used.
			
\begin{figure}[htb]
				\begin{center}
					\includegraphics[width=0.75\linewidth,page=1]{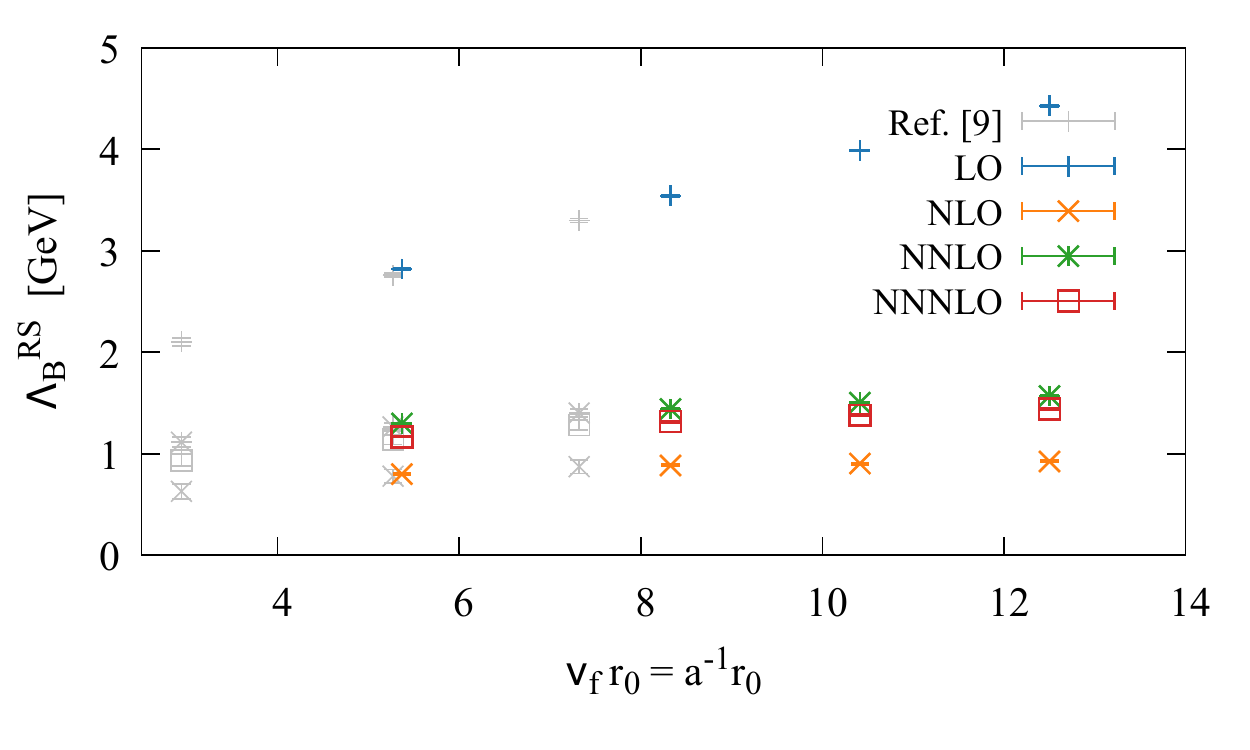}
				\end{center}
				\caption{\label{FIG001}$\Lambda_B^\text{RS}(1/a)$ for our four lattice spacings at LO, NLO, NNLO and NNNLO.}
			\end{figure}

			In Figure~\ref{FIG001} we show $\Lambda_B^\text{RS}(1/a)$ for our four lattice spacings at LO, NLO, NNLO and NNNLO (colored data points; note that at LO $\Lambda_B^\text{RS}(1/a) = \Lambda_B^L(a)$, i.e.\ lattice and RS masses are identical). 
			The corresponding numerical values are collected in Table~\ref{tab:LambdaRS_a}.
			For comparison we also show results from Ref.~\cite{Bali:2003jq} (gray data points), where lattice data from Ref.~\cite{Foster:1998wu} at coarser lattice spacings was used. 
			Our converted results show the same convergence behavior as the results from Ref.~\cite{Bali:2003jq} and the two sets of data points seem to be consistent with each other.
			
			\begin{table}[htb]\centering
				\begin {tabular}{cC>{\centering \arraybackslash }m{6.4em}>{\centering \arraybackslash }m{6.4em}>{\centering \arraybackslash }m{6.4em}}%
\toprule $a$ in fm&$\Lambda _B^\text {RS}(1/a) = \Lambda _B^L(a) $ LO&$\Lambda _B^\text {RS}(1/a)$\hspace {1cm} NLO&$\Lambda _B^\text {RS}(1/a)$ \hspace {1cm}NNLO&$\Lambda _B^\text {RS}(1/a)$ \hspace {1cm}NNNLO\\\midrule %
\pgfutilensuremath {0.093}&\pgfmathprintnumber [fixed,fixed zerofill,precision=3]{2.821101}(\pgfmathprintnumber [fixed,fixed zerofill,precision=0]{4.614e0})&\pgfmathprintnumber [fixed,fixed zerofill,precision=3]{0.797566}(\pgfmathprintnumber [fixed,fixed zerofill,precision=0]{4.614e0})&\pgfmathprintnumber [fixed,fixed zerofill,precision=3]{1.297897}(\pgfmathprintnumber [fixed,fixed zerofill,precision=0]{4.614e0})&\pgfmathprintnumber [fixed,fixed zerofill,precision=3]{1.166510}(\pgfmathprintnumber [fixed,fixed zerofill,precision=0]{4.614e0})\\%
\pgfutilensuremath {0.060}&\pgfmathprintnumber [fixed,fixed zerofill,precision=3]{3.541043}(\pgfmathprintnumber [fixed,fixed zerofill,precision=0]{6.7839996e0})&\pgfmathprintnumber [fixed,fixed zerofill,precision=3]{0.883174}(\pgfmathprintnumber [fixed,fixed zerofill,precision=0]{6.7839996e0})&\pgfmathprintnumber [fixed,fixed zerofill,precision=3]{1.439632}(\pgfmathprintnumber [fixed,fixed zerofill,precision=0]{6.7839996e0})&\pgfmathprintnumber [fixed,fixed zerofill,precision=3]{1.315900}(\pgfmathprintnumber [fixed,fixed zerofill,precision=0]{6.7839996e0})\\%
\pgfutilensuremath {0.048}&\pgfmathprintnumber [fixed,fixed zerofill,precision=3]{3.989623}(\pgfmathprintnumber [fixed,fixed zerofill,precision=0]{6.1929993e0})&\pgfmathprintnumber [fixed,fixed zerofill,precision=3]{0.901806}(\pgfmathprintnumber [fixed,fixed zerofill,precision=0]{6.1929993e0})&\pgfmathprintnumber [fixed,fixed zerofill,precision=3]{1.502394}(\pgfmathprintnumber [fixed,fixed zerofill,precision=0]{6.1929993e0})&\pgfmathprintnumber [fixed,fixed zerofill,precision=3]{1.378328}(\pgfmathprintnumber [fixed,fixed zerofill,precision=0]{6.1929993e0})\\%
\pgfutilensuremath {0.040}&\pgfmathprintnumber [fixed,fixed zerofill,precision=3]{4.428578}(\pgfmathprintnumber [fixed,fixed zerofill,precision=0]{8.0970001e0})&\pgfmathprintnumber [fixed,fixed zerofill,precision=3]{0.923024}(\pgfmathprintnumber [fixed,fixed zerofill,precision=0]{8.0970001e0})&\pgfmathprintnumber [fixed,fixed zerofill,precision=3]{1.567831}(\pgfmathprintnumber [fixed,fixed zerofill,precision=0]{8.0970001e0})&\pgfmathprintnumber [fixed,fixed zerofill,precision=3]{1.441866}(\pgfmathprintnumber [fixed,fixed zerofill,precision=0]{8.0970001e0})\\\bottomrule %
\end {tabular}%

				\caption{$\Lambda_B^\text{RS}(1/a)$ in GeV for our four lattice spacings at LO, NLO, NNLO and NNNLO. The errors are purely statistical.}
				\label{tab:LambdaRS_a}
			\end{table}
			
			To obtain $\Lambda_B^\text{RS}$ at the scale $\nu_f=2.5 / r_0 \approx 1 \, \text{GeV}$, we continue following Ref.~\cite{Bali:2003jq}.
			In a first step, we fit the NNNLO expression \eqref{eq:expanded_conversion_formula} with $\nu_f = 1/a_\text{min} = 1 / 0.040 \, \text{fm}$ to our four lattice data points $\Lambda_B^L(a)$, where the only fit parameter is $\Lambda_B^\text{RS}(\nu_f = 1 / 0.040 \, \text{fm})$.
			Since $\nu_f \neq 1/a$ for the data points from ensembles $A$, $B$ and $C$, there are now non-vanishing logarithms in Eq.~\eqref{eq:expanded_conversion_formula}.
			We obtain 
			\begin{equation}\label{eq:LambdaRS_smallesta}
				\Lambda_B^\text{RS} (\nu_f=1/0.040\,\text{fm} = 12.5 / r_0) = 1.463(3)\,\text{GeV} .
			\end{equation}
			$\chi^2_\text{red}=4.36$ indicates that our four lattice spacings together with the perturbative conversion procedure do not lead to four fully consistent results for $\Lambda_B^\text{RS}(\nu_f=1/0.040\,\text{fm})$ within our rather small statistical errors.
			The discrepancies could originate either in the sizable separation of scales $\nu_f \neq 1/a$ and the corresponding large logarithms, the truncation of the perturbative series or in lattice discretization errors, which are expected to be proportional to $a^2$.
			To account for this tension we use the difference to the result from an analogous fit excluding the lattice gluelump mass $\Lambda_B^L(a = 0.093 \, \text{fm})$, which gives $\Lambda_B^\text{RS} (\nu_f=1/0.040\,\text{fm}) =  1.460(4)\,\text{GeV}$ with  $\chi^2_\text{red}=5.89$, as part of the final systematic error (see the discussion at the end of this section).
			Moreover, we consider an additional $a^2$-term in Eq.~\eqref{eq:expanded_conversion_formula} and carry out another fit including all four lattice gluelump masses, which yields $\Lambda_B^\text{RS} (\nu_f=1/0.040\,\text{fm}) =  1.454(6)\,\text{GeV}$ with  $\chi^2_\text{red}=4.94$. Again we include the difference to the result \eqref{eq:LambdaRS_smallesta} in the final systematic error.
			We note that a straightforward conversion of the lattice data point at our smallest lattice spacing, as done for Figure~\ref{FIG001} and Table~\ref{tab:LambdaRS_a} , gives $\Lambda_B^\text{RS} (\nu_f=1/0.040\,\text{fm}) =  1.442(8)\,\text{GeV}$, which is slightly lower.
			
			In a second step, the result at $\nu_f = 1/0.040\,\text{fm} = 12.5 / r_0$ is propagated to the scale \\ $\nu_f^{\prime} = 2.5 / r_0 \approx 1\,\text{GeV}$ using
			\begin{equation}\label{eq:LambdaRS_running}
				\Lambda_B^\text{RS} (\nu_f^{\prime}=2.5/r_0) = \Lambda_B^\text{RS} (\nu_f= 12.5 / r_0) + \left(\delta \Lambda_B^\text{RS,PV}(\nu_f) -\delta \Lambda_B^\text{RS,PV}(\nu_f^{\prime})\right) .
			\end{equation}
			To avoid errors from widely separated scales $\nu_f^{\prime}$ and $\nu_f$ the Principal Value (PV) prescription in the RS scheme is used to compute $\delta \Lambda_B^\text{RS,PV}$.
			The key equation is Eq.~(61) in Ref.~\cite{Bali:2003jq}, where we replace $N_m$ by $N_{\Lambda}$ using $N_{\Lambda}=-1.37(9)$ from Ref.~\cite{Bali:2013qla}. We obtain 
			\begin{equation}\label{eq:LambdaRS_1GeV}
				\Lambda_B^\text{RS} (\nu_f = 2.5 / r_0 \approx 1\,\text{GeV}) = 0.857(3)\,\text{GeV}.
			\end{equation}
			This result is lower than the result of Ref.~\cite{Bali:2003jq}, $\Lambda_B^\text{RS} (\nu_f=2.5 / r_0 \approx 1\,\text{GeV}) = 0.912(12)\,\text{GeV}$, which is based on lattice gluelump masses from Ref.~\cite{Foster:1998wu}.
			The error quoted in Eq.~\eqref{eq:LambdaRS_1GeV} is a statistical bootstrap error, which does not include systematic uncertainties. 
			As expected it is much smaller than its counterpart from Ref.~\cite{Bali:2003jq}, roughly by a factor of $4$, because we provide more accurate lattice gluelump masses as input.
			
			Finally, we discuss systematic errors and compare them to Ref.~\cite{Bali:2003jq}.
			We use the five-loop running coupling instead of the four-loop running coupling with a more precise $\Lambda_{\overline{\text{MS}}}$ value \cite{FlavourLatticeAveragingGroupFLAG:2021npn}, which reduces the systematic error associated with the uncertainty of $\Lambda_{\overline{\text{MS}}}$ from $0.04\,\text{GeV}$ to $0.03\,\text{GeV}$.
			Systematic errors already discussed in the context of our result (\ref{eq:LambdaRS_smallesta}) above translate to $\approx 0.003\,\text{GeV}$ (separation of scales and large logarithms) and $\approx 0.01\,\text{GeV}$ (discretization errors), respectively.
			The perturbative error, which Ref.~\cite{Bali:2003jq} estimates as the difference between the NNLO and NNNLO result, is in our case
			$\approx 0.03\,\text{GeV}$.
			Additionally, there is a perturbative error coming from the $10\%$ uncertainty in $N_{V_s} - N_{V_o}$, i.e.\ a contribution of $\approx 0.07\,\text{GeV}$ \cite{Bali:2003jq}.
			All these systematic errors, which are estimated in exactly the same way as in Ref.~\cite{Bali:2003jq}, add up to $0.143\,\text{GeV}$ compared to $0.205\,\text{GeV}$ quoted in Ref.~\cite{Bali:2003jq}. Our final result is
			\begin{equation}
				\label{EQN645} \Lambda_B^\text{RS} (\nu_f = 2.5 / r_0 \approx 1\,\text{GeV}) = 0.857(3)(143) \,\text{GeV} ,
			\end{equation}
			where the first error is statistical and the second error is systematic.
			Clearly, the systematic error is much larger than the statistical error associated with the lattice gluelump masses.
			Consequently, improvements on the perturbative side seem to be necessary to increase the precision of the $1^{+-}$ gluelump mass in the RS scheme.
			
			For completeness we note that Ref.~\cite{Bali:2003jq} also includes a determination of $\Lambda_B^\text{RS} (\nu_f = 2.5/r_0 \approx 1\,\text{GeV}) $ via the (hybrid) static potentials $\Sigma_g^+,\Pi_u$ and  $\Sigma_u^-$ resulting in \\ $\Lambda_B^\text{RS} (\nu_f = 2.5/r_0 \approx 1\,\text{GeV}) = [0.888 \pm 0.039(\text{latt.}) \pm 0.083(\text{th.}) \pm 0.032(\Lambda_{\overline{\text{MS}}})]\,\text{GeV}$, which is consistent with our result (\ref{EQN645}).

% ********************
% ********************
% ********************
% ********************

	$\quad$
	
	%\newpage
	\clearpage
	
	\section{\label{sec:conclusions}Summary and outlook}
	
	We have carried out a comprehensive up-to-date lattice gauge theory computation of the gluelump spectrum in pure SU(3) gauge theory.
	We have considered ground states for 20 $\mathcal{R}^{PC}$ representations and provide both the corresponding masses and mass splittings.
	For the latter we have studied the continuum limits using extrapolations based on lattice data from four ensembles with rather fine lattice spacings.
	Our computations complement and improve on existing work, in particular on Ref.~\cite{Foster:1998wu}:
	\begin{itemize}
		\item We use lattice spacings as small as $a = 0.040 \, \text{fm}$, which is significantly smaller than the smallest lattice spacing from Ref.~\cite{Foster:1998wu}, $a = 0.068 \, \text{fm}$.
		
		\item Our continuum extrapolations of gluelump mass splittings are based on fits to lattice data from ensembles with four different lattice spacings, where on each ensemble computations with unsmeared and with HYP2 temporal links were performed.
		
		\item We have computed gluelump masses for 20 $\mathcal{R}^{PC}$ representations and have studied the continuum limits of the corresponding 19 gluelump mass splittings with the $T_1^{+-}$ gluelump mass as reference, whereas previously only 10 ground state masses and continuum limits of 5 mass splittings were provided.
		
		\item The assignment of continuum total angular momentum $J$ to the lattice results on gluelump masses is extensively discussed.
	\end{itemize}
	
	Our results on gluelump masses also complement and extend our recent results on hybrid static potentials \cite{Capitani:2018rox,Schlosser:2021wnr}, since gluelump masses can be interpreted as the $r \rightarrow 0$ limit of hybrid static potentials.
	
	Moreover, we have repeated a perturbative analysis and determination of the $1^{+-}$ gluelump mass in the RS scheme from Ref.~\cite{Bali:2003jq} using our improved lattice gluelump data as input.
	From this analysis it is obvious that the remaining error of this RS gluelump mass is currently dominated by perturbation theory and not by the accuracy of lattice gluelump masses.
	We expect that this will motivate experts from the field of perturbation theory to improve the perturbative equations entering RS gluelump mass determinations.
	Such an improvement might be within reach, in particular in view of closely related perturbative advances reported in the literature, e.g.\ the determination of the coefficients $c_n^{(8,0)}$, $n = 0,\ldots,19$ (see Refs.\ \cite{Bali:2013qla,Bali:2013pla}) appearing in Eq.\ (\ref{EQN001}) up to order $\alpha_L^{20}$.
	
	A remaining problem on the lattice gauge theory side concerns several of the higher gluelump states, where the assignment of the correct continuum total angular momentum $J$ is not clear, or where states with similar mass, but different $J$ appear in the same cubic representation and might mix (see the detailed discussion in Section~\ref{sec:spinidentification}).
	We plan to continue our work in this direction, by implementing several operators for each $\mathcal{R}^{PC}$ representation, which resemble possibly competing continuum angular momenta $J$.
	After diagonalizing the corresponding correlation matrices, e.g.\ by solving generalized eigenvalue problems, we expect that a clear assignment of continuum $J$ values is possible.

	% ********************
	% ********************
	% ********************
	% ********************
	% ********************

	%\newpage
	\clearpage

	\section*{Acknowledgments}
	
	We thank Christian Reisinger for providing his multilevel code.
	We acknowledge interesting and useful discussions with Nora Brambilla, Antonio Pineda and Joan Soto.
	
	M.W.\ acknowledges support by the Heisenberg Programme of the Deutsche Forschungsgemeinschaft (DFG, German Research Foundation) -- project number 399217702.
	
	Calculations on the GOETHE-HLR and on the FUCHS-CSC high-performance computers of the Frankfurt University were conducted for this research.
	We thank HPC-Hessen, funded by the State Ministry of Higher Education, Research and the Arts, for programming advice.
	
		% ********************
		% ********************
		% ********************
		% ********************
		% ********************

	%	\newpage
		
		\appendix

	\section{Summary of lattice field theory results}
	
		% ********************
		
		\subsection{\label{app:lattice_gluelump_masses}Lattice gluelump masses for all ensembles and unsmeared and HYP2 smeared temporal links}
		
		\begin{table}[htb]
			\centering
			{\scriptsize
				\begin {tabular}{cllllllll}%
\toprule $\mathcal {R}^{PC}$&$m_{\mathcal {R}^{PC}}^{A,\text {none}}\, a$&$m_{\mathcal {R}^{PC}}^{B,\text {none}}\, a$&$m_{\mathcal {R}^{PC}}^{C,\text {none}}\, a$&$m_{\mathcal {R}^{PC}}^{D,\text {none}}\, a$&$m_{\mathcal {R}^{PC}}^{A,{\text {HYP2}}}\, a$&$m_{\mathcal {R}^{PC}}^{B,{\text {HYP2}}}\, a$&$m_{\mathcal {R}^{PC}}^{C,{\text {HYP2}}}\, a$&$m_{\mathcal {R}^{PC}}^{D,{\text {HYP2}}}\, a$\\\midrule %
$T_1^{++}$&\pgfmathprintnumber [fixed,fixed zerofill,precision=3]{2.14411760811e+00}(\pgfmathprintnumber [fixed,fixed zerofill,precision=0]{4.3807312e1 })&\pgfmathprintnumber [fixed,fixed zerofill,precision=3]{1.63316961119e+00}(\pgfmathprintnumber [fixed,fixed zerofill,precision=0]{1.7050385e1 })&\pgfmathprintnumber [fixed,fixed zerofill,precision=3]{1.45072287328e+00}(\pgfmathprintnumber [fixed,fixed zerofill,precision=0]{3.396222e0 })&\pgfmathprintnumber [fixed,fixed zerofill,precision=3]{1.27879501500e+00}(\pgfmathprintnumber [fixed,fixed zerofill,precision=0]{6.3227737e0 })&\pgfmathprintnumber [fixed,fixed zerofill,precision=3]{1.59764187060e+00}(\pgfmathprintnumber [fixed,fixed zerofill,precision=0]{3.9470566e1})&\pgfmathprintnumber [fixed,fixed zerofill,precision=3]{1.15495440274e+00}(\pgfmathprintnumber [fixed,fixed zerofill,precision=0]{6.700032e0})&\pgfmathprintnumber [fixed,fixed zerofill,precision=3]{9.79754924121e-01}(\pgfmathprintnumber [fixed,fixed zerofill,precision=0]{3.3562225e0})&\pgfmathprintnumber [fixed,fixed zerofill,precision=3]{8.28304174645e-01}(\pgfmathprintnumber [fixed,fixed zerofill,precision=0]{6.0830032e0})\\%
\rowcolor [gray]{.8}$T_1^{+-}$&\pgfmathprintnumber [fixed,fixed zerofill,precision=3]{1.33190446543e+00}(\pgfmathprintnumber [fixed,fixed zerofill,precision=0]{2.2199646e0 })&\pgfmathprintnumber [fixed,fixed zerofill,precision=3]{1.07774148529e+00}(\pgfmathprintnumber [fixed,fixed zerofill,precision=0]{2.0461624e0 })&\pgfmathprintnumber [fixed,fixed zerofill,precision=3]{9.71009515125e-01}(\pgfmathprintnumber [fixed,fixed zerofill,precision=0]{1.5022705e0 })&\pgfmathprintnumber [fixed,fixed zerofill,precision=3]{8.97840115443e-01}(\pgfmathprintnumber [fixed,fixed zerofill,precision=0]{1.6568146e0 })&\pgfmathprintnumber [fixed,fixed zerofill,precision=3]{7.70513938190e-01}(\pgfmathprintnumber [fixed,fixed zerofill,precision=0]{2.0796631e0})&\pgfmathprintnumber [fixed,fixed zerofill,precision=3]{5.79623487213e-01}(\pgfmathprintnumber [fixed,fixed zerofill,precision=0]{1.9502747e0})&\pgfmathprintnumber [fixed,fixed zerofill,precision=3]{4.99576528843e-01}(\pgfmathprintnumber [fixed,fixed zerofill,precision=0]{1.453891e0})&\pgfmathprintnumber [fixed,fixed zerofill,precision=3]{4.47590811633e-01}(\pgfmathprintnumber [fixed,fixed zerofill,precision=0]{1.574295e0})\\%
$T_1^{-+}$&\pgfmathprintnumber [fixed,fixed zerofill,precision=3]{1.93567085261e+00}(\pgfmathprintnumber [fixed,fixed zerofill,precision=0]{6.7079895e0 })&\pgfmathprintnumber [fixed,fixed zerofill,precision=3]{1.46410757862e+00}(\pgfmathprintnumber [fixed,fixed zerofill,precision=0]{9.030957e0 })&\pgfmathprintnumber [fixed,fixed zerofill,precision=3]{1.29196271590e+00}(\pgfmathprintnumber [fixed,fixed zerofill,precision=0]{4.8536102e0 })&\pgfmathprintnumber [fixed,fixed zerofill,precision=3]{1.17341583945e+00}(\pgfmathprintnumber [fixed,fixed zerofill,precision=0]{2.9474564e0 })&\pgfmathprintnumber [fixed,fixed zerofill,precision=3]{1.37794831629e+00}(\pgfmathprintnumber [fixed,fixed zerofill,precision=0]{6.3520782e0})&\pgfmathprintnumber [fixed,fixed zerofill,precision=3]{9.66349784567e-01}(\pgfmathprintnumber [fixed,fixed zerofill,precision=0]{8.6654785e0})&\pgfmathprintnumber [fixed,fixed zerofill,precision=3]{8.12952489588e-01}(\pgfmathprintnumber [fixed,fixed zerofill,precision=0]{8.1309097e0})&\pgfmathprintnumber [fixed,fixed zerofill,precision=3]{6.98691687918e-01}(\pgfmathprintnumber [fixed,fixed zerofill,precision=0]{1.1857834e1})\\%
$T_1^{--}$&\pgfmathprintnumber [fixed,fixed zerofill,precision=3]{1.47354234296e+00}(\pgfmathprintnumber [fixed,fixed zerofill,precision=0]{8.7880219e0 })&\pgfmathprintnumber [fixed,fixed zerofill,precision=3]{1.19539345536e+00}(\pgfmathprintnumber [fixed,fixed zerofill,precision=0]{2.8455795e0 })&\pgfmathprintnumber [fixed,fixed zerofill,precision=3]{1.06245116782e+00}(\pgfmathprintnumber [fixed,fixed zerofill,precision=0]{2.5842697e0 })&\pgfmathprintnumber [fixed,fixed zerofill,precision=3]{9.69789195762e-01}(\pgfmathprintnumber [fixed,fixed zerofill,precision=0]{2.0615875e0 })&\pgfmathprintnumber [fixed,fixed zerofill,precision=3]{9.06816121602e-01}(\pgfmathprintnumber [fixed,fixed zerofill,precision=0]{1.6240845e1})&\pgfmathprintnumber [fixed,fixed zerofill,precision=3]{6.98182475564e-01}(\pgfmathprintnumber [fixed,fixed zerofill,precision=0]{2.8058655e0})&\pgfmathprintnumber [fixed,fixed zerofill,precision=3]{5.91972524057e-01}(\pgfmathprintnumber [fixed,fixed zerofill,precision=0]{2.5699768e0})&\pgfmathprintnumber [fixed,fixed zerofill,precision=3]{5.19821800249e-01}(\pgfmathprintnumber [fixed,fixed zerofill,precision=0]{2.0617844e0})\\%
$T_2^{++}$&\pgfmathprintnumber [fixed,fixed zerofill,precision=3]{2.07083735334e+00}(\pgfmathprintnumber [fixed,fixed zerofill,precision=0]{9.0786102e0 })&\pgfmathprintnumber [fixed,fixed zerofill,precision=3]{1.55954509128e+00}(\pgfmathprintnumber [fixed,fixed zerofill,precision=0]{1.3558273e1 })&\pgfmathprintnumber [fixed,fixed zerofill,precision=3]{1.38248323372e+00}(\pgfmathprintnumber [fixed,fixed zerofill,precision=0]{3.3810165e0 })&\pgfmathprintnumber [fixed,fixed zerofill,precision=3]{1.24785668615e+00}(\pgfmathprintnumber [fixed,fixed zerofill,precision=0]{3.6249527e0 })&\pgfmathprintnumber [fixed,fixed zerofill,precision=3]{1.51302354783e+00}(\pgfmathprintnumber [fixed,fixed zerofill,precision=0]{8.8220795e0})&\pgfmathprintnumber [fixed,fixed zerofill,precision=3]{1.06404755664e+00}(\pgfmathprintnumber [fixed,fixed zerofill,precision=0]{1.3050659e1})&\pgfmathprintnumber [fixed,fixed zerofill,precision=3]{9.11744628494e-01}(\pgfmathprintnumber [fixed,fixed zerofill,precision=0]{3.3431503e0})&\pgfmathprintnumber [fixed,fixed zerofill,precision=3]{7.97749612802e-01}(\pgfmathprintnumber [fixed,fixed zerofill,precision=0]{3.5961136e0})\\%
$T_2^{+-}$&\pgfmathprintnumber [fixed,fixed zerofill,precision=3]{1.73498597924e+00}(\pgfmathprintnumber [fixed,fixed zerofill,precision=0]{9.1356155e0 })&\pgfmathprintnumber [fixed,fixed zerofill,precision=3]{1.36001361453e+00}(\pgfmathprintnumber [fixed,fixed zerofill,precision=0]{4.7235367e0 })&\pgfmathprintnumber [fixed,fixed zerofill,precision=3]{1.19767115466e+00}(\pgfmathprintnumber [fixed,fixed zerofill,precision=0]{4.5322098e0 })&\pgfmathprintnumber [fixed,fixed zerofill,precision=3]{1.08731218603e+00}(\pgfmathprintnumber [fixed,fixed zerofill,precision=0]{3.9483322e0 })&\pgfmathprintnumber [fixed,fixed zerofill,precision=3]{1.18050550477e+00}(\pgfmathprintnumber [fixed,fixed zerofill,precision=0]{8.2013199e0})&\pgfmathprintnumber [fixed,fixed zerofill,precision=3]{8.60348490218e-01}(\pgfmathprintnumber [fixed,fixed zerofill,precision=0]{4.512352e0})&\pgfmathprintnumber [fixed,fixed zerofill,precision=3]{7.26177137344e-01}(\pgfmathprintnumber [fixed,fixed zerofill,precision=0]{4.2886276e0})&\pgfmathprintnumber [fixed,fixed zerofill,precision=3]{6.37018430908e-01}(\pgfmathprintnumber [fixed,fixed zerofill,precision=0]{3.8382278e0})\\%
$T_2^{-+}$&\pgfmathprintnumber [fixed,fixed zerofill,precision=3]{2.03025710985e+00}(\pgfmathprintnumber [fixed,fixed zerofill,precision=0]{8.235466e0 })&\pgfmathprintnumber [fixed,fixed zerofill,precision=3]{1.48912213555e+00}(\pgfmathprintnumber [fixed,fixed zerofill,precision=0]{2.6262558e1 })&\pgfmathprintnumber [fixed,fixed zerofill,precision=3]{1.35109522740e+00}(\pgfmathprintnumber [fixed,fixed zerofill,precision=0]{5.8495392e0 })&\pgfmathprintnumber [fixed,fixed zerofill,precision=3]{1.22241967021e+00}(\pgfmathprintnumber [fixed,fixed zerofill,precision=0]{3.614122e0 })&\pgfmathprintnumber [fixed,fixed zerofill,precision=3]{1.46971255254e+00}(\pgfmathprintnumber [fixed,fixed zerofill,precision=0]{7.8747437e0})&\pgfmathprintnumber [fixed,fixed zerofill,precision=3]{1.02879214621e+00}(\pgfmathprintnumber [fixed,fixed zerofill,precision=0]{1.142012e1})&\pgfmathprintnumber [fixed,fixed zerofill,precision=3]{8.79610581194e-01}(\pgfmathprintnumber [fixed,fixed zerofill,precision=0]{5.7044281e0})&\pgfmathprintnumber [fixed,fixed zerofill,precision=3]{7.68337080317e-01}(\pgfmathprintnumber [fixed,fixed zerofill,precision=0]{5.6846817e0})\\%
$T_2^{--}$&\pgfmathprintnumber [fixed,fixed zerofill,precision=3]{1.57629950578e+00}(\pgfmathprintnumber [fixed,fixed zerofill,precision=0]{2.4233368e0 })&\pgfmathprintnumber [fixed,fixed zerofill,precision=3]{1.21063897220e+00}(\pgfmathprintnumber [fixed,fixed zerofill,precision=0]{1.4036697e1 })&\pgfmathprintnumber [fixed,fixed zerofill,precision=3]{1.09610388864e+00}(\pgfmathprintnumber [fixed,fixed zerofill,precision=0]{2.833783e0 })&\pgfmathprintnumber [fixed,fixed zerofill,precision=3]{1.00076625773e+00}(\pgfmathprintnumber [fixed,fixed zerofill,precision=0]{2.33228e0 })&\pgfmathprintnumber [fixed,fixed zerofill,precision=3]{1.01934299831e+00}(\pgfmathprintnumber [fixed,fixed zerofill,precision=0]{2.3263565e0})&\pgfmathprintnumber [fixed,fixed zerofill,precision=3]{7.17214207026e-01}(\pgfmathprintnumber [fixed,fixed zerofill,precision=0]{1.2426865e1})&\pgfmathprintnumber [fixed,fixed zerofill,precision=3]{6.16701947176e-01}(\pgfmathprintnumber [fixed,fixed zerofill,precision=0]{5.6209763e0})&\pgfmathprintnumber [fixed,fixed zerofill,precision=3]{5.51242239861e-01}(\pgfmathprintnumber [fixed,fixed zerofill,precision=0]{2.3225906e0})\\%
$A_1^{++}$&\pgfmathprintnumber [fixed,fixed zerofill,precision=3]{1.75312906389e+00}(\pgfmathprintnumber [fixed,fixed zerofill,precision=0]{7.5018661e0 })&\pgfmathprintnumber [fixed,fixed zerofill,precision=3]{1.37062751497e+00}(\pgfmathprintnumber [fixed,fixed zerofill,precision=0]{5.1774414e0 })&\pgfmathprintnumber [fixed,fixed zerofill,precision=3]{1.20077239513e+00}(\pgfmathprintnumber [fixed,fixed zerofill,precision=0]{5.5684738e0 })&\pgfmathprintnumber [fixed,fixed zerofill,precision=3]{1.09853006586e+00}(\pgfmathprintnumber [fixed,fixed zerofill,precision=0]{3.4460388e0 })&\pgfmathprintnumber [fixed,fixed zerofill,precision=3]{1.19402633106e+00}(\pgfmathprintnumber [fixed,fixed zerofill,precision=0]{7.1031342e0})&\pgfmathprintnumber [fixed,fixed zerofill,precision=3]{8.72981821413e-01}(\pgfmathprintnumber [fixed,fixed zerofill,precision=0]{5.1132233e0})&\pgfmathprintnumber [fixed,fixed zerofill,precision=3]{7.30119929951e-01}(\pgfmathprintnumber [fixed,fixed zerofill,precision=0]{5.5981583e0})&\pgfmathprintnumber [fixed,fixed zerofill,precision=3]{6.48379307091e-01}(\pgfmathprintnumber [fixed,fixed zerofill,precision=0]{3.4288757e0})\\%
$A_1^{+-}$&\pgfmathprintnumber [fixed,fixed zerofill,precision=3]{2.27577171509e+00}(\pgfmathprintnumber [fixed,fixed zerofill,precision=0]{2.7087677e1 })&\pgfmathprintnumber [fixed,fixed zerofill,precision=3]{1.74821959166e+00}(\pgfmathprintnumber [fixed,fixed zerofill,precision=0]{5.830075e0 })&\pgfmathprintnumber [fixed,fixed zerofill,precision=3]{1.48599428950e+00}(\pgfmathprintnumber [fixed,fixed zerofill,precision=0]{7.7650314e0 })&\pgfmathprintnumber [fixed,fixed zerofill,precision=3]{1.35116937132e+00}(\pgfmathprintnumber [fixed,fixed zerofill,precision=0]{4.382344e0 })&\pgfmathprintnumber [fixed,fixed zerofill,precision=3]{1.71831020941e+00}(\pgfmathprintnumber [fixed,fixed zerofill,precision=0]{2.5702301e1})&\pgfmathprintnumber [fixed,fixed zerofill,precision=3]{1.25106674180e+00}(\pgfmathprintnumber [fixed,fixed zerofill,precision=0]{5.6643173e0})&\pgfmathprintnumber [fixed,fixed zerofill,precision=3]{1.01707695437e+00}(\pgfmathprintnumber [fixed,fixed zerofill,precision=0]{7.4870636e0})&\pgfmathprintnumber [fixed,fixed zerofill,precision=3]{8.75149295116e-01}(\pgfmathprintnumber [fixed,fixed zerofill,precision=0]{1.6227661e1})\\%
$A_1^{-+}$&\pgfmathprintnumber [fixed,fixed zerofill,precision=3]{2.15946592895e+00}(\pgfmathprintnumber [fixed,fixed zerofill,precision=0]{1.0785477e2 })&\pgfmathprintnumber [fixed,fixed zerofill,precision=3]{1.79382829944e+00}(\pgfmathprintnumber [fixed,fixed zerofill,precision=0]{5.5322617e0 })&\pgfmathprintnumber [fixed,fixed zerofill,precision=3]{1.55148884154e+00}(\pgfmathprintnumber [fixed,fixed zerofill,precision=0]{7.4091858e0 })&\pgfmathprintnumber [fixed,fixed zerofill,precision=3]{1.37022432308e+00}(\pgfmathprintnumber [fixed,fixed zerofill,precision=0]{8.72202e0 })&\pgfmathprintnumber [fixed,fixed zerofill,precision=3]{1.77584637152e+00}(\pgfmathprintnumber [fixed,fixed zerofill,precision=0]{2.3766312e1})&\pgfmathprintnumber [fixed,fixed zerofill,precision=3]{1.27503093936e+00}(\pgfmathprintnumber [fixed,fixed zerofill,precision=0]{1.4003418e1})&\pgfmathprintnumber [fixed,fixed zerofill,precision=3]{1.08170441997e+00}(\pgfmathprintnumber [fixed,fixed zerofill,precision=0]{7.2030304e0})&\pgfmathprintnumber [fixed,fixed zerofill,precision=3]{9.18884401544e-01}(\pgfmathprintnumber [fixed,fixed zerofill,precision=0]{8.621785e0})\\%
$A_1^{--}$&\pgfmathprintnumber [fixed,fixed zerofill,precision=3]{1.96599991904e+00}(\pgfmathprintnumber [fixed,fixed zerofill,precision=0]{1.029654e1 })&\pgfmathprintnumber [fixed,fixed zerofill,precision=3]{1.41561543326e+00}(\pgfmathprintnumber [fixed,fixed zerofill,precision=0]{2.739354e1 })&\pgfmathprintnumber [fixed,fixed zerofill,precision=3]{1.31365302202e+00}(\pgfmathprintnumber [fixed,fixed zerofill,precision=0]{4.1212646e0 })&\pgfmathprintnumber [fixed,fixed zerofill,precision=3]{1.21050076142e+00}(\pgfmathprintnumber [fixed,fixed zerofill,precision=0]{3.7915741e1 })&\pgfmathprintnumber [fixed,fixed zerofill,precision=3]{1.40699147379e+00}(\pgfmathprintnumber [fixed,fixed zerofill,precision=0]{9.5820358e0})&\pgfmathprintnumber [fixed,fixed zerofill,precision=3]{9.64117113415e-01}(\pgfmathprintnumber [fixed,fixed zerofill,precision=0]{1.3618774e1})&\pgfmathprintnumber [fixed,fixed zerofill,precision=3]{8.43731596783e-01}(\pgfmathprintnumber [fixed,fixed zerofill,precision=0]{4.0268524e0})&\pgfmathprintnumber [fixed,fixed zerofill,precision=3]{7.30656236004e-01}(\pgfmathprintnumber [fixed,fixed zerofill,precision=0]{3.439154e0})\\%
$A_2^{++}$&\pgfmathprintnumber [fixed,fixed zerofill,precision=3]{2.35051261153e+00}(\pgfmathprintnumber [fixed,fixed zerofill,precision=0]{6.3025879e0 })&\pgfmathprintnumber [fixed,fixed zerofill,precision=3]{1.70012387977e+00}(\pgfmathprintnumber [fixed,fixed zerofill,precision=0]{1.5428299e1 })&\pgfmathprintnumber [fixed,fixed zerofill,precision=3]{1.50047632419e+00}(\pgfmathprintnumber [fixed,fixed zerofill,precision=0]{8.4570465e0 })&\pgfmathprintnumber [fixed,fixed zerofill,precision=3]{1.32796120941e+00}(\pgfmathprintnumber [fixed,fixed zerofill,precision=0]{7.6060135e0 })&\pgfmathprintnumber [fixed,fixed zerofill,precision=3]{1.79287731148e+00}(\pgfmathprintnumber [fixed,fixed zerofill,precision=0]{6.0018417e0})&\pgfmathprintnumber [fixed,fixed zerofill,precision=3]{1.20184336300e+00}(\pgfmathprintnumber [fixed,fixed zerofill,precision=0]{1.4356537e1})&\pgfmathprintnumber [fixed,fixed zerofill,precision=3]{1.02884346735e+00}(\pgfmathprintnumber [fixed,fixed zerofill,precision=0]{8.1567093e0})&\pgfmathprintnumber [fixed,fixed zerofill,precision=3]{8.77940106011e-01}(\pgfmathprintnumber [fixed,fixed zerofill,precision=0]{7.467714e0})\\%
$A_2^{+-}$&\pgfmathprintnumber [fixed,fixed zerofill,precision=3]{1.88720467992e+00}(\pgfmathprintnumber [fixed,fixed zerofill,precision=0]{8.0171875e0 })&\pgfmathprintnumber [fixed,fixed zerofill,precision=3]{1.37129950231e+00}(\pgfmathprintnumber [fixed,fixed zerofill,precision=0]{2.2528839e1 })&\pgfmathprintnumber [fixed,fixed zerofill,precision=3]{1.28826540909e+00}(\pgfmathprintnumber [fixed,fixed zerofill,precision=0]{6.001912e0 })&\pgfmathprintnumber [fixed,fixed zerofill,precision=3]{1.13316357211e+00}(\pgfmathprintnumber [fixed,fixed zerofill,precision=0]{1.7462921e1 })&\pgfmathprintnumber [fixed,fixed zerofill,precision=3]{1.30627500113e+00}(\pgfmathprintnumber [fixed,fixed zerofill,precision=0]{2.039653e1})&\pgfmathprintnumber [fixed,fixed zerofill,precision=3]{8.81296893703e-01}(\pgfmathprintnumber [fixed,fixed zerofill,precision=0]{1.9961792e1})&\pgfmathprintnumber [fixed,fixed zerofill,precision=3]{7.95244296712e-01}(\pgfmathprintnumber [fixed,fixed zerofill,precision=0]{1.158606e1})&\pgfmathprintnumber [fixed,fixed zerofill,precision=3]{6.76166652929e-01}(\pgfmathprintnumber [fixed,fixed zerofill,precision=0]{1.5623962e1})\\%
$A_2^{-+}$&\pgfmathprintnumber [fixed,fixed zerofill,precision=3]{2.06897080448e+00}(\pgfmathprintnumber [fixed,fixed zerofill,precision=0]{1.2638626e1 })&\pgfmathprintnumber [fixed,fixed zerofill,precision=3]{1.57984810906e+00}(\pgfmathprintnumber [fixed,fixed zerofill,precision=0]{3.8082642e0 })&\pgfmathprintnumber [fixed,fixed zerofill,precision=3]{1.35060315095e+00}(\pgfmathprintnumber [fixed,fixed zerofill,precision=0]{8.160617e0 })&\pgfmathprintnumber [fixed,fixed zerofill,precision=3]{1.20541335379e+00}(\pgfmathprintnumber [fixed,fixed zerofill,precision=0]{8.722966e0 })&\pgfmathprintnumber [fixed,fixed zerofill,precision=3]{1.51191036541e+00}(\pgfmathprintnumber [fixed,fixed zerofill,precision=0]{1.1612396e1})&\pgfmathprintnumber [fixed,fixed zerofill,precision=3]{1.06889564188e+00}(\pgfmathprintnumber [fixed,fixed zerofill,precision=0]{7.9673096e0})&\pgfmathprintnumber [fixed,fixed zerofill,precision=3]{8.78778924204e-01}(\pgfmathprintnumber [fixed,fixed zerofill,precision=0]{7.9806686e0})&\pgfmathprintnumber [fixed,fixed zerofill,precision=3]{7.57051877968e-01}(\pgfmathprintnumber [fixed,fixed zerofill,precision=0]{8.5012619e0})\\%
$A_2^{--}$&-&-&-&-&\pgfmathprintnumber [fixed,fixed zerofill,precision=3]{1.54613415546e+00}(\pgfmathprintnumber [fixed,fixed zerofill,precision=0]{9.1410782e1})&\pgfmathprintnumber [fixed,fixed zerofill,precision=3]{1.26626190045e+00}(\pgfmathprintnumber [fixed,fixed zerofill,precision=0]{1.4538254e1})&\pgfmathprintnumber [fixed,fixed zerofill,precision=3]{1.05876149959e+00}(\pgfmathprintnumber [fixed,fixed zerofill,precision=0]{8.0174973e0})&\pgfmathprintnumber [fixed,fixed zerofill,precision=3]{9.09626708834e-01}(\pgfmathprintnumber [fixed,fixed zerofill,precision=0]{8.4497131e0})\\%
$E^{++}$&\pgfmathprintnumber [fixed,fixed zerofill,precision=3]{1.91685047873e+00}(\pgfmathprintnumber [fixed,fixed zerofill,precision=0]{8.8025635e0 })&\pgfmathprintnumber [fixed,fixed zerofill,precision=3]{1.47685515999e+00}(\pgfmathprintnumber [fixed,fixed zerofill,precision=0]{2.9849564e0 })&\pgfmathprintnumber [fixed,fixed zerofill,precision=3]{1.25549882811e+00}(\pgfmathprintnumber [fixed,fixed zerofill,precision=0]{1.1154755e1 })&\pgfmathprintnumber [fixed,fixed zerofill,precision=3]{1.16201980616e+00}(\pgfmathprintnumber [fixed,fixed zerofill,precision=0]{2.824472e0 })&\pgfmathprintnumber [fixed,fixed zerofill,precision=3]{1.35944636655e+00}(\pgfmathprintnumber [fixed,fixed zerofill,precision=0]{8.4412094e0})&\pgfmathprintnumber [fixed,fixed zerofill,precision=3]{9.77587616603e-01}(\pgfmathprintnumber [fixed,fixed zerofill,precision=0]{2.9536057e0})&\pgfmathprintnumber [fixed,fixed zerofill,precision=3]{7.84851140159e-01}(\pgfmathprintnumber [fixed,fixed zerofill,precision=0]{1.1051071e1})&\pgfmathprintnumber [fixed,fixed zerofill,precision=3]{7.11216103224e-01}(\pgfmathprintnumber [fixed,fixed zerofill,precision=0]{2.8105087e0})\\%
$E^{+-}$&\pgfmathprintnumber [fixed,fixed zerofill,precision=3]{1.72581181326e+00}(\pgfmathprintnumber [fixed,fixed zerofill,precision=0]{1.2138245e1 })&\pgfmathprintnumber [fixed,fixed zerofill,precision=3]{1.35648299974e+00}(\pgfmathprintnumber [fixed,fixed zerofill,precision=0]{3.3259888e0 })&\pgfmathprintnumber [fixed,fixed zerofill,precision=3]{1.18093859130e+00}(\pgfmathprintnumber [fixed,fixed zerofill,precision=0]{5.6327835e0 })&\pgfmathprintnumber [fixed,fixed zerofill,precision=3]{1.08072941958e+00}(\pgfmathprintnumber [fixed,fixed zerofill,precision=0]{2.2301117e0 })&\pgfmathprintnumber [fixed,fixed zerofill,precision=3]{1.16479495340e+00}(\pgfmathprintnumber [fixed,fixed zerofill,precision=0]{1.0994461e1})&\pgfmathprintnumber [fixed,fixed zerofill,precision=3]{8.58359131210e-01}(\pgfmathprintnumber [fixed,fixed zerofill,precision=0]{3.2414764e0})&\pgfmathprintnumber [fixed,fixed zerofill,precision=3]{7.09807346201e-01}(\pgfmathprintnumber [fixed,fixed zerofill,precision=0]{5.5107681e0})&\pgfmathprintnumber [fixed,fixed zerofill,precision=3]{6.30629628514e-01}(\pgfmathprintnumber [fixed,fixed zerofill,precision=0]{2.2226334e0})\\%
$E^{-+}$&\pgfmathprintnumber [fixed,fixed zerofill,precision=3]{2.01427287567e+00}(\pgfmathprintnumber [fixed,fixed zerofill,precision=0]{1.0055832e1 })&\pgfmathprintnumber [fixed,fixed zerofill,precision=3]{1.52087026610e+00}(\pgfmathprintnumber [fixed,fixed zerofill,precision=0]{1.5132126e1 })&\pgfmathprintnumber [fixed,fixed zerofill,precision=3]{1.25453605068e+00}(\pgfmathprintnumber [fixed,fixed zerofill,precision=0]{2.4673264e1 })&\pgfmathprintnumber [fixed,fixed zerofill,precision=3]{1.20908652992e+00}(\pgfmathprintnumber [fixed,fixed zerofill,precision=0]{3.3574799e0 })&\pgfmathprintnumber [fixed,fixed zerofill,precision=3]{1.46016325385e+00}(\pgfmathprintnumber [fixed,fixed zerofill,precision=0]{9.3345322e0})&\pgfmathprintnumber [fixed,fixed zerofill,precision=3]{1.02736889137e+00}(\pgfmathprintnumber [fixed,fixed zerofill,precision=0]{1.474971e1})&\pgfmathprintnumber [fixed,fixed zerofill,precision=3]{7.85426868255e-01}(\pgfmathprintnumber [fixed,fixed zerofill,precision=0]{2.4099258e1})&\pgfmathprintnumber [fixed,fixed zerofill,precision=3]{7.58580178294e-01}(\pgfmathprintnumber [fixed,fixed zerofill,precision=0]{3.33629e0})\\%
$E^{--}$&\pgfmathprintnumber [fixed,fixed zerofill,precision=3]{1.56319249806e+00}(\pgfmathprintnumber [fixed,fixed zerofill,precision=0]{6.4352997e0 })&\pgfmathprintnumber [fixed,fixed zerofill,precision=3]{1.22656007476e+00}(\pgfmathprintnumber [fixed,fixed zerofill,precision=0]{6.6392105e0 })&\pgfmathprintnumber [fixed,fixed zerofill,precision=3]{1.08863590662e+00}(\pgfmathprintnumber [fixed,fixed zerofill,precision=0]{8.802327e0 })&\pgfmathprintnumber [fixed,fixed zerofill,precision=3]{1.00965921303e+00}(\pgfmathprintnumber [fixed,fixed zerofill,precision=0]{2.1778366e0 })&\pgfmathprintnumber [fixed,fixed zerofill,precision=3]{1.00590251230e+00}(\pgfmathprintnumber [fixed,fixed zerofill,precision=0]{6.0137238e0})&\pgfmathprintnumber [fixed,fixed zerofill,precision=3]{7.27305652979e-01}(\pgfmathprintnumber [fixed,fixed zerofill,precision=0]{6.2628525e0})&\pgfmathprintnumber [fixed,fixed zerofill,precision=3]{6.18624188598e-01}(\pgfmathprintnumber [fixed,fixed zerofill,precision=0]{8.2859222e0})&\pgfmathprintnumber [fixed,fixed zerofill,precision=3]{5.62914869935e-01}(\pgfmathprintnumber [fixed,fixed zerofill,precision=0]{4.5628113e0})\\\bottomrule %
\end {tabular}%

			}
			\caption{\label{tab:amRPC}Lattice gluelump masses $m^{e,s}_{\mathcal{R}^{PC}}\, a$ in units of the lattice spacing obtained from fits to effective mass plateaus (see Section~\ref{sec:masses_at_finite_a}). The row corresponding to the lightest gluelump with $\mathcal{R}^{PC} = T_1^{+-}$ is shaded in gray.}
		\end{table}

		% ********************
		
		\subsection{\label{app:gluelump_mass_splittings}Gluelump mass splittings for all ensembles and unsmeared and HYP2 smeared temporal links}
		
		\begin{table}[htb]
			\centering
			{\scriptsize
			\begin {tabular}{cllll}%
\toprule $\mathcal {R}^{PC}$&$\Delta m_{\mathcal {R}^{PC}}^{A,\text {none}}\, a$&$\Delta m_{\mathcal {R}^{PC}}^{B,\text {none}}\, a$&$\Delta m_{\mathcal {R}^{PC}}^{C,\text {none}}\, a$&$\Delta m_{\mathcal {R}^{PC}}^{D,\text {none}}\, a$\\\midrule %
$T_1^{++}$&\pgfmathprintnumber [fixed,fixed zerofill,precision=3]{8.12213142680e-01}(\pgfmathprintnumber [fixed,fixed zerofill,precision=0]{4.3739227e1 })&\pgfmathprintnumber [fixed,fixed zerofill,precision=3]{5.55428125900e-01}(\pgfmathprintnumber [fixed,fixed zerofill,precision=0]{1.6911026e1 })&\pgfmathprintnumber [fixed,fixed zerofill,precision=3]{4.79713358155e-01}(\pgfmathprintnumber [fixed,fixed zerofill,precision=0]{3.4990509e0 })&\pgfmathprintnumber [fixed,fixed zerofill,precision=3]{3.80954899557e-01}(\pgfmathprintnumber [fixed,fixed zerofill,precision=0]{6.3792923e0 })\\%
\rowcolor [gray]{1.0}$T_1^{+-}$&0&0&0&0\\%
$T_1^{-+}$&\pgfmathprintnumber [fixed,fixed zerofill,precision=3]{6.03766387180e-01}(\pgfmathprintnumber [fixed,fixed zerofill,precision=0]{6.7974838e0 })&\pgfmathprintnumber [fixed,fixed zerofill,precision=3]{3.86366093330e-01}(\pgfmathprintnumber [fixed,fixed zerofill,precision=0]{9.102156e0 })&\pgfmathprintnumber [fixed,fixed zerofill,precision=3]{3.20953200775e-01}(\pgfmathprintnumber [fixed,fixed zerofill,precision=0]{4.7533157e0 })&\pgfmathprintnumber [fixed,fixed zerofill,precision=3]{2.75575724007e-01}(\pgfmathprintnumber [fixed,fixed zerofill,precision=0]{3.1490524e0 })\\%
$T_1^{--}$&\pgfmathprintnumber [fixed,fixed zerofill,precision=3]{1.41637877530e-01}(\pgfmathprintnumber [fixed,fixed zerofill,precision=0]{8.7710892e0 })&\pgfmathprintnumber [fixed,fixed zerofill,precision=3]{1.17651970070e-01}(\pgfmathprintnumber [fixed,fixed zerofill,precision=0]{3.068689e0 })&\pgfmathprintnumber [fixed,fixed zerofill,precision=3]{9.14416526950e-02}(\pgfmathprintnumber [fixed,fixed zerofill,precision=0]{2.9121979e0 })&\pgfmathprintnumber [fixed,fixed zerofill,precision=3]{7.19490803190e-02}(\pgfmathprintnumber [fixed,fixed zerofill,precision=0]{2.4098251e0 })\\%
$T_2^{++}$&\pgfmathprintnumber [fixed,fixed zerofill,precision=3]{7.38932887910e-01}(\pgfmathprintnumber [fixed,fixed zerofill,precision=0]{9.1249283e0 })&\pgfmathprintnumber [fixed,fixed zerofill,precision=3]{4.81803605990e-01}(\pgfmathprintnumber [fixed,fixed zerofill,precision=0]{1.368869e1 })&\pgfmathprintnumber [fixed,fixed zerofill,precision=3]{4.11473718595e-01}(\pgfmathprintnumber [fixed,fixed zerofill,precision=0]{3.3943115e0 })&\pgfmathprintnumber [fixed,fixed zerofill,precision=3]{3.50016570707e-01}(\pgfmathprintnumber [fixed,fixed zerofill,precision=0]{3.9614456e0 })\\%
$T_2^{+-}$&\pgfmathprintnumber [fixed,fixed zerofill,precision=3]{4.03081513810e-01}(\pgfmathprintnumber [fixed,fixed zerofill,precision=0]{9.0630249e0 })&\pgfmathprintnumber [fixed,fixed zerofill,precision=3]{2.82272129240e-01}(\pgfmathprintnumber [fixed,fixed zerofill,precision=0]{4.9112549e0 })&\pgfmathprintnumber [fixed,fixed zerofill,precision=3]{2.26661639535e-01}(\pgfmathprintnumber [fixed,fixed zerofill,precision=0]{4.3833603e0 })&\pgfmathprintnumber [fixed,fixed zerofill,precision=3]{1.89472070587e-01}(\pgfmathprintnumber [fixed,fixed zerofill,precision=0]{4.0034332e0 })\\%
$T_2^{-+}$&\pgfmathprintnumber [fixed,fixed zerofill,precision=3]{6.98352644420e-01}(\pgfmathprintnumber [fixed,fixed zerofill,precision=0]{8.1565399e0 })&\pgfmathprintnumber [fixed,fixed zerofill,precision=3]{4.11380650260e-01}(\pgfmathprintnumber [fixed,fixed zerofill,precision=0]{2.6128876e1 })&\pgfmathprintnumber [fixed,fixed zerofill,precision=3]{3.80085712275e-01}(\pgfmathprintnumber [fixed,fixed zerofill,precision=0]{5.7492386e0 })&\pgfmathprintnumber [fixed,fixed zerofill,precision=3]{3.24579554767e-01}(\pgfmathprintnumber [fixed,fixed zerofill,precision=0]{3.6977936e0 })\\%
$T_2^{--}$&\pgfmathprintnumber [fixed,fixed zerofill,precision=3]{2.44395040350e-01}(\pgfmathprintnumber [fixed,fixed zerofill,precision=0]{2.5610565e0 })&\pgfmathprintnumber [fixed,fixed zerofill,precision=3]{1.32897486910e-01}(\pgfmathprintnumber [fixed,fixed zerofill,precision=0]{1.3619278e1 })&\pgfmathprintnumber [fixed,fixed zerofill,precision=3]{1.25094373515e-01}(\pgfmathprintnumber [fixed,fixed zerofill,precision=0]{2.480542e0 })&\pgfmathprintnumber [fixed,fixed zerofill,precision=3]{1.02926142287e-01}(\pgfmathprintnumber [fixed,fixed zerofill,precision=0]{2.2642532e0 })\\%
$A_1^{++}$&\pgfmathprintnumber [fixed,fixed zerofill,precision=3]{4.21224598460e-01}(\pgfmathprintnumber [fixed,fixed zerofill,precision=0]{7.5689133e0 })&\pgfmathprintnumber [fixed,fixed zerofill,precision=3]{2.92886029680e-01}(\pgfmathprintnumber [fixed,fixed zerofill,precision=0]{5.4207275e0 })&\pgfmathprintnumber [fixed,fixed zerofill,precision=3]{2.29762880005e-01}(\pgfmathprintnumber [fixed,fixed zerofill,precision=0]{5.3184906e0 })&\pgfmathprintnumber [fixed,fixed zerofill,precision=3]{2.00689950417e-01}(\pgfmathprintnumber [fixed,fixed zerofill,precision=0]{3.7604584e0 })\\%
$A_1^{+-}$&\pgfmathprintnumber [fixed,fixed zerofill,precision=3]{9.43867249660e-01}(\pgfmathprintnumber [fixed,fixed zerofill,precision=0]{2.7027771e1 })&\pgfmathprintnumber [fixed,fixed zerofill,precision=3]{6.70478106370e-01}(\pgfmathprintnumber [fixed,fixed zerofill,precision=0]{6.007367e0 })&\pgfmathprintnumber [fixed,fixed zerofill,precision=3]{5.14984774375e-01}(\pgfmathprintnumber [fixed,fixed zerofill,precision=0]{7.7106155e0 })&\pgfmathprintnumber [fixed,fixed zerofill,precision=3]{4.53329255877e-01}(\pgfmathprintnumber [fixed,fixed zerofill,precision=0]{4.4553696e0 })\\%
$A_1^{-+}$&\pgfmathprintnumber [fixed,fixed zerofill,precision=3]{8.27561463520e-01}(\pgfmathprintnumber [fixed,fixed zerofill,precision=0]{1.0785431e2 })&\pgfmathprintnumber [fixed,fixed zerofill,precision=3]{7.16086814150e-01}(\pgfmathprintnumber [fixed,fixed zerofill,precision=0]{5.7954163e0 })&\pgfmathprintnumber [fixed,fixed zerofill,precision=3]{5.80479326415e-01}(\pgfmathprintnumber [fixed,fixed zerofill,precision=0]{7.4143784e0 })&\pgfmathprintnumber [fixed,fixed zerofill,precision=3]{4.72384207637e-01}(\pgfmathprintnumber [fixed,fixed zerofill,precision=0]{8.6462799e0 })\\%
$A_1^{--}$&\pgfmathprintnumber [fixed,fixed zerofill,precision=3]{6.34095453610e-01}(\pgfmathprintnumber [fixed,fixed zerofill,precision=0]{1.0189804e1 })&\pgfmathprintnumber [fixed,fixed zerofill,precision=3]{3.37873947970e-01}(\pgfmathprintnumber [fixed,fixed zerofill,precision=0]{2.7286545e1 })&\pgfmathprintnumber [fixed,fixed zerofill,precision=3]{3.42643506895e-01}(\pgfmathprintnumber [fixed,fixed zerofill,precision=0]{4.2168823e0 })&\pgfmathprintnumber [fixed,fixed zerofill,precision=3]{3.12660645977e-01}(\pgfmathprintnumber [fixed,fixed zerofill,precision=0]{3.8022354e1 })\\%
$A_2^{++}$&\pgfmathprintnumber [fixed,fixed zerofill,precision=3]{1.01860814610e+00}(\pgfmathprintnumber [fixed,fixed zerofill,precision=0]{6.4950974e0 })&\pgfmathprintnumber [fixed,fixed zerofill,precision=3]{6.22382394480e-01}(\pgfmathprintnumber [fixed,fixed zerofill,precision=0]{1.5333038e1 })&\pgfmathprintnumber [fixed,fixed zerofill,precision=3]{5.29466809065e-01}(\pgfmathprintnumber [fixed,fixed zerofill,precision=0]{8.4685684e0 })&\pgfmathprintnumber [fixed,fixed zerofill,precision=3]{4.30121093967e-01}(\pgfmathprintnumber [fixed,fixed zerofill,precision=0]{7.517279e0 })\\%
$A_2^{+-}$&\pgfmathprintnumber [fixed,fixed zerofill,precision=3]{5.55300214490e-01}(\pgfmathprintnumber [fixed,fixed zerofill,precision=0]{7.9065231e0 })&\pgfmathprintnumber [fixed,fixed zerofill,precision=3]{2.93558017020e-01}(\pgfmathprintnumber [fixed,fixed zerofill,precision=0]{2.2339706e1 })&\pgfmathprintnumber [fixed,fixed zerofill,precision=3]{3.17255893965e-01}(\pgfmathprintnumber [fixed,fixed zerofill,precision=0]{6.01111e0 })&\pgfmathprintnumber [fixed,fixed zerofill,precision=3]{2.35323456667e-01}(\pgfmathprintnumber [fixed,fixed zerofill,precision=0]{1.7358841e1 })\\%
$A_2^{-+}$&\pgfmathprintnumber [fixed,fixed zerofill,precision=3]{7.37066339050e-01}(\pgfmathprintnumber [fixed,fixed zerofill,precision=0]{1.2568573e1 })&\pgfmathprintnumber [fixed,fixed zerofill,precision=3]{5.02106623770e-01}(\pgfmathprintnumber [fixed,fixed zerofill,precision=0]{4.1737854e0 })&\pgfmathprintnumber [fixed,fixed zerofill,precision=3]{3.79593635825e-01}(\pgfmathprintnumber [fixed,fixed zerofill,precision=0]{8.0719574e0 })&\pgfmathprintnumber [fixed,fixed zerofill,precision=3]{3.07573238347e-01}(\pgfmathprintnumber [fixed,fixed zerofill,precision=0]{8.8025223e0 })\\%
$A_2^{--}$&-&-&-&-\\%
$E^{++}$&\pgfmathprintnumber [fixed,fixed zerofill,precision=3]{5.84946013300e-01}(\pgfmathprintnumber [fixed,fixed zerofill,precision=0]{8.925888e0 })&\pgfmathprintnumber [fixed,fixed zerofill,precision=3]{3.99113674700e-01}(\pgfmathprintnumber [fixed,fixed zerofill,precision=0]{3.3842575e0 })&\pgfmathprintnumber [fixed,fixed zerofill,precision=3]{2.84489312985e-01}(\pgfmathprintnumber [fixed,fixed zerofill,precision=0]{1.1040405e1 })&\pgfmathprintnumber [fixed,fixed zerofill,precision=3]{2.64179690717e-01}(\pgfmathprintnumber [fixed,fixed zerofill,precision=0]{3.051622e0 })\\%
$E^{+-}$&\pgfmathprintnumber [fixed,fixed zerofill,precision=3]{3.93907347830e-01}(\pgfmathprintnumber [fixed,fixed zerofill,precision=0]{1.2157944e1 })&\pgfmathprintnumber [fixed,fixed zerofill,precision=3]{2.78741514450e-01}(\pgfmathprintnumber [fixed,fixed zerofill,precision=0]{3.7382202e0 })&\pgfmathprintnumber [fixed,fixed zerofill,precision=3]{2.09929076175e-01}(\pgfmathprintnumber [fixed,fixed zerofill,precision=0]{5.8229248e0 })&\pgfmathprintnumber [fixed,fixed zerofill,precision=3]{1.82889304137e-01}(\pgfmathprintnumber [fixed,fixed zerofill,precision=0]{2.615062e0 })\\%
$E^{-+}$&\pgfmathprintnumber [fixed,fixed zerofill,precision=3]{6.82368410240e-01}(\pgfmathprintnumber [fixed,fixed zerofill,precision=0]{9.9672012e0 })&\pgfmathprintnumber [fixed,fixed zerofill,precision=3]{4.43128780810e-01}(\pgfmathprintnumber [fixed,fixed zerofill,precision=0]{1.5120377e1 })&\pgfmathprintnumber [fixed,fixed zerofill,precision=3]{2.83526535555e-01}(\pgfmathprintnumber [fixed,fixed zerofill,precision=0]{2.4568741e1 })&\pgfmathprintnumber [fixed,fixed zerofill,precision=3]{3.11246414477e-01}(\pgfmathprintnumber [fixed,fixed zerofill,precision=0]{3.6968109e0 })\\%
$E^{--}$&\pgfmathprintnumber [fixed,fixed zerofill,precision=3]{2.31288032630e-01}(\pgfmathprintnumber [fixed,fixed zerofill,precision=0]{6.1061386e0 })&\pgfmathprintnumber [fixed,fixed zerofill,precision=3]{1.48818589470e-01}(\pgfmathprintnumber [fixed,fixed zerofill,precision=0]{6.2952408e0 })&\pgfmathprintnumber [fixed,fixed zerofill,precision=3]{1.17626391495e-01}(\pgfmathprintnumber [fixed,fixed zerofill,precision=0]{8.6443619e0 })&\pgfmathprintnumber [fixed,fixed zerofill,precision=3]{1.11819097587e-01}(\pgfmathprintnumber [fixed,fixed zerofill,precision=0]{2.4216217e0 })\\\bottomrule %
\end {tabular}%

			\begin {tabular}{cllll}%
\toprule $\mathcal {R}^{PC}$&$\Delta m_{\mathcal {R}^{PC}}^{A,{\text {HYP2}}}\, a$&$\Delta m_{\mathcal {R}^{PC}}^{B,{\text {HYP2}}}\, a$&$\Delta m_{\mathcal {R}^{PC}}^{C,{\text {HYP2}}}\, a$&$\Delta m_{\mathcal {R}^{PC}}^{D,{\text {HYP2}}}\, a$\\\midrule %
$T_1^{++}$&\pgfmathprintnumber [fixed,fixed zerofill,precision=3]{8.27127932410e-01}(\pgfmathprintnumber [fixed,fixed zerofill,precision=0]{3.9367004e1})&\pgfmathprintnumber [fixed,fixed zerofill,precision=3]{5.75330915527e-01}(\pgfmathprintnumber [fixed,fixed zerofill,precision=0]{6.7761093e0})&\pgfmathprintnumber [fixed,fixed zerofill,precision=3]{4.80178395278e-01}(\pgfmathprintnumber [fixed,fixed zerofill,precision=0]{3.4500885e0})&\pgfmathprintnumber [fixed,fixed zerofill,precision=3]{3.80713363012e-01}(\pgfmathprintnumber [fixed,fixed zerofill,precision=0]{6.1434998e0})\\%
\rowcolor [gray]{1.0}$T_1^{+-}$&0&0&0&0\\%
$T_1^{-+}$&\pgfmathprintnumber [fixed,fixed zerofill,precision=3]{6.07434378100e-01}(\pgfmathprintnumber [fixed,fixed zerofill,precision=0]{6.431987e0})&\pgfmathprintnumber [fixed,fixed zerofill,precision=3]{3.86726297354e-01}(\pgfmathprintnumber [fixed,fixed zerofill,precision=0]{8.7510818e0})&\pgfmathprintnumber [fixed,fixed zerofill,precision=3]{3.13375960745e-01}(\pgfmathprintnumber [fixed,fixed zerofill,precision=0]{8.0584e0})&\pgfmathprintnumber [fixed,fixed zerofill,precision=3]{2.51100876285e-01}(\pgfmathprintnumber [fixed,fixed zerofill,precision=0]{1.1828766e1})\\%
$T_1^{--}$&\pgfmathprintnumber [fixed,fixed zerofill,precision=3]{1.36302183412e-01}(\pgfmathprintnumber [fixed,fixed zerofill,precision=0]{1.6146957e1})&\pgfmathprintnumber [fixed,fixed zerofill,precision=3]{1.18558988351e-01}(\pgfmathprintnumber [fixed,fixed zerofill,precision=0]{2.9658997e0})&\pgfmathprintnumber [fixed,fixed zerofill,precision=3]{9.23959952140e-02}(\pgfmathprintnumber [fixed,fixed zerofill,precision=0]{2.8685913e0})&\pgfmathprintnumber [fixed,fixed zerofill,precision=3]{7.22309886160e-02}(\pgfmathprintnumber [fixed,fixed zerofill,precision=0]{2.324649e0})\\%
$T_2^{++}$&\pgfmathprintnumber [fixed,fixed zerofill,precision=3]{7.42509609640e-01}(\pgfmathprintnumber [fixed,fixed zerofill,precision=0]{8.8163193e0})&\pgfmathprintnumber [fixed,fixed zerofill,precision=3]{4.84424069427e-01}(\pgfmathprintnumber [fixed,fixed zerofill,precision=0]{1.3160034e1})&\pgfmathprintnumber [fixed,fixed zerofill,precision=3]{4.12168099651e-01}(\pgfmathprintnumber [fixed,fixed zerofill,precision=0]{3.3749908e0})&\pgfmathprintnumber [fixed,fixed zerofill,precision=3]{3.50158801169e-01}(\pgfmathprintnumber [fixed,fixed zerofill,precision=0]{3.8895798e0})\\%
$T_2^{+-}$&\pgfmathprintnumber [fixed,fixed zerofill,precision=3]{4.09991566580e-01}(\pgfmathprintnumber [fixed,fixed zerofill,precision=0]{8.0840286e0})&\pgfmathprintnumber [fixed,fixed zerofill,precision=3]{2.80725003005e-01}(\pgfmathprintnumber [fixed,fixed zerofill,precision=0]{4.6435669e0})&\pgfmathprintnumber [fixed,fixed zerofill,precision=3]{2.26600608501e-01}(\pgfmathprintnumber [fixed,fixed zerofill,precision=0]{4.0816422e0})&\pgfmathprintnumber [fixed,fixed zerofill,precision=3]{1.89427619275e-01}(\pgfmathprintnumber [fixed,fixed zerofill,precision=0]{3.8082443e0})\\%
$T_2^{-+}$&\pgfmathprintnumber [fixed,fixed zerofill,precision=3]{6.99198614350e-01}(\pgfmathprintnumber [fixed,fixed zerofill,precision=0]{7.7734116e0})&\pgfmathprintnumber [fixed,fixed zerofill,precision=3]{4.49168658997e-01}(\pgfmathprintnumber [fixed,fixed zerofill,precision=0]{1.1233185e1})&\pgfmathprintnumber [fixed,fixed zerofill,precision=3]{3.80034052351e-01}(\pgfmathprintnumber [fixed,fixed zerofill,precision=0]{5.6073364e0})&\pgfmathprintnumber [fixed,fixed zerofill,precision=3]{3.20746268684e-01}(\pgfmathprintnumber [fixed,fixed zerofill,precision=0]{5.8183502e0})\\%
$T_2^{--}$&\pgfmathprintnumber [fixed,fixed zerofill,precision=3]{2.48829060120e-01}(\pgfmathprintnumber [fixed,fixed zerofill,precision=0]{2.343341e0})&\pgfmathprintnumber [fixed,fixed zerofill,precision=3]{1.37590719813e-01}(\pgfmathprintnumber [fixed,fixed zerofill,precision=0]{1.1966202e1})&\pgfmathprintnumber [fixed,fixed zerofill,precision=3]{1.17125418333e-01}(\pgfmathprintnumber [fixed,fixed zerofill,precision=0]{5.3402908e0})&\pgfmathprintnumber [fixed,fixed zerofill,precision=3]{1.03651428228e-01}(\pgfmathprintnumber [fixed,fixed zerofill,precision=0]{2.2318436e0})\\%
$A_1^{++}$&\pgfmathprintnumber [fixed,fixed zerofill,precision=3]{4.23512392870e-01}(\pgfmathprintnumber [fixed,fixed zerofill,precision=0]{7.1256424e0})&\pgfmathprintnumber [fixed,fixed zerofill,precision=3]{2.93358334200e-01}(\pgfmathprintnumber [fixed,fixed zerofill,precision=0]{5.3584534e0})&\pgfmathprintnumber [fixed,fixed zerofill,precision=3]{2.30543401108e-01}(\pgfmathprintnumber [fixed,fixed zerofill,precision=0]{5.3484253e0})&\pgfmathprintnumber [fixed,fixed zerofill,precision=3]{2.00788495458e-01}(\pgfmathprintnumber [fixed,fixed zerofill,precision=0]{3.6414993e0})\\%
$A_1^{+-}$&\pgfmathprintnumber [fixed,fixed zerofill,precision=3]{9.47796271220e-01}(\pgfmathprintnumber [fixed,fixed zerofill,precision=0]{2.5577118e1})&\pgfmathprintnumber [fixed,fixed zerofill,precision=3]{6.71443254587e-01}(\pgfmathprintnumber [fixed,fixed zerofill,precision=0]{5.8674698e0})&\pgfmathprintnumber [fixed,fixed zerofill,precision=3]{5.17500425527e-01}(\pgfmathprintnumber [fixed,fixed zerofill,precision=0]{7.4447845e0})&\pgfmathprintnumber [fixed,fixed zerofill,precision=3]{4.27558483483e-01}(\pgfmathprintnumber [fixed,fixed zerofill,precision=0]{1.6330292e1})\\%
$A_1^{-+}$&\pgfmathprintnumber [fixed,fixed zerofill,precision=3]{1.00533243333e+00}(\pgfmathprintnumber [fixed,fixed zerofill,precision=0]{2.3818222e1})&\pgfmathprintnumber [fixed,fixed zerofill,precision=3]{6.95407452147e-01}(\pgfmathprintnumber [fixed,fixed zerofill,precision=0]{1.395874e1})&\pgfmathprintnumber [fixed,fixed zerofill,precision=3]{5.82127891127e-01}(\pgfmathprintnumber [fixed,fixed zerofill,precision=0]{7.1843689e0})&\pgfmathprintnumber [fixed,fixed zerofill,precision=3]{4.71293589911e-01}(\pgfmathprintnumber [fixed,fixed zerofill,precision=0]{8.5424591e0})\\%
$A_1^{--}$&\pgfmathprintnumber [fixed,fixed zerofill,precision=3]{6.36477535600e-01}(\pgfmathprintnumber [fixed,fixed zerofill,precision=0]{9.4584503e0})&\pgfmathprintnumber [fixed,fixed zerofill,precision=3]{3.84493626202e-01}(\pgfmathprintnumber [fixed,fixed zerofill,precision=0]{1.354338e1})&\pgfmathprintnumber [fixed,fixed zerofill,precision=3]{3.44155067940e-01}(\pgfmathprintnumber [fixed,fixed zerofill,precision=0]{4.1144394e0})&\pgfmathprintnumber [fixed,fixed zerofill,precision=3]{2.83065424371e-01}(\pgfmathprintnumber [fixed,fixed zerofill,precision=0]{3.4642624e0})\\%
$A_2^{++}$&\pgfmathprintnumber [fixed,fixed zerofill,precision=3]{1.02236337329e+00}(\pgfmathprintnumber [fixed,fixed zerofill,precision=0]{6.1990738e0})&\pgfmathprintnumber [fixed,fixed zerofill,precision=3]{6.22219875787e-01}(\pgfmathprintnumber [fixed,fixed zerofill,precision=0]{1.4242706e1})&\pgfmathprintnumber [fixed,fixed zerofill,precision=3]{5.29266938507e-01}(\pgfmathprintnumber [fixed,fixed zerofill,precision=0]{8.1693222e0})&\pgfmathprintnumber [fixed,fixed zerofill,precision=3]{4.30349294378e-01}(\pgfmathprintnumber [fixed,fixed zerofill,precision=0]{7.3886292e0})\\%
$A_2^{+-}$&\pgfmathprintnumber [fixed,fixed zerofill,precision=3]{5.35761062940e-01}(\pgfmathprintnumber [fixed,fixed zerofill,precision=0]{2.0216415e1})&\pgfmathprintnumber [fixed,fixed zerofill,precision=3]{3.01673406490e-01}(\pgfmathprintnumber [fixed,fixed zerofill,precision=0]{1.9866653e1})&\pgfmathprintnumber [fixed,fixed zerofill,precision=3]{2.95667767869e-01}(\pgfmathprintnumber [fixed,fixed zerofill,precision=0]{1.1409775e1})&\pgfmathprintnumber [fixed,fixed zerofill,precision=3]{2.28575841296e-01}(\pgfmathprintnumber [fixed,fixed zerofill,precision=0]{1.5466675e1})\\%
$A_2^{-+}$&\pgfmathprintnumber [fixed,fixed zerofill,precision=3]{7.41396427220e-01}(\pgfmathprintnumber [fixed,fixed zerofill,precision=0]{1.1484283e1})&\pgfmathprintnumber [fixed,fixed zerofill,precision=3]{4.89272154667e-01}(\pgfmathprintnumber [fixed,fixed zerofill,precision=0]{8.0241074e0})&\pgfmathprintnumber [fixed,fixed zerofill,precision=3]{3.79202395361e-01}(\pgfmathprintnumber [fixed,fixed zerofill,precision=0]{7.8760849e0})&\pgfmathprintnumber [fixed,fixed zerofill,precision=3]{3.09461066335e-01}(\pgfmathprintnumber [fixed,fixed zerofill,precision=0]{8.5339905e0})\\%
$A_2^{--}$&\pgfmathprintnumber [fixed,fixed zerofill,precision=3]{7.75620217270e-01}(\pgfmathprintnumber [fixed,fixed zerofill,precision=0]{9.1298813e1})&\pgfmathprintnumber [fixed,fixed zerofill,precision=3]{6.86638413237e-01}(\pgfmathprintnumber [fixed,fixed zerofill,precision=0]{1.448935e1})&\pgfmathprintnumber [fixed,fixed zerofill,precision=3]{5.59184970747e-01}(\pgfmathprintnumber [fixed,fixed zerofill,precision=0]{8.0318268e0})&\pgfmathprintnumber [fixed,fixed zerofill,precision=3]{4.62035897201e-01}(\pgfmathprintnumber [fixed,fixed zerofill,precision=0]{8.722055e0})\\%
$E^{++}$&\pgfmathprintnumber [fixed,fixed zerofill,precision=3]{5.88932428360e-01}(\pgfmathprintnumber [fixed,fixed zerofill,precision=0]{8.5486511e0})&\pgfmathprintnumber [fixed,fixed zerofill,precision=3]{3.97964129390e-01}(\pgfmathprintnumber [fixed,fixed zerofill,precision=0]{3.2921371e0})&\pgfmathprintnumber [fixed,fixed zerofill,precision=3]{2.85274611316e-01}(\pgfmathprintnumber [fixed,fixed zerofill,precision=0]{1.0919083e1})&\pgfmathprintnumber [fixed,fixed zerofill,precision=3]{2.63625291591e-01}(\pgfmathprintnumber [fixed,fixed zerofill,precision=0]{2.961203e0})\\%
$E^{+-}$&\pgfmathprintnumber [fixed,fixed zerofill,precision=3]{3.94281015210e-01}(\pgfmathprintnumber [fixed,fixed zerofill,precision=0]{1.0988876e1})&\pgfmathprintnumber [fixed,fixed zerofill,precision=3]{2.78735643997e-01}(\pgfmathprintnumber [fixed,fixed zerofill,precision=0]{3.6242249e0})&\pgfmathprintnumber [fixed,fixed zerofill,precision=3]{2.10230817358e-01}(\pgfmathprintnumber [fixed,fixed zerofill,precision=0]{5.6763824e0})&\pgfmathprintnumber [fixed,fixed zerofill,precision=3]{1.83038816881e-01}(\pgfmathprintnumber [fixed,fixed zerofill,precision=0]{2.5111694e0})\\%
$E^{-+}$&\pgfmathprintnumber [fixed,fixed zerofill,precision=3]{6.89649315660e-01}(\pgfmathprintnumber [fixed,fixed zerofill,precision=0]{9.1710938e0})&\pgfmathprintnumber [fixed,fixed zerofill,precision=3]{4.47745404157e-01}(\pgfmathprintnumber [fixed,fixed zerofill,precision=0]{1.4788086e1})&\pgfmathprintnumber [fixed,fixed zerofill,precision=3]{2.85850339412e-01}(\pgfmathprintnumber [fixed,fixed zerofill,precision=0]{2.3994934e1})&\pgfmathprintnumber [fixed,fixed zerofill,precision=3]{3.10989366661e-01}(\pgfmathprintnumber [fixed,fixed zerofill,precision=0]{3.6439056e0})\\%
$E^{--}$&\pgfmathprintnumber [fixed,fixed zerofill,precision=3]{2.35388574110e-01}(\pgfmathprintnumber [fixed,fixed zerofill,precision=0]{5.6065002e0})&\pgfmathprintnumber [fixed,fixed zerofill,precision=3]{1.47682165766e-01}(\pgfmathprintnumber [fixed,fixed zerofill,precision=0]{5.9574493e0})&\pgfmathprintnumber [fixed,fixed zerofill,precision=3]{1.19047659755e-01}(\pgfmathprintnumber [fixed,fixed zerofill,precision=0]{8.1148819e0})&\pgfmathprintnumber [fixed,fixed zerofill,precision=3]{1.15324058302e-01}(\pgfmathprintnumber [fixed,fixed zerofill,precision=0]{4.4013107e0})\\\bottomrule %
\end {tabular}%

			\caption{\label{tab:Delta amRPC}Gluelump mass splittings $\Delta m^{e,s}_{\mathcal{R}^{PC}}\, a$ in units of the lattice spacing obtained by subtracting the lattice gluelump masses from Table~\ref{tab:amRPC} (see Section~\ref{sec:continuumextrapolation_effmassfit}). $\Delta m^{e,s}_{T_1^{+-}} = 0$ by definition (see Eq.~\eqref{EQN102}), because we use $m^{e,s}_{T_1^{+-}}$ as reference mass.}}
		\end{table}

		\clearpage
		
		% ************************************************
		% ************************************************
		% ************************************************

	\bibliographystyle{utphys.bst}
	\bibliography{gluelumps.bib}

\providecommand{\href}[2]{#2}\begingroup\raggedright\begin{thebibliography}{10}

\bibitem{Berwein:2015vca}
M.~Berwein, N.~Brambilla, J.~Tarr\'us~Castell\`a, and A.~Vairo, ``{Quarkonium
  Hybrids with Nonrelativistic Effective Field Theories},''
  \href{http://dx.doi.org/10.1103/PhysRevD.92.114019}{{\em Phys. Rev. D}
  {\bfseries 92} no.~11, (2015) 114019},
  \href{http://arxiv.org/abs/1510.04299}{{\ttfamily arXiv:1510.04299
  [hep-ph]}}.

\bibitem{ATLAS:2019duq}
{\bfseries ATLAS} Collaboration, ``{Generation and Simulation of $R$-Hadrons in
  the ATLAS Experiment},''  {\bfseries SUSY 2019} (2019) .

\bibitem{Karl:1999wq}
G.~Karl and J.~E. Paton, ``{Gluelump spectrum in the bag model},''
  \href{http://dx.doi.org/10.1103/PhysRevD.60.034015}{{\em Phys. Rev. D}
  {\bfseries 60} (1999) 034015},
  \href{http://arxiv.org/abs/hep-ph/9904407}{{\ttfamily arXiv:hep-ph/9904407}}.

\bibitem{Guo:2007sm}
P.~Guo, A.~P. Szczepaniak, G.~Galata, A.~Vassallo, and E.~Santopinto,
  ``{Gluelump spectrum from Coulomb gauge QCD},''
  \href{http://dx.doi.org/10.1103/PhysRevD.77.056005}{{\em Phys. Rev. D}
  {\bfseries 77} (2008) 056005},
  \href{http://arxiv.org/abs/0707.3156}{{\ttfamily arXiv:0707.3156 [hep-ph]}}.

\bibitem{Buisseret:2008pd}
F.~Buisseret, ``{Gluelump model with transverse constituent gluons},''
  \href{http://dx.doi.org/10.1140/epja/i2008-10663-9}{{\em Eur. Phys. J. A}
  {\bfseries 38} (2008) 233--238},
  \href{http://arxiv.org/abs/0808.2399}{{\ttfamily arXiv:0808.2399 [hep-ph]}}.

\bibitem{Simonov:2000ky}
Y.~A. Simonov, ``{Gluelump spectrum in the QCD string model},''
  \href{http://dx.doi.org/10.1016/S0550-3213(00)00592-7}{{\em Nucl. Phys. B}
  {\bfseries 592} (2001) 350--368},
  \href{http://arxiv.org/abs/hep-ph/0003114}{{\ttfamily arXiv:hep-ph/0003114}}.

\bibitem{Mathieu:2005wc}
V.~Mathieu, C.~Semay, and F.~Brau, ``{Casimir scaling, glueballs and hybrid
  gluelumps},'' \href{http://dx.doi.org/10.1140/epja/i2005-10251-7}{{\em Eur.
  Phys. J. A} {\bfseries 27} (2006) 225--230},
  \href{http://arxiv.org/abs/hep-ph/0511210}{{\ttfamily arXiv:hep-ph/0511210}}.

\bibitem{Foster:1998wu}
{\bfseries UKQCD} Collaboration, M.~Foster and C.~Michael, ``{Hadrons with a
  heavy color adjoint particle},''
  \href{http://dx.doi.org/10.1103/PhysRevD.59.094509}{{\em Phys. Rev. D}
  {\bfseries 59} (1999) 094509},
  \href{http://arxiv.org/abs/hep-lat/9811010}{{\ttfamily
  arXiv:hep-lat/9811010}}.

\bibitem{Bali:2003jq}
G.~S. Bali and A.~Pineda, ``{QCD phenomenology of static sources and gluonic
  excitations at short distances},''
  \href{http://dx.doi.org/10.1103/PhysRevD.69.094001}{{\em Phys. Rev. D}
  {\bfseries 69} (2004) 094001},
  \href{http://arxiv.org/abs/hep-ph/0310130}{{\ttfamily arXiv:hep-ph/0310130}}.

\bibitem{Marsh:2013xsa}
K.~Marsh and R.~Lewis, ``{A lattice QCD study of generalized gluelumps},''
  \href{http://dx.doi.org/10.1103/PhysRevD.89.014502}{{\em Phys. Rev. D}
  {\bfseries 89} no.~1, (2014) 014502},
  \href{http://arxiv.org/abs/1309.1627}{{\ttfamily arXiv:1309.1627 [hep-lat]}}.

\bibitem{Capitani:2018rox}
S.~Capitani, O.~Philipsen, C.~Reisinger, C.~Riehl, and M.~Wagner, ``{Precision
  computation of hybrid static potentials in SU(3) lattice gauge theory},''
  \href{http://dx.doi.org/10.1103/PhysRevD.99.034502}{{\em Phys. Rev. D}
  {\bfseries 99} no.~3, (2019) 034502},
  \href{http://arxiv.org/abs/1811.11046}{{\ttfamily arXiv:1811.11046
  [hep-lat]}}.

\bibitem{Schlosser:2021wnr}
C.~Schlosser and M.~Wagner, ``{Hybrid static potentials in SU(3) lattice gauge
  theory at small quark-antiquark separations},''
  \href{http://dx.doi.org/10.1103/PhysRevD.105.054503}{{\em Phys. Rev. D}
  {\bfseries 105} no.~5, (2022) 054503},
  \href{http://arxiv.org/abs/2111.00741}{{\ttfamily arXiv:2111.00741
  [hep-lat]}}.

\bibitem{Jansen:2008si}
{\bfseries ETM} Collaboration, K.~Jansen, C.~Michael, A.~Shindler, and
  M.~Wagner, ``{The Static-light meson spectrum from twisted mass lattice
  QCD},'' \href{http://dx.doi.org/10.1088/1126-6708/2008/12/058}{{\em JHEP}
  {\bfseries 12} (2008) 058},
\href{http://arxiv.org/abs/0810.1843}{{\ttfamily arXiv:0810.1843 [hep-lat]}}.
%%CITATION = ARXIV:0810.1843;%%.

\bibitem{JH2022}
J.~Herr, ``{Computation of gluelump masses for hybrid static potentials in
  SU(3) lattice gauge theory using the multilevel algorithm},'' {\em MSc thesis
  at Goethe University Frankfurt} (2022) .

\bibitem{Philipsen:2014mra}
O.~Philipsen, C.~Pinke, A.~Sciarra, and M.~Bach, ``{CL$^2$QCD - Lattice QCD
  based on OpenCL},'' \href{http://dx.doi.org/10.22323/1.214.0038}{{\em PoS}
  {\bfseries LATTICE2014} (2014) 038},
  \href{http://arxiv.org/abs/1411.5219}{{\ttfamily arXiv:1411.5219 [hep-lat]}}.

\bibitem{Necco:2001xg}
S.~Necco and R.~Sommer, ``{The N(f) = 0 heavy quark potential from short to
  intermediate distances},''
  \href{http://dx.doi.org/10.1016/S0550-3213(01)00582-X}{{\em Nucl. Phys. B}
  {\bfseries 622} (2002) 328--346},
  \href{http://arxiv.org/abs/hep-lat/0108008}{{\ttfamily
  arXiv:hep-lat/0108008}}.

\bibitem{Luscher:2001up}
M.~L\"uscher and P.~Weisz, ``{Locality and exponential error reduction in
  numerical lattice gauge theory},''
  \href{http://dx.doi.org/10.1088/1126-6708/2001/09/010}{{\em JHEP} {\bfseries
  09} (2001) 010}, \href{http://arxiv.org/abs/hep-lat/0108014}{{\ttfamily
  arXiv:hep-lat/0108014}}.

\bibitem{Brambilla:2021wqs}
N.~Brambilla, V.~Leino, O.~Philipsen, C.~Reisinger, A.~Vairo, and M.~Wagner,
  ``{Lattice gauge theory computation of the static force},''
  \href{http://arxiv.org/abs/2106.01794}{{\ttfamily arXiv:2106.01794
  [hep-lat]}}.

\bibitem{Hasenfratz:2001tw}
A.~Hasenfratz, R.~Hoffmann, and F.~Knechtli, ``{The Static potential with
  hypercubic blocking},''
  \href{http://dx.doi.org/10.1016/S0920-5632(01)01733-9}{{\em Nucl. Phys. B
  Proc. Suppl.} {\bfseries 106} (2002) 418--420},
  \href{http://arxiv.org/abs/hep-lat/0110168}{{\ttfamily
  arXiv:hep-lat/0110168}}.

\bibitem{DellaMorte:2003mn}
{\bfseries ALPHA} Collaboration, M.~Della~Morte, S.~Durr, J.~Heitger, H.~Molke,
  J.~Rolf, A.~Shindler, and R.~Sommer, ``{Lattice HQET with exponentially
  improved statistical precision},''
  \href{http://dx.doi.org/10.1016/j.physletb.2005.03.017}{{\em Phys. Lett. B}
  {\bfseries 581} (2004) 93--98},
  \href{http://arxiv.org/abs/hep-lat/0307021}{{\ttfamily
  arXiv:hep-lat/0307021}}. [Erratum: Phys.Lett.B 612, 313--314 (2005)].

\bibitem{DellaMorte:2005nwx}
M.~Della~Morte, A.~Shindler, and R.~Sommer, ``{On lattice actions for static
  quarks},'' \href{http://dx.doi.org/10.1088/1126-6708/2005/08/051}{{\em JHEP}
  {\bfseries 08} (2005) 051},
  \href{http://arxiv.org/abs/hep-lat/0506008}{{\ttfamily
  arXiv:hep-lat/0506008}}.

\bibitem{Johnson:1982yq}
R.~C. Johnson, ``{Angular momentum on a lattice},''
  \href{http://dx.doi.org/10.1016/0370-2693(82)90134-4}{{\em Phys. Lett. B}
  {\bfseries 114} (1982) 147--151}.

\bibitem{Brambilla:2018pyn}
N.~Brambilla, W.~K. Lai, J.~Segovia, J.~Tarr\'us~Castell\`a, and A.~Vairo,
  ``{Spin structure of heavy-quark hybrids},''
  \href{http://dx.doi.org/10.1103/PhysRevD.99.014017}{{\em Phys. Rev. D}
  {\bfseries 99} no.~1, (2019) 014017},
  \href{http://arxiv.org/abs/1805.07713}{{\ttfamily arXiv:1805.07713
  [hep-ph]}}. [Erratum: Phys.Rev.D 101, 099902 (2020)].

\bibitem{Pineda:2002se}
A.~Pineda, ``{The Static potential: Lattice versus perturbation theory in a
  renormalon based approach},''
  \href{http://dx.doi.org/10.1088/0954-3899/29/2/313}{{\em J. Phys. G}
  {\bfseries 29} (2003) 371--385},
  \href{http://arxiv.org/abs/hep-ph/0208031}{{\ttfamily arXiv:hep-ph/0208031}}.

\bibitem{Bali:2013pla}
G.~S. Bali, C.~Bauer, A.~Pineda, and C.~Torrero, ``{Perturbative expansion of
  the energy of static sources at large orders in four-dimensional SU(3) gauge
  theory},'' \href{http://dx.doi.org/10.1103/PhysRevD.87.094517}{{\em Phys.
  Rev. D} {\bfseries 87} (2013) 094517},
  \href{http://arxiv.org/abs/1303.3279}{{\ttfamily arXiv:1303.3279 [hep-lat]}}.

\bibitem{Bali:2013qla}
G.~S. Bali, C.~Bauer, and A.~Pineda, ``{The static quark self-energy at
  $O(\alpha^{20})$ in perturbation theory},''
  \href{http://dx.doi.org/10.22323/1.187.0371}{{\em PoS} {\bfseries
  LATTICE2013} (2014) 371}, \href{http://arxiv.org/abs/1311.0114}{{\ttfamily
  arXiv:1311.0114 [hep-lat]}}.

\bibitem{Baikov:2016tgj}
P.~A. Baikov, K.~G. Chetyrkin, and J.~H. K\"uhn, ``{Five-Loop Running of the
  QCD coupling constant},''
  \href{http://dx.doi.org/10.1103/PhysRevLett.118.082002}{{\em Phys. Rev.
  Lett.} {\bfseries 118} no.~8, (2017) 082002},
  \href{http://arxiv.org/abs/1606.08659}{{\ttfamily arXiv:1606.08659
  [hep-ph]}}.

\bibitem{FlavourLatticeAveragingGroupFLAG:2021npn}
{\bfseries Flavour Lattice Averaging Group (FLAG)} Collaboration, Y.~Aoki {\em
  et~al.}, ``{FLAG Review 2021},''
  \href{http://dx.doi.org/10.1140/epjc/s10052-022-10536-1}{{\em Eur. Phys. J.
  C} {\bfseries 82} no.~10, (2022) 869},
  \href{http://arxiv.org/abs/2111.09849}{{\ttfamily arXiv:2111.09849
  [hep-lat]}}.

\end{thebibliography}\endgroup
\end{document}